\documentclass[aps,reprint,twocolumn,pra,groupedaddress,floatfix]{revtex4-1}
\usepackage{amsmath,amssymb,amsfonts}
\usepackage{mathrsfs}
\usepackage{stackengine}
\usepackage{soul}
\usepackage{graphicx}
\usepackage{color}
\begin{document}
\title{Tailoring inhomogeneous $\mathcal{PT}$-symmetric fiber Bragg grating spectra}
\author{S. Vignesh Raja}
\email{vickyneeshraja@gmail.com}
\author{A. Govindarajan}
\email{govin.nld@gmail.com}
\author{A. Mahalingam}
\email{drmaha@annauniv.edu}
\author{M. Lakshmanan}
\email{lakshman.cnld@gmail.com}
\affiliation{$^{*,\ddagger}$Department of Physics, Anna University, Chennai - 600 025, India}
\affiliation{$^{\dagger,\mathsection}$Centre for Nonlinear Dynamics, School of Physics, Bharathidasan University, Tiruchirappalli - 620 024, India}
\begin{abstract}
The unique spectral behavior exhibited by a class of non-uniform Bragg periodic structures, namely chirped and apodized fiber Bragg gratings (FBGs) influenced by parity and time reversal ($\mathcal{PT}$) symmetry, is presented.  The interplay between the $\mathcal{PT}$-symmetry and nonuniformities brings exceptional functionalities in the broken $\mathcal{PT}$-symmetric phase such as wavelength selective amplification and single-mode lasing for a wide range of variations in gain-loss.  We observe that the device is no more passive and it undergoes a series of transitions from asymmetric reflection to unidirectional invisibility and multi-mode amplification as a consequence of variation in the imaginary part of the strength of modulation in different apodization profiles, namely Gaussian and raised cosine, at the given value of chirping. The chirping affords bandwidth control as well as control over the magnitude of the reflected (transmitted) light. Likewise, apodization offers additional functionality in the form of suppression of uncontrolled lasing behavior in the broken $\mathcal{PT}$-symmetric regime besides moderating the reflected signals outside the band edges of the spectra. 
	\end{abstract}
	\maketitle
\section{Introduction}
The possibility of inscribing newly engineered Bragg structures in the core of an optical fiber never seems to get exhausted despite a large number of scientific contributions dealing with their fundamental aspects (see \cite{erdogan1997fiber,giles1997lightwave,hill1997fiber,othonos1997fiber} and references therein).
  From an application point of view, fiber Bragg gratings have emerged as an inevitable component in modern telecommunication arena as their spectral characteristics can be tailor-made at ease to facilitate functionalities like spectral filtering, wavelength routing \cite{hill1997fiber}, gain equalizers \cite{othonos1997fiber}, pulse compression \cite{litchinitser1997fiber}, optical add-drop multiplexing \cite{baumann1996compact}, optical delay lines \cite{lenz2001optical}, and so forth.

The existence of group delay ripples and significant reflectivity at the side lobes of the spectra are some of the major concerns associated with uniform gratings which potentially lead to other detrimental effects such as cross talk and effective bandwidth reduction in dense wavelength division multiplexing (DWDM) systems \cite{erdogan1997fiber}. A uniform FBG features abrupt change in the modulation index at the interfaces between the grating and the background material. Hence, the boundaries of such FBGs behave similar to a Fabry-P\'{e}rot like cavity and thus resulting in the above detrimental effects \cite{erdogan1997fiber,litchinitser1997fiber,rebola2002performance,ennser1998optimization}. By employing an apodization technique, the index modulation is allowed to vanish smoothly at each of the interfaces rather than varying it in an abrupt way in contrast to the uniform gratings \cite{litchinitser1997fiber}. As a result, reflections in the side lobes are severely reduced \cite{rebola2002performance} and hence  the overall spectral response gets highly improved \cite{giles1997lightwave,  komukai1997efficient}. On the other hand, chirping customizes the dispersive characteristics of the propagating light by broadening the pulse profile. In a chirped FBG, the Bragg wavelength of the system varies as a function of the propagation distance ($z$) \cite{ouellette1987dispersion} as a consequence of the variation in the grating period with respect to position $z$ \cite{othonos1997fiber, erdogan1997fiber}.   The incoming  optical field experiences an enhancement or  reduction in the spatial frequency of the grating as in the case of positively and negatively chirped FBGs, respectively \cite{maywar1998effect}. Dispersion of a chirped grating without apodization becomes detrimental on high-bit rate DWDM systems due to the presence of ripples along the spectra.  A mix of suitable apodization profile and judiciously tailored chirping parameter into a single FBG can blend these constructive characteristics and thus deliver a bandwidth controlled smooth spectrum \cite{pastor1996design, ennser1998optimization}. Hence chirped and apodized FBGs are widely used in fiber lasers \cite{cheng2008single} and dispersion compensation applications \cite{hill1994chirped,eggleton1997implications}. 

The $\mathcal{PT}$-symmetric notion brings about some remarkable optical behaviors in a fiber Bragg grating which include coexistence of perfect absorption and lasing \cite{huang2014type}, unidirectional invisibility \cite{lin2011unidirectional}, nonreciprocal light propagation dynamics \cite{kulishov2005nonreciprocal} and so on. The  inherent loss in an optical system can thus be beneficial rather than being catastrophic with the inclusion of $\mathcal{PT}$-symmetric potentials \cite{govindarajan2018tailoring,lupu2013switching, Govindarajan:19}. It has also paved the way for some fascinating works which essentially confirm that supplying an appropriate amount of loss is necessary to preserve the $\mathcal{PT}$-symmetric condition $n (z) = n^* (-z)$ \cite{hahn2016unidirectional, razzari2012optics}. In the literature, there are two distinct directions in which the studies on $\mathcal{PT}$-symmetric systems are carried out. Primarily, the first one is focused on studying the dynamics of the system under various physical conditions (see \cite{ozdemir2019parity,el2007theory,el2018non,kottos2010optical,longhi2018parity,makris2008beam,ruter2010observation, sukhorukov2010nonlinear, bludov2013stable, miroshnichenko2011nonlinearly, ramezani2012bypassing} and references therein). Studies on applications of $\mathcal{PT}$-symmetric system is the other direction of interest \cite{Phang:13, Phang:14, Phang2015, baum2015parity, PhysRevA.100.033838, PhysRevA.100.053806}. Invoking the notion of $\mathcal{PT}$ symmetry in traditional structures is considered to be a trending subject matter in optics as they are more fruitful than the conventional systems owing to the management of intrinsic loss of the materials rather than neglecting it. As Lupu \emph{et al.} have pointed out, the notion of $\mathcal{PT}$-symmetry opens up an alternative route to overcome some of the critical problems prevailing in the current hybrid integration optical technologies to build tunable and reconfigurable devices \cite{PhysRevA.91.053825}. In the context of nonuniform $\mathcal{PT}$-symmetric FBG, incorporation of the apodization techniques with the aid of duty cycle methods was proposed recently \cite{lupu2016tailoring}.  It is well known that apodization without chirping might lead to undesirable truncation of the spectral response which in turn reduces the effective length of the grating. Also, the nonuniform dc index change offered by an apodized FBG without chirping is highly undesirable since it can induce crosstalk between adjacent channels of  closely spaced DWDM systems. If the grating period is allowed to vary along the propagation direction (chirping) in the presence of apodization then it is possible to obtain strong sidelobe suppression, 
flat-top amplitude response, and low delay distortion which are made possible by employing a chirped and apodized FBG. As of now, there seem to exist no scientific contributions which analyzes the effects of different apodization profiles on the spectral features of  $\mathcal{PT}$-symmetric chirped FBGs. Having stated the need to study the proposed system, we aim at exploring the multi-functional capabilities  of the system such as directional dependent broadband dispersion compensator, optical delay lines and optical demultiplexer in the unbroken $\mathcal{PT}$-symmetric regime, tunable mode selective lasing in the broken $\mathcal{PT}$-symmetric regime.   To set up such applications, we first examine  typical grating characteristics such as reflection, transmission, group delay and dispersion in a chirped $\mathcal{PT}$-symmetric FBG in the presence of Gaussian and raised-cosine apodization profiles.

The paper is structured as follows. Section \ref{Sec:2} illustrates the mathematical model of the system. Sections \ref{Sec:3} and \ref{Sec:4} present the light propagation characteristics and applications in the unbroken $\mathcal{PT}$-symmetric regime, respectively. Section \ref{Sec:5} demonstrates the lasing behavior in the broken $\mathcal{PT}$-symmetric regime. Section \ref{Sec:6} brings out the exceptional point dynamics of the system. 
The article is concluded in Sec. \ref{Sec:8}.
\section{mathematical model}

\begin{figure}[t]
	\centering
	\includegraphics[width=1\linewidth]{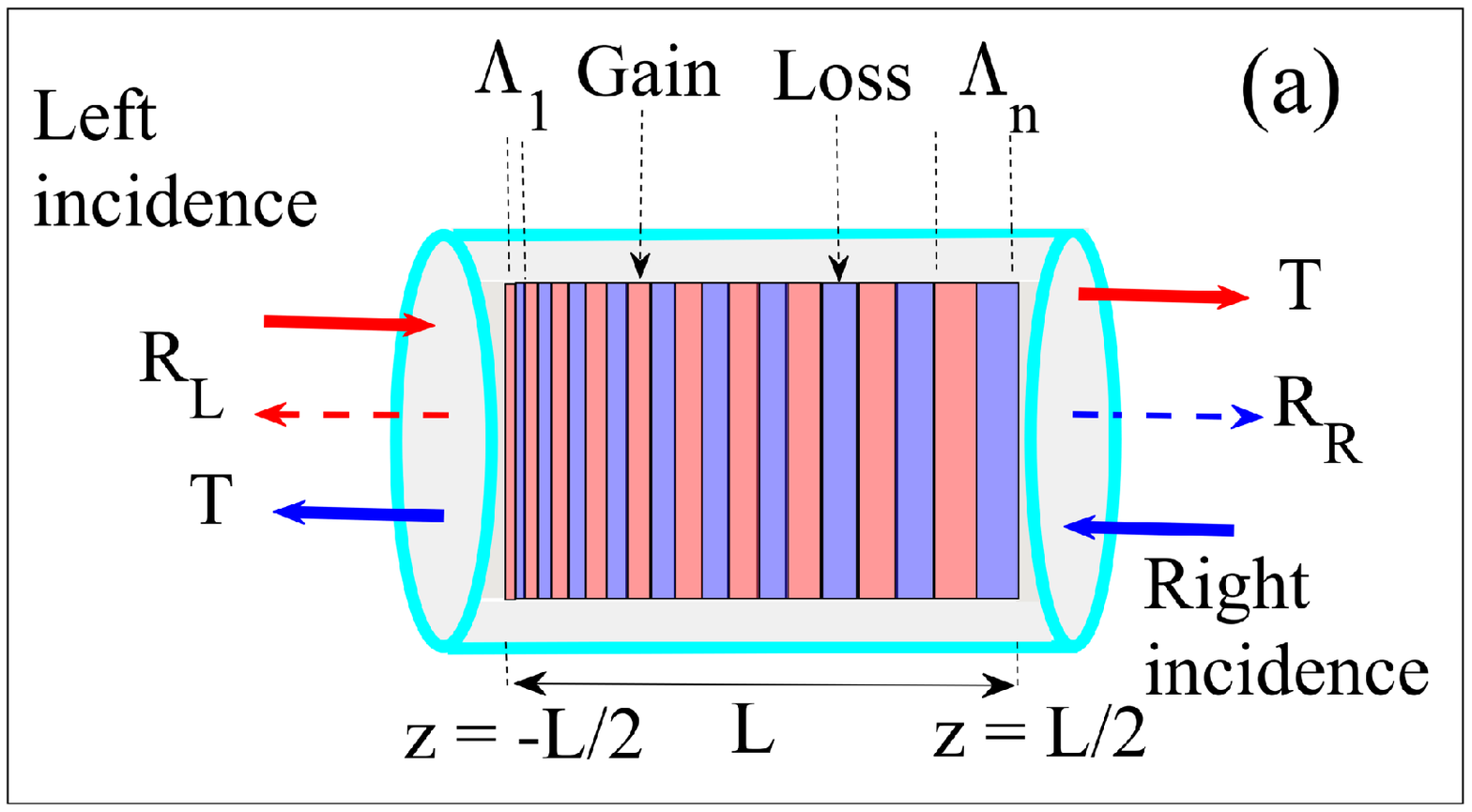}\\\includegraphics[width=1\linewidth]{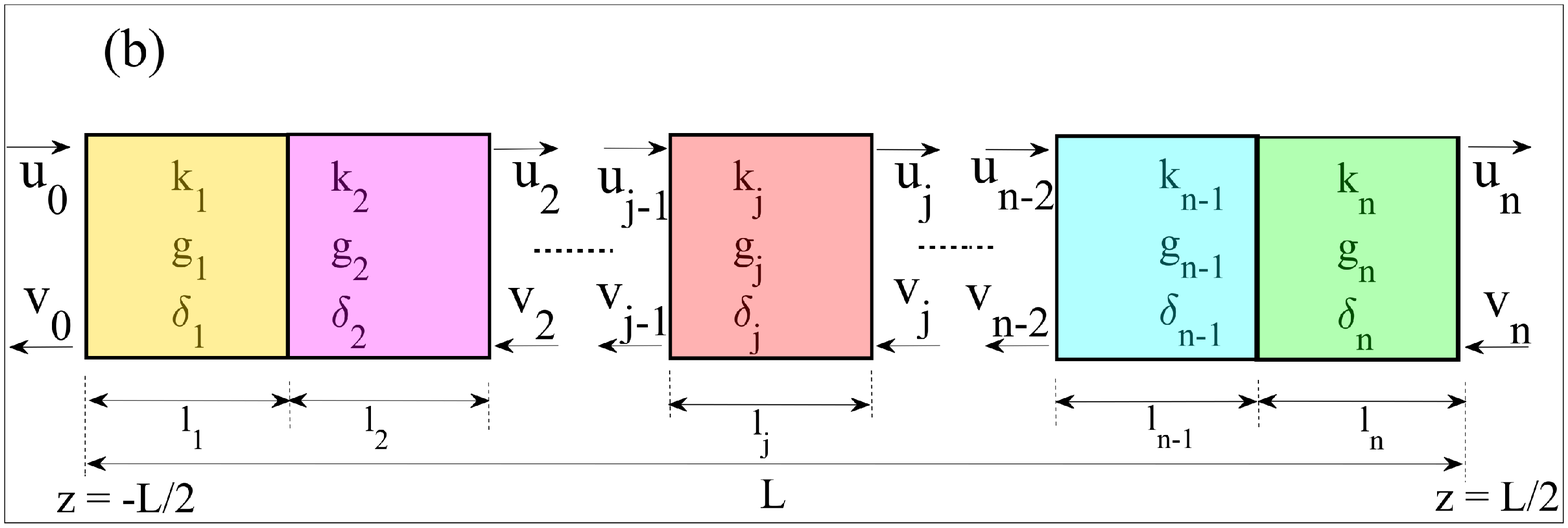}\\\includegraphics[width=0.75\linewidth]{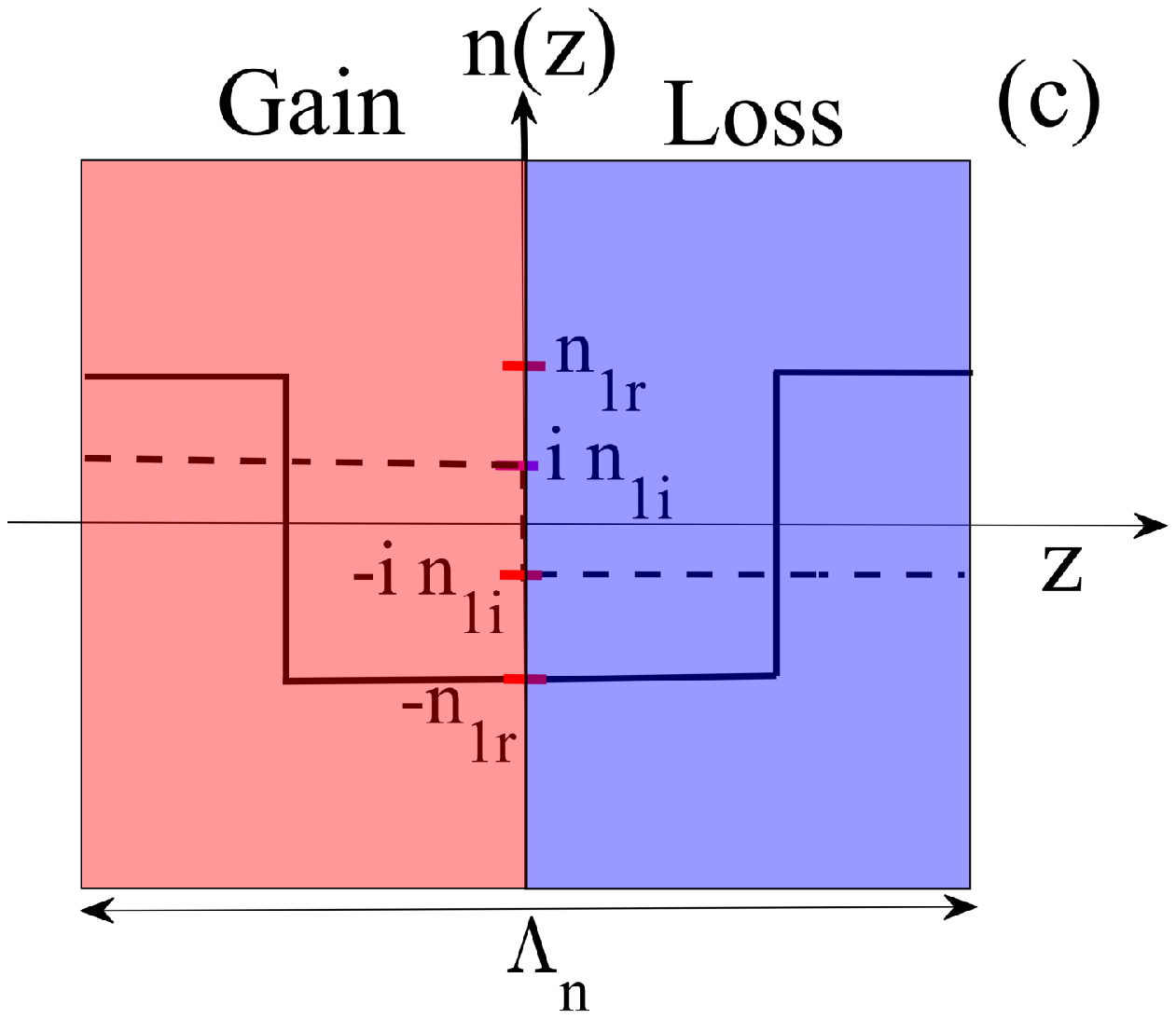}
	\caption{(a) Schematic of a chirped $\mathcal{PT}$-symmetric FBG (b) Segmentation of the nonuniform $\mathcal{PT}$-symmetric FBG into $j$ number of uniform and periodic sections. The length, period, coupling, gain-loss, detuning parameter are given by $l_j$, $\lambda_j$, $k_j$, $g_j$, $\delta_j$. The input, transmitted, reflected signal at a section $j$ are represented by $u_{j-1}$, $u_j$, and $v_{j-1}$. (c) Represents one unit cell of the $\mathcal{PT}$-FBG \cite{Phang:14}, where the solid and dotted lines represent the modulation of real ($n_{1r}$) and imaginary part ($n_{1i}$)  of refractive index profile, respectively.}
	\label{fig0}    
\end{figure}

\label{Sec:2}
The refractive index profile of a chirped and apodized $\mathcal{PT}$-symmetric fiber Bragg grating (CAPT-FBG) of length $L$ and grating period $\Lambda$ (see Fig. \ref{fig0}) is written as \cite{lupu2016tailoring, miri2012bragg}
\begin{align}
\nonumber n(z)=n_{0}+ n_{1R}\cos\left(\frac{2\pi}{\Lambda}z+\phi(z)\right) f(z)\\+  in_{1I}\sin\left(\frac{2\pi}{\Lambda}z+\phi(z)\right) f(z) \label{Eq:Norm1}
\end{align}

In Eq. (\ref{Eq:Norm1}) the first term on the right hand side represents the refractive index of the background material, $n_{1R}$ and $n_{1I}$ stand for the real and imaginary parts of modulation strength of Bragg grating, respectively. Also, $\phi(z)$ represents the slowly varying grating phase  \cite{ouellette1987dispersion,ennser1998optimization} and it should be even function of $z$ for the system to be $\mathcal{PT}$-symmetric and $f(z)$ denotes the apodization profile \cite{lupu2016tailoring}. Though the chirping is responsible to make the FBG non-uniform in addition to the apodization by breaking the periodicity, it should be noted that each unit cell will certainly obey the $\mathcal{PT}$-symmetric condition ($n(z)=n^*(-z)$) locally for a given grating period ($\Lambda_n$) as shown in Fig. \ref{fig0}(c). The mathematical form of the Gaussian apodization profile is given by \cite{erdogan1997fiber, rebola2002performance}
 \begin{align}
f(z)=\exp \left[-G(z/L)^2\right]
 \label{Eq:Norm2}
 \end{align}
 where G stands for the Gauss width parameter \cite{erdogan1997fiber,pastor1996design,8345713,rebola2002performance}.
 
 For a raised-cosine profile one has the representation \cite{lupu2016tailoring, erdogan1997fiber} 
 \begin{align}
 f(z)= \cfrac{1+\cos\left(\pi z/L\right)}{2}
  \label{Eq:Norm3}
 \end{align}

For an unapodized grating $f (z) = 1$. The coupled mode equations which describe the dynamics of the optical fields evolving from the proposed system is found to be \cite{lupu2016tailoring,kulishov2005nonreciprocal}
\begin{gather}
+i\frac{du}{dz}+\delta(z) u+\left(k(z)+g(z)\right)v=0,\\
-i\frac{dv}{dz}+\delta(z) v+\left(k(z)-g(z)\right)u=0,
\label{Eq:Norm4}
\end{gather}
where $u$ and $v$ are the forward and backward propagating optical fields inside the grating. The coupled mode equations are solved by conventional transfer matrix method (piece-wise uniform routine) having $n$ number of uniform sections (see Fig. \ref{fig0}). The coupling, gain/loss and the detuning parameters accordingly are defined by \cite{erdogan1997fiber,miri2012bragg}
\begin{gather} \nonumber k(z)=\pi n_{1R}(z)/\lambda, \quad g(z)=\pi n_{1I}(z)/\lambda,\\ \delta(z)=\delta_{0}-\cfrac{1}{2}
\cfrac{d\phi}{dz} =\delta_{0}+\frac{4 \pi n_{\mathrm{0}} z}{\lambda_{b}^{2}} \frac{d \lambda_{b}}{d z}
\label{Eq:Norm5}
\end{gather}
 where $\delta_{0}=2 \pi n_{0}\left(\cfrac{1}{\lambda}-\cfrac{1}{\lambda_{b}}\right)$. The derivative term  $d \lambda_{b}/d z$ represents the variation in the Bragg wavelength ($\lambda_{b}=2n_{0}\Lambda (z)$) proportional to grating position or simply chirp (C), whereas $\lambda$ symbolizes the operating wavelength. 
 
Let $j=1, 2,3...,n$ represents the $j^{th}$ section of the grating with length and period of each section taken to be $l_j$ and $\lambda_j$ respectively, such that $\sum _{j=1}^{n}  l_j = L$. Hence, the transmitted and reflected amplitudes  of the traversing field at any section $j$ (after propagating $j-1$ sections) are taken to be $u_j$ and $v_j$, respectively \cite{erdogan1997fiber}. Assuming $u_0=1$ and $v_0=0$ and taking $M_j$ to be the transfer matrix that relates the propagation through an individual section $j$ with the previous section $j-1$, it can be mathematically expressed as
 \begin{gather}
 \left[\begin{array}{c}
 	u_j\\
 	v_j
 \end{array}\right]=M_j\left[\begin{array}{c}
 u_{j-1}\\
 v_{j-1}
 \end{array}\right].
 \label{Eq:Norm6}
 \end{gather}
 For a $\mathcal{PT}$-symmetric FBG, $M_j$ is given by Eq. (\ref{Eq:Norm6}) with

 \begin{gather}
 \left[M_j\right]=\left[\begin{array}{cc}
   m_{11} & m_{12}\\
   m_{21} & m_{22}
 \end{array}\right]
 \label{Eq:Norm7}
 \end{gather}
 
 where the matrix elements of $M_j$ are given by 
\begin{gather}
\nonumber m_{11} = m_{22}^* =\cosh(\hat{\sigma_j} l_j )+i\left(\cfrac{\delta_j}{\hat{\sigma_j}}\right)\sinh(\hat{\sigma_j} l_j),
\\\nonumber m_{12}=i\left(\cfrac{k_j+g_j}{\hat{\sigma_j}}\right)\sinh(\hat{\sigma_j} l_j),
\\\nonumber m_{21}=-i\left(\cfrac{k_j-g_j}{\hat{\sigma_j}}\right)\sinh(\hat{\sigma_j} l_j),
\label{Eq:Norm8}
\end{gather}
where
$\hat{\sigma_j}=\sqrt{\left(k_j^2-g_j^2-\delta_j^2\right)}$ represents the eigenvalues of the matrix $M_j$ and the local values of coupling, gain/loss, and detuning parameters in the $j^{th}$ section are given by $k_j$, $g_j$, $\delta_j$. The transfer matrix $M$ which describes the output fields of the grating of length $L$ is the multiplication of matrices of the $n$ individual sections each having a length {$l_j$}. The output field in terms of the incident field is written as
\begin{gather}
\left[\begin{array}{c}
u_n\\
v_n
\end{array}\right]=M\left[\begin{array}{c}
u_0\\
v_0
\end{array}\right],
\label{Eq:Norm9}
\end{gather}
where $M=\left[\begin{array}{cc}
M_{11} & M_{12}\\
M_{21} & M_{22}
\end{array}\right]=M_n*M_{n-1}*.....M_j*....M_1. $

The reflection amplitudes of the grating for the left, right incidence, and transmission amplitudes are specified as \cite{lin2011unidirectional}
\begin{gather}
%$R=\left|\cfrac{v(0)}{u_{0}}\right|^2
\nonumber r_L=- M_{21}/M_{22},\quad r_R=M_{12}/M_{22},\\
t_L =t_R = t = 1/M_{22}.
\label{Eq:Norm10}
\end{gather}

The corresponding reflection and transmission coefficients read as
\begin{gather}
    R_{L}=|r_{L}|^2, \quad R_{R} = |r_{R}|^2, \quad T=|t|^{2}.
\end{gather}

It is important to note that the $\mathcal{PT}$-symmetric structures possess real eigenvalues below a phase-transition point and so they can be regarded as closed systems \cite{bender1998real}. Nevertheless, they function like open systems at the broken $\mathcal{PT}$-symmetric phase as a consequence of  the complex nature of the eigenvalues of the non-Hermitian Hamiltonians  \cite{ozdemir2019parity, mostafazadeh2002pseudo}. Hence they do not obey the conservation relation of the form $T + R = 1$ \cite{mostafazadeh2013invisibility}. But the generalized conservation relation for the $\mathcal{PT}$-symmetric structures is given by $|T-1|=\sqrt{R_{L} R_{R}}$ \cite{ge2012conservation}. 

Further, we like to emphasize on the group delay and dispersion characteristics of the system. The phase ($\theta$) of the reflected and transmitted fields which dictates the group delay ($\tau$) and dispersion characteristics (D) of the $\mathcal{PT}$-symmetric FBG is given by

\begin{gather}
\nonumber\theta_{R_L} = \arctan(-M_{21}/M_{22}), \quad\theta_{R_R} = \arctan(M_{12}/M_{22}), \\ \theta_{T} = \arctan(1/M_{22}).
\label{Eq:Norm12}
\end{gather}

The time delay offered by the group velocity  ($v_g$) of the propagating optical fields can be found if the phase given in Eq. (\ref{Eq:Norm12}) is further expanded around a local value of frequency, say $\omega_0$. The gradient of ${d \theta}/{d \omega}$ is the group delay which is proportional to the operating frequency ($\omega$) and it reads as \cite{erdogan1997fiber}

\begin{gather}
\tau_{L,R,T} = d \theta/d \omega = -\frac{\lambda^{2}}{2 \pi c} \frac{d \theta_{R_L,R_R,T}}{d \lambda},
\label{Eq:Norm13}
\end{gather}

where c represents the speed of light in free space. The dispersion offered by the system is proportional to the derivative of delay ($\tau$) with respect to the operating wavelength ($\lambda$) and it can also be expressed in terms of the second derivative of the phase with respect to the frequency ($d^2 \theta$/$d\omega^2$) as

\begin{gather}
D_{L,R,T} = -\frac{2 \pi c}{\lambda^{2}} \frac{d^{2} \theta}{d \omega^{2}} = \frac{d \tau_{L,R,T}}{d \lambda}.
\label{Eq:Norm14}
\end{gather} 

Note that the delay and dispersion are measured in terms of picoseconds and picoseconds/nanometer, respectively. We may also mention that, works dealing with the time delay characteristics in a $\mathcal{PT}$-symmetric system are not many \cite{lin2011unidirectional, kulishov2005nonreciprocal}. Even these contributions are limited to the uniform grating structures. We believe that the  present work is the first work pertaining to the investigation of delay and dispersion features of a nonuniform $\mathcal{PT}$-symmetric FBG. In our studies, we will first investigate the dynamics of the nonuniform $\mathcal{PT}$-symmetric FBG in the unbroken regime followed by the dynamics in the broken $\mathcal{PT}$-symmetric regime and finally the dynamics at the exceptional point. 

\section{Effect of nonuniformities on the spectra of an unbroken $\mathcal{PT}$-symmetric FBG}
\label{Sec:3}
It is to be remembered that the device works in unbroken $\mathcal{PT}$-symmetric regime if it satisfies the condition $n_{1R}>n_{1I}$.  Throughout the analysis, we use the following system parameters of a standard FBG \cite{erdogan1997fiber} as $L = 10$ mm, $n_0 = 1.45 $, $n_{1R}= 10^{-3}$, and $\lambda_b = 1550$ nm. Before we proceed to study the grating characteristics of the proposed system in detail, it is important to investigate the effect of chirping, apodization, and $\mathcal{PT}$-symmetry individually. In the absence of $\mathcal{PT}$-symmetry ($n_{1I} = 0$) the reflection coefficient (R) is found to be the same for both left ($R_L$) and right  light incidence ($R_R$) and the relation between R and transmission (T) is given by $T + R = 1$ as shown in Figs. \ref{fig1}(a), \ref{fig1}(c), and \ref{fig1}(e). The effect of $\mathcal{PT}$-symmetry on the linear spectra is well-known and that it gives rise to enhanced and reduced reflection for right and left light incidence, respectively, as seen in Figs. \ref{fig1}(b), \ref{fig1}(d), and \ref{fig1}(f). These conclusions are true irrespective of the nature of apodization profile and the chirping. From Figs. \ref{fig1}(a) and \ref{fig1}(b), one can visualize that in the absence of apodization, chirping produces unwanted ripples in the stopband of the grating. Also, it induces a stronger reflection in the sidelobes. In the presence of Gaussian apodization, the reflection (transmission) in the side lobes is reduced but not suppressed completely as seen in Figs. \ref{fig1}(c) and \ref{fig1}(d). However the ripples in the stopband are strongly reduced. To obtain a smooth, ripple-less, and side lobe suppressed spectra one can adapt the raised-cosine apodization as illustrated in Figs. \ref{fig1}(e), and \ref{fig1}(f).
\subsection{Gaussian apodization profile}
\begin{figure}[t]
	\centering
	\includegraphics[width=0.5\linewidth]{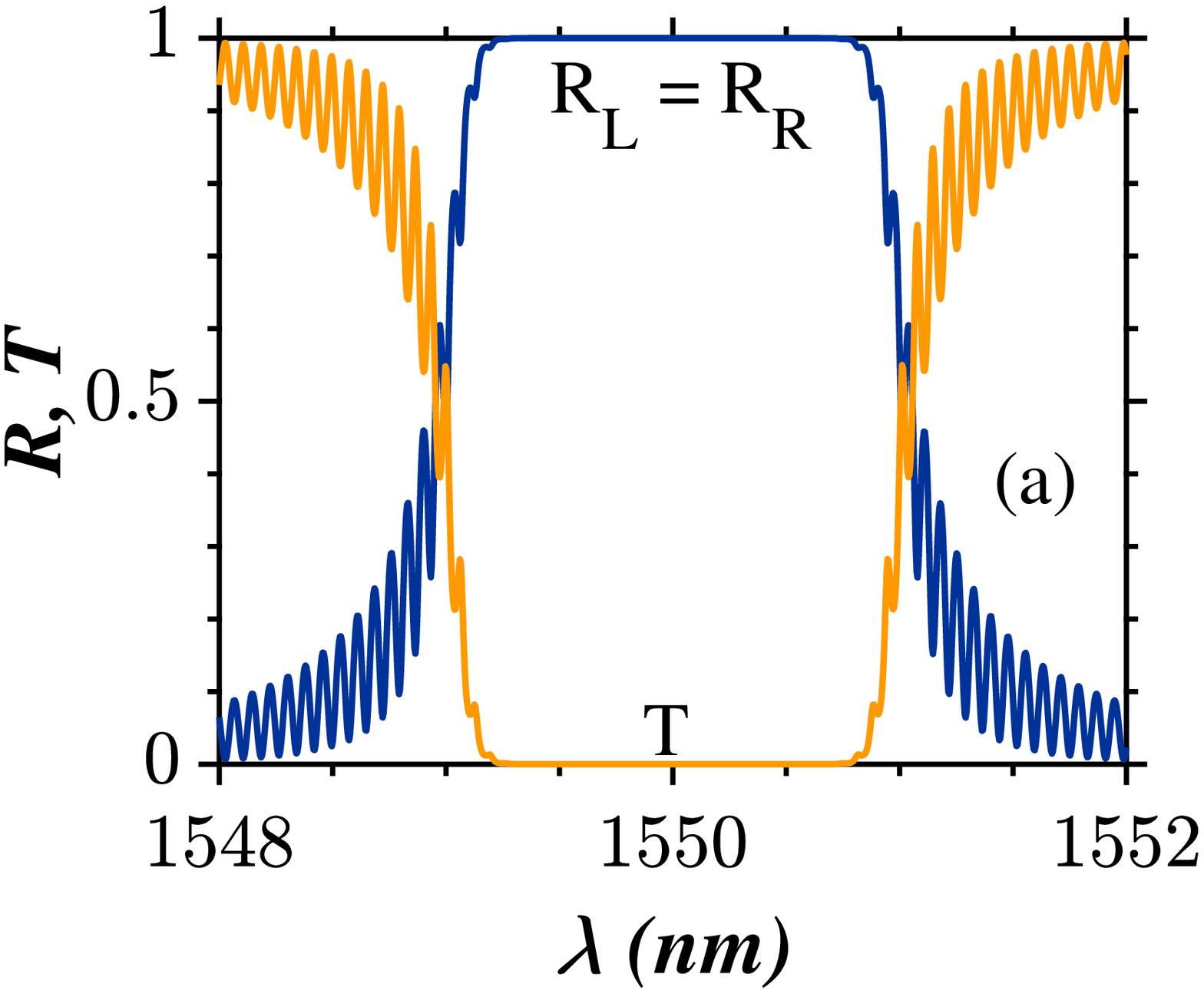}\includegraphics[width=0.5\linewidth]{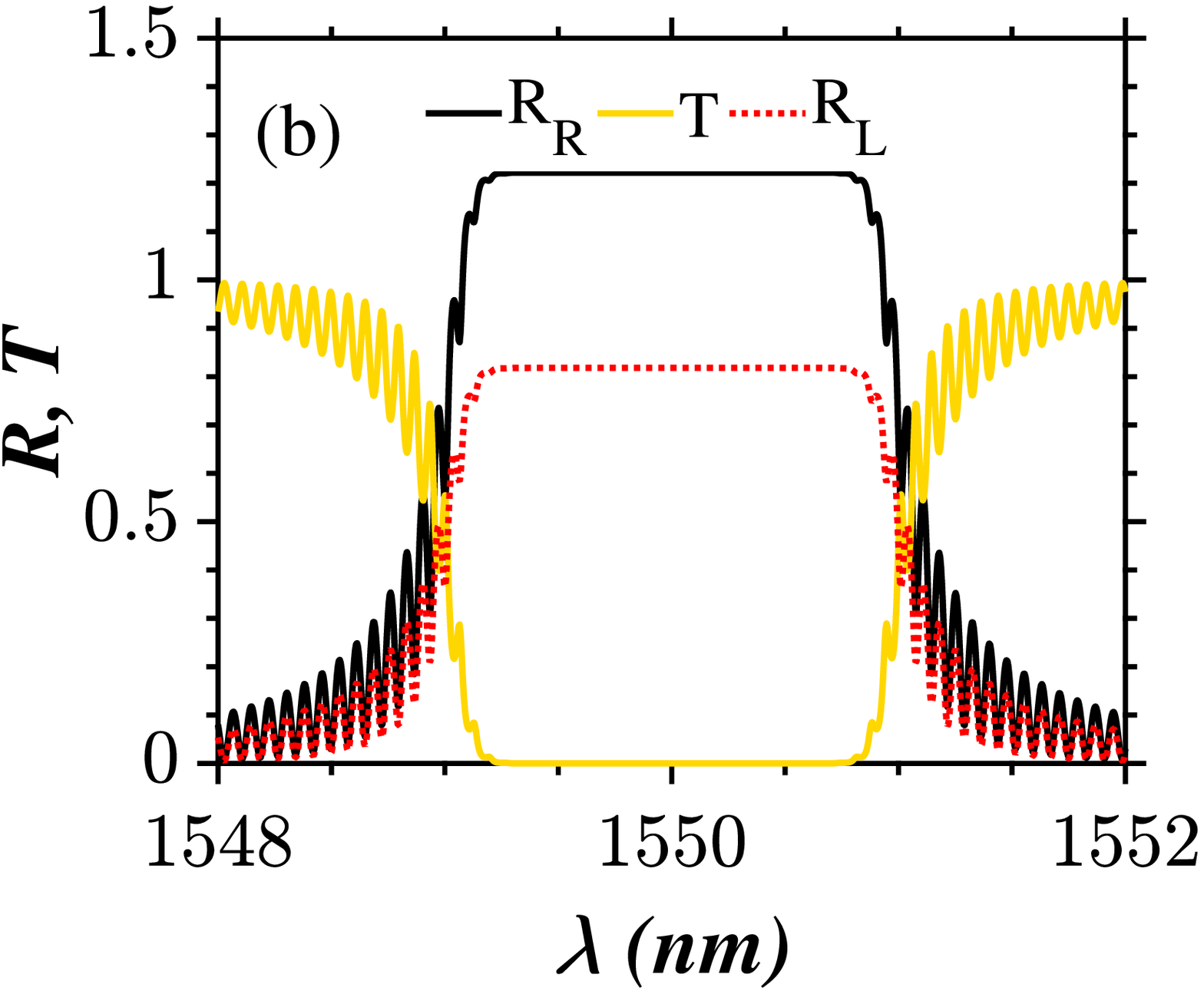}\\\includegraphics[width=0.5\linewidth]{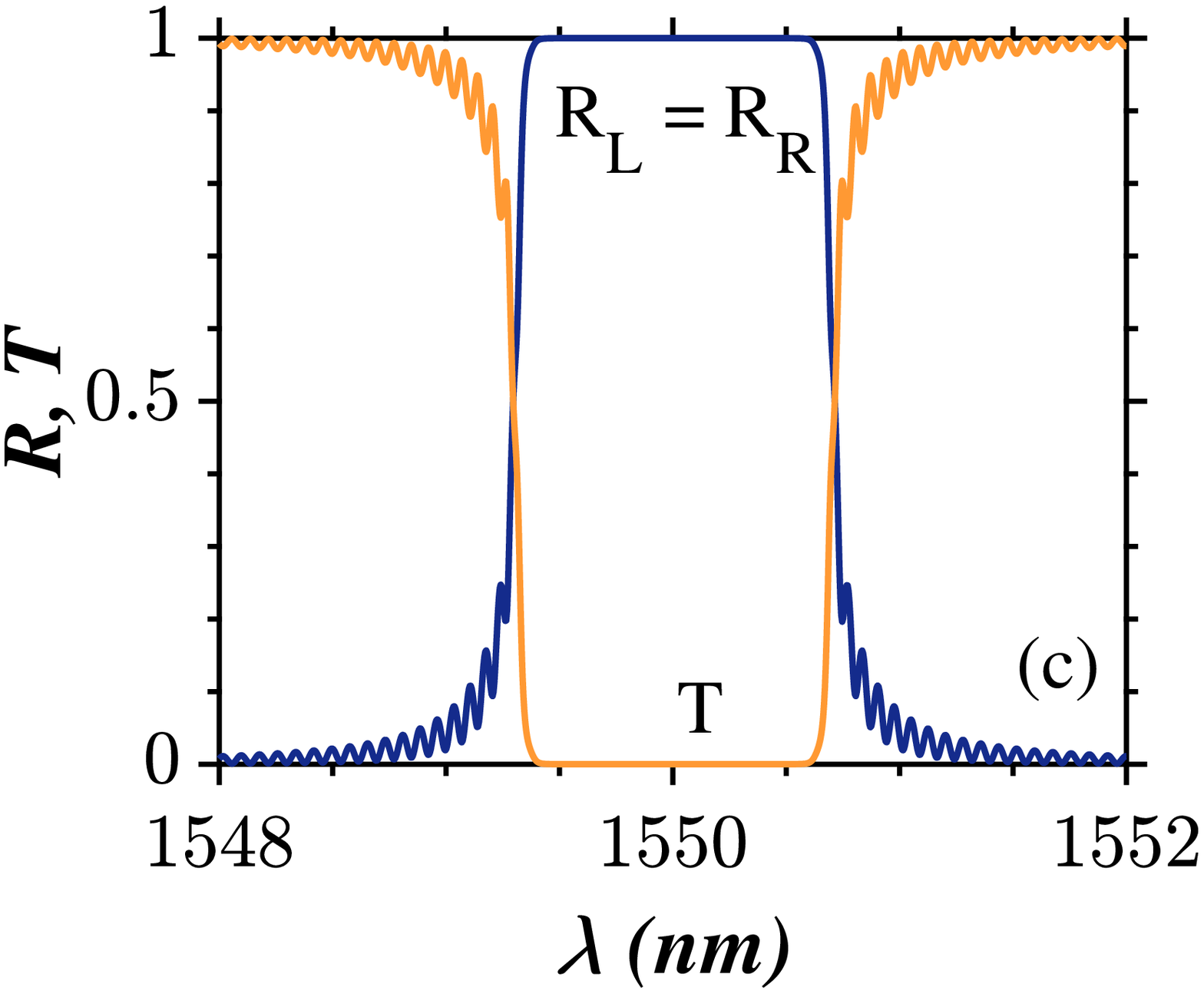}\includegraphics[width=0.5\linewidth]{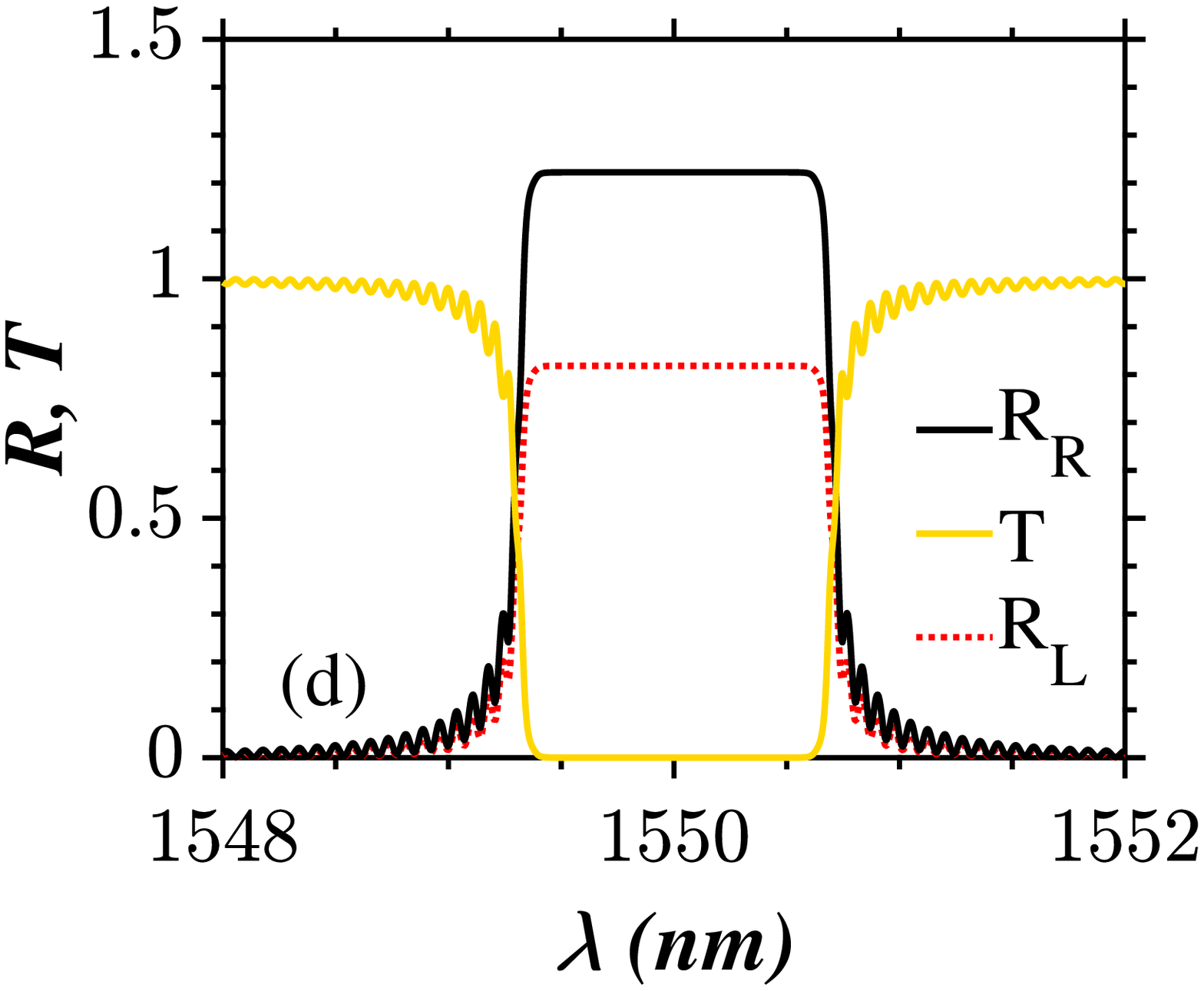}\\\includegraphics[width=0.5\linewidth]{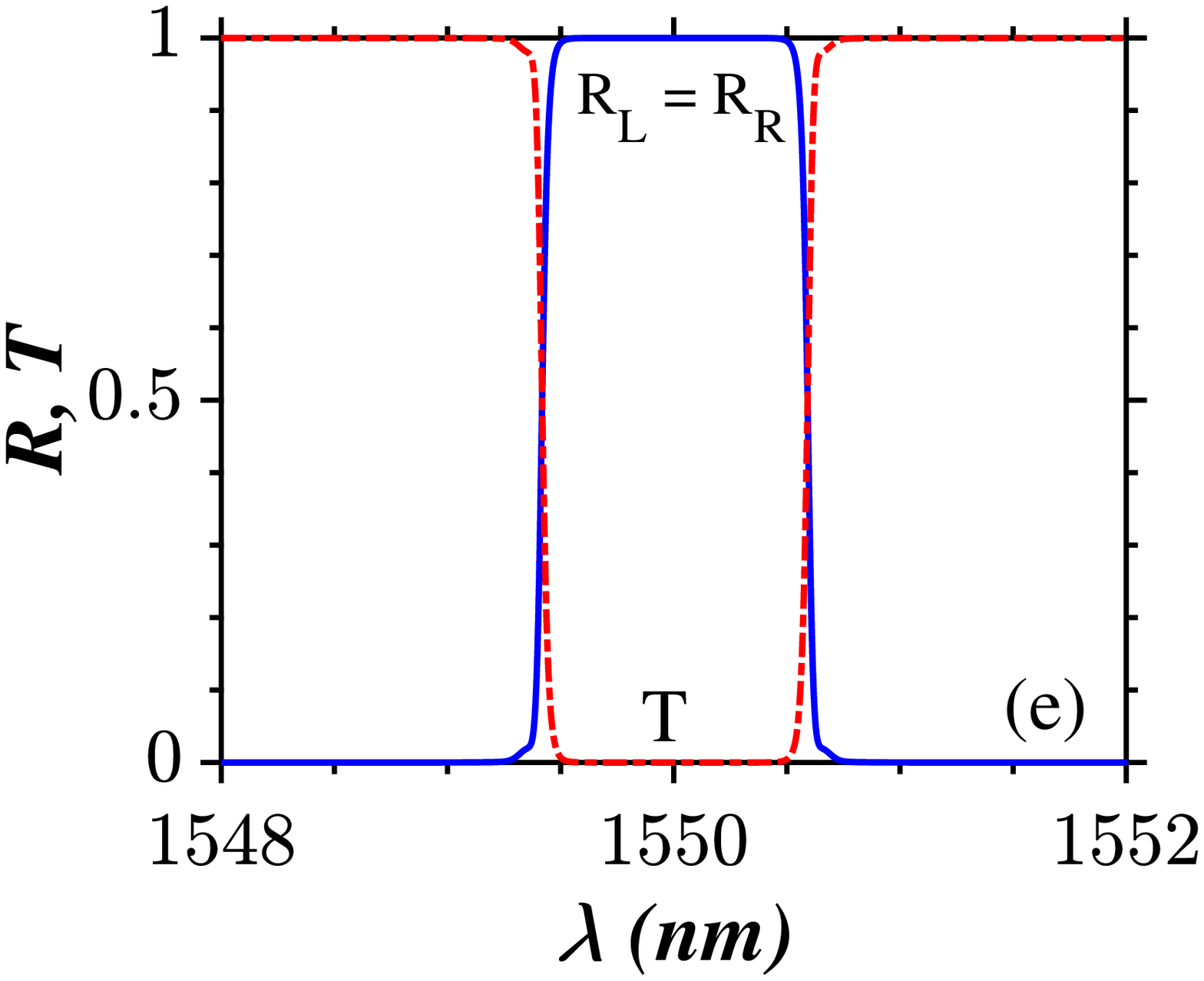}\includegraphics[width=0.5\linewidth]{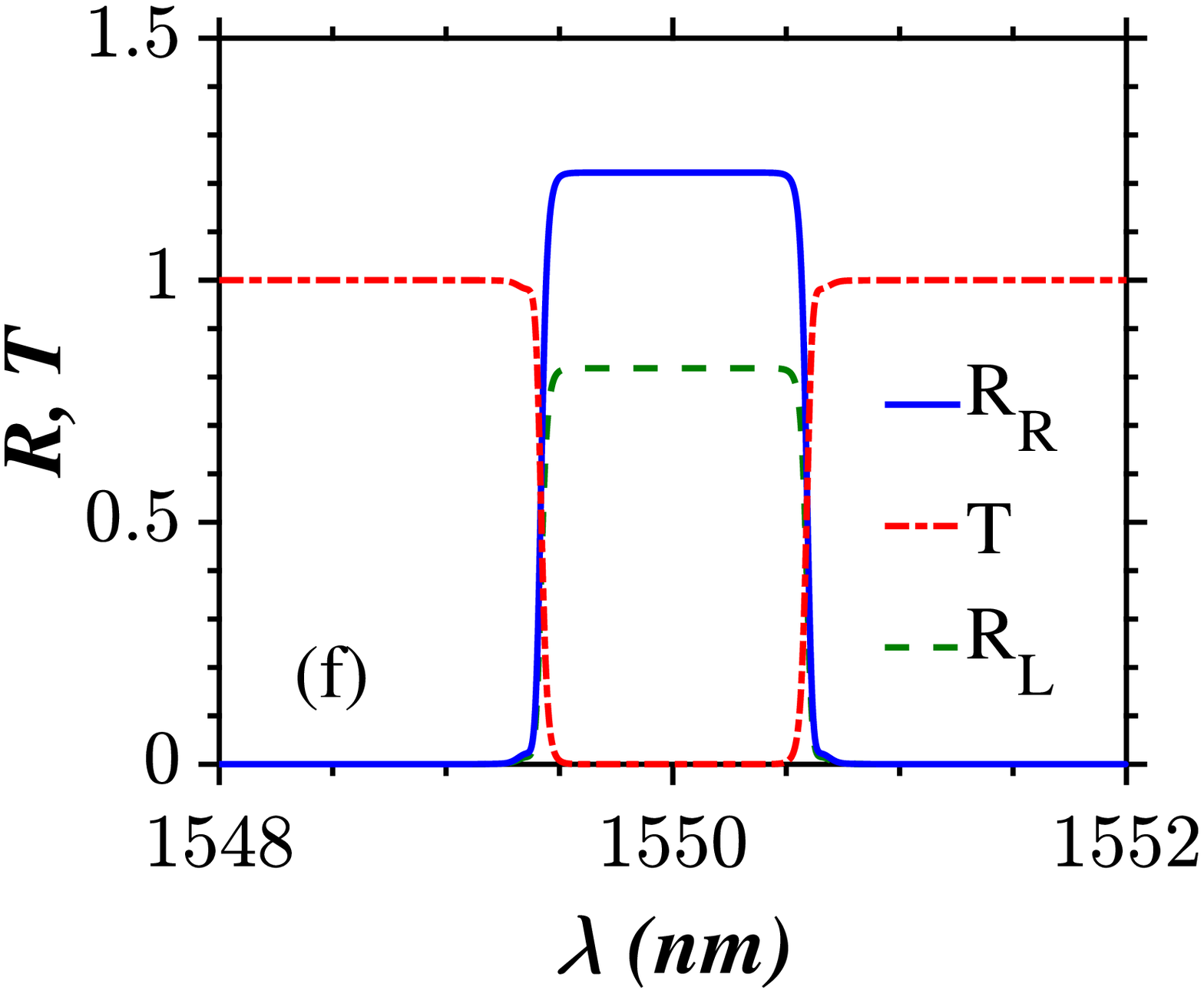}
	\caption{ Reflection and transmission spectra of chirped ($C = 0.125$ nm/cm) $\mathcal{PT}$-symmetric FBG. Plots in the left and right panels are simulated in the absence ($n_{1I} = 0$) and presence ($n_{1I}$ = 0.0001 ) of $\mathcal{PT}$-symmetry, respectively.  The upper, middle, and lower panels portray the spectra of a chirped  FBG, chirped FBG with a Gaussian apodization ($G = 4$), and chirped FBG with raised-cosine apodization, respectively.}
	\label{fig1}    
\end{figure}

The different amplitudes of reflection (transmission) coefficients in the side lobes  arise essentially due to the interplay among $\mathcal{PT}$-symmetry, chirping, and Gaussian apodization as seen in Figs. \ref{fig2}(a) -- \ref{fig2}(d). With an increase in the value of $n_{1I}$, the sidelobes show growth in the reflectivity. For instance, in $R_R$ it is measured to be 0.2311 in Fig. \ref{fig2}(a), whereas in Fig. \ref{fig2}(b) it is found to be 0.4958.
From these figures, we can also infer that any increase in the value of $n_{1I}$ intensifies the light reflected from the rear side ($R_R$) compared to Fig. \ref{fig1}(d). However, there is a fall in the amount of light reflected from the front end ($R_L$). Thus we can conclude that the asymmetric light propagation behavior depends on the $\mathcal{PT}$-symmetry and it is portrayed in Fig. \ref{fig2}(e). The growth in the reflection coefficient for right incidence ($R_R$) occurs as a result of the constructive interplay between the modes of the optical fields propagating along with the gain segments of the $\mathcal{PT}$-symmetric grating \cite{kulishov2005nonreciprocal}. When the input signal is incident on the other side of the grating the reflection ($R_L$) decreases, as any increase in loss ($-g$) decreases the coupling ($k$) between the two propagating waves. 
Finally, we look at the role of chirping on the light propagation characteristics of the device. In the absence of apodization, it may produce an adverse effect like the introduction of ripples in the spectra. However, it plays a positive role in controlling the bandwidth of the spectra in the presence of apodization. With a decrease in the value of chirping, the full width half maximum (FWHM) of the spectrum reduces. This is illustrated in Figs. \ref{fig2}(c) and (f). On the other hand, if the value of chirping is high ($C = 0.5$ nm/cm), the FWHM is increased as observed in  Figs. \ref{fig2}(d) and \ref{fig2}(f). Hence, one can enlarge or taper the nature of bandwidth by merely manipulating the chirping parameter along with the judiciously chosen apodization profile.

\subsection{ Raised-cosine apodization profile}
\begin{figure}[hthb]
	\centering
	\includegraphics[width=0.5\linewidth]{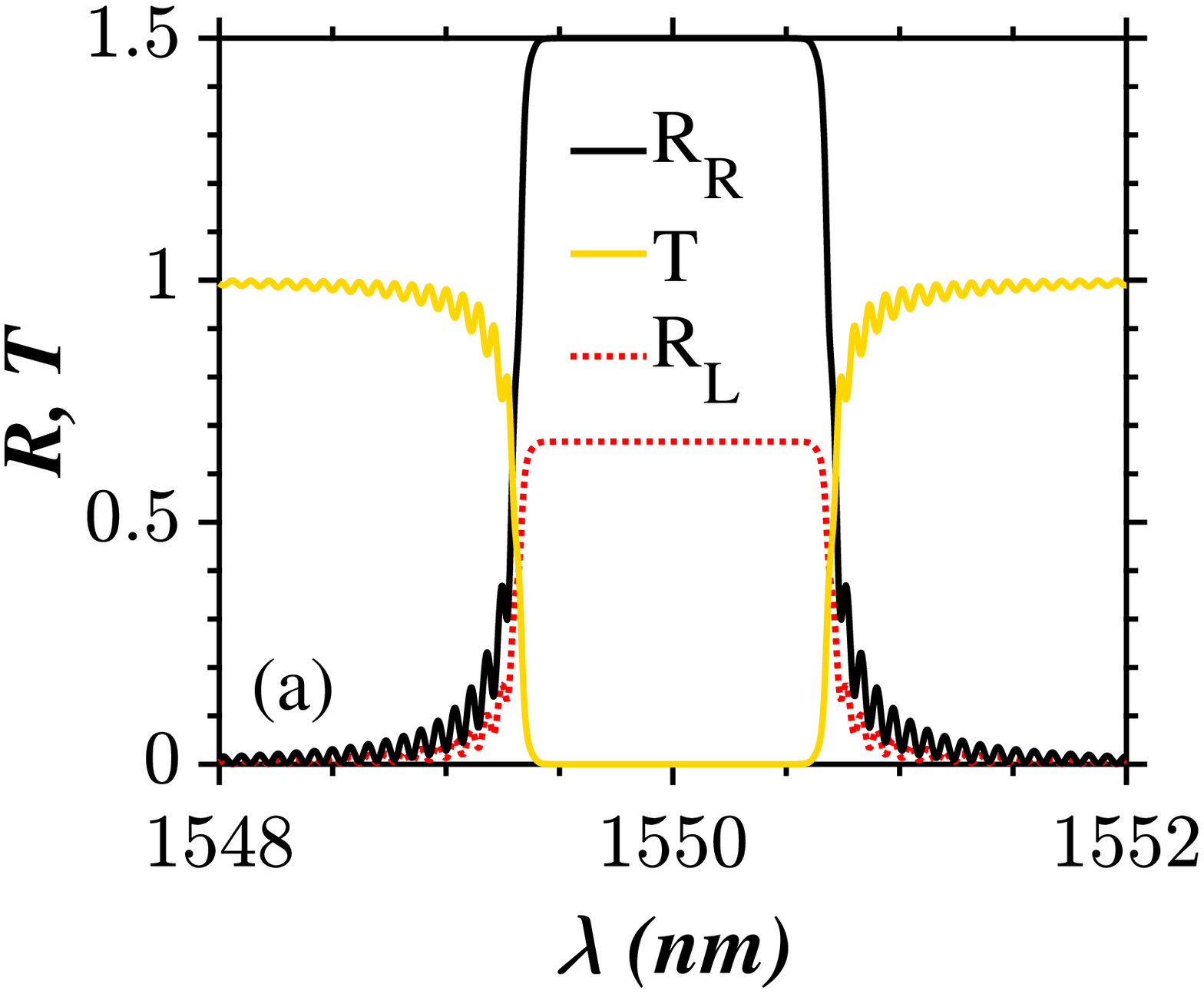}\includegraphics[width=0.5\linewidth]{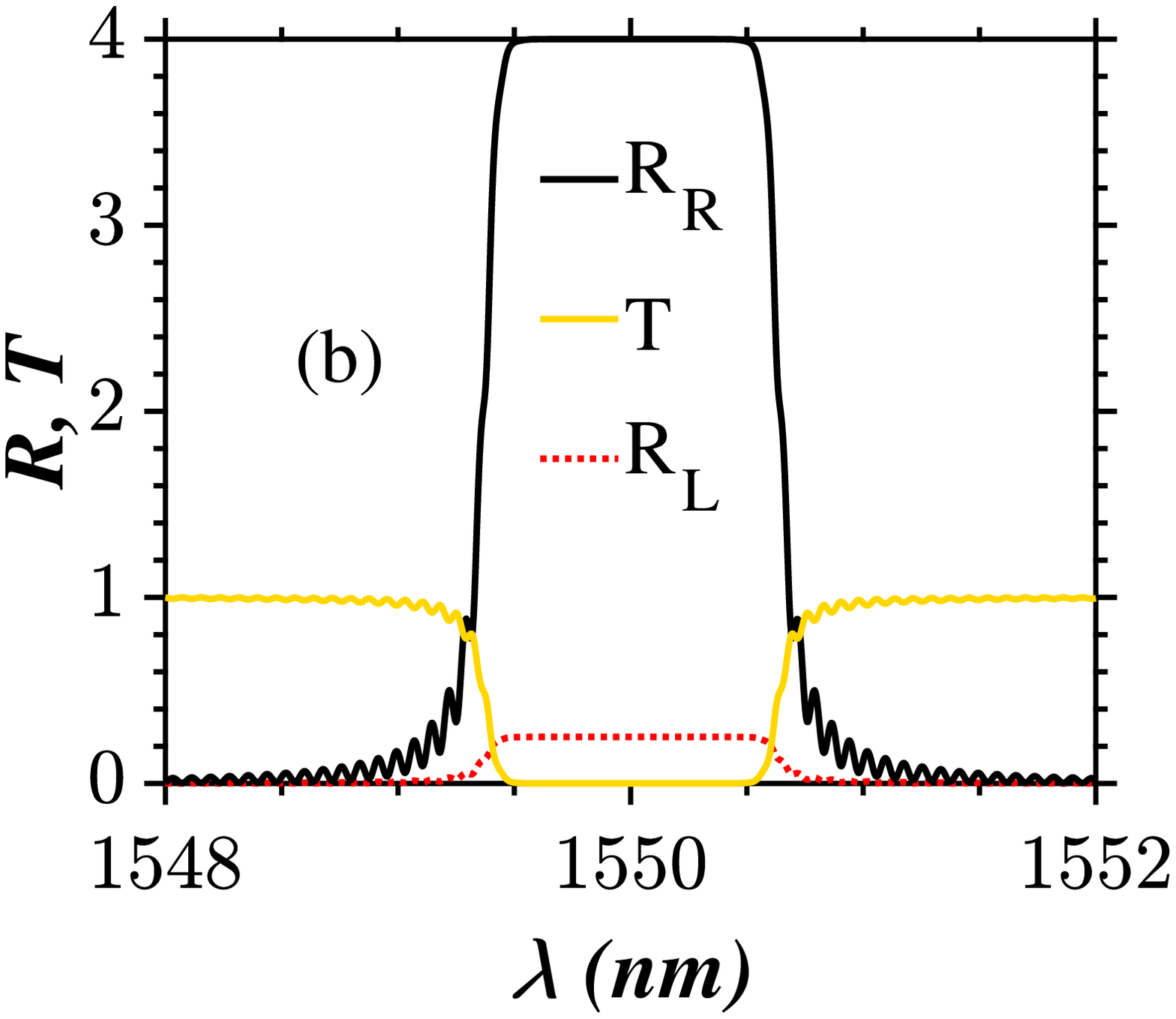}\\\includegraphics[width=0.5\linewidth]{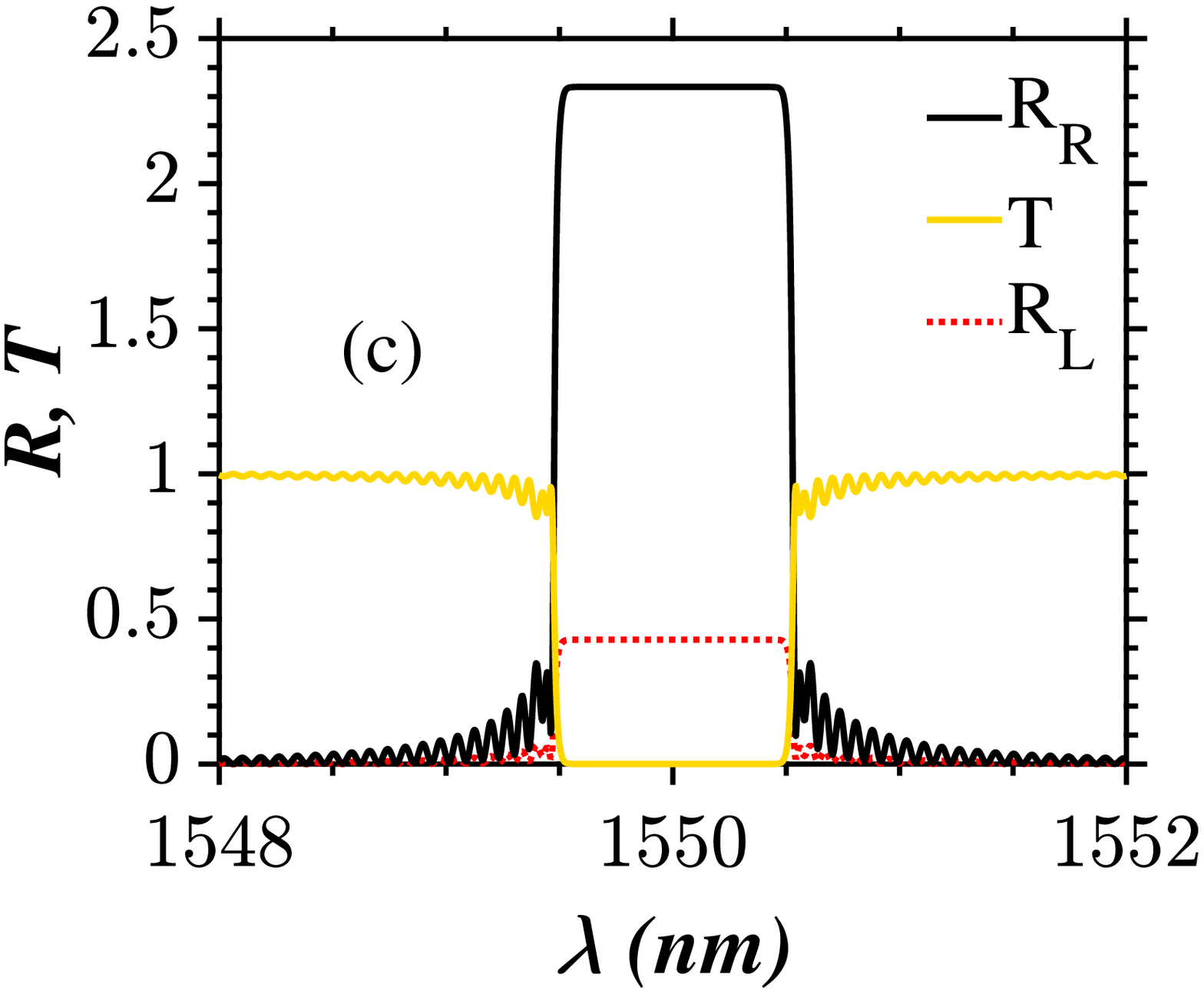}\includegraphics[width=0.5\linewidth]{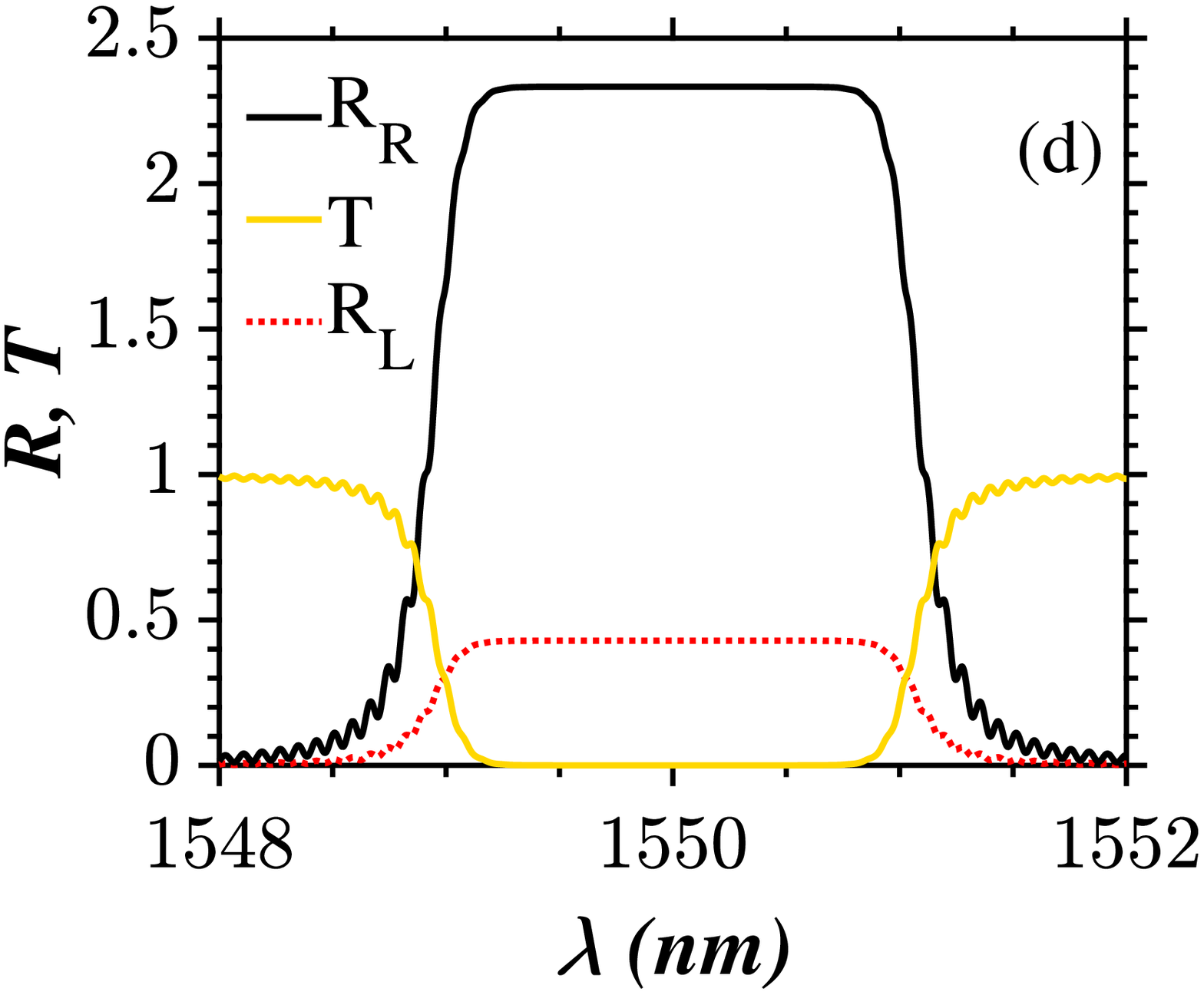}\\\includegraphics[width=0.5\linewidth]{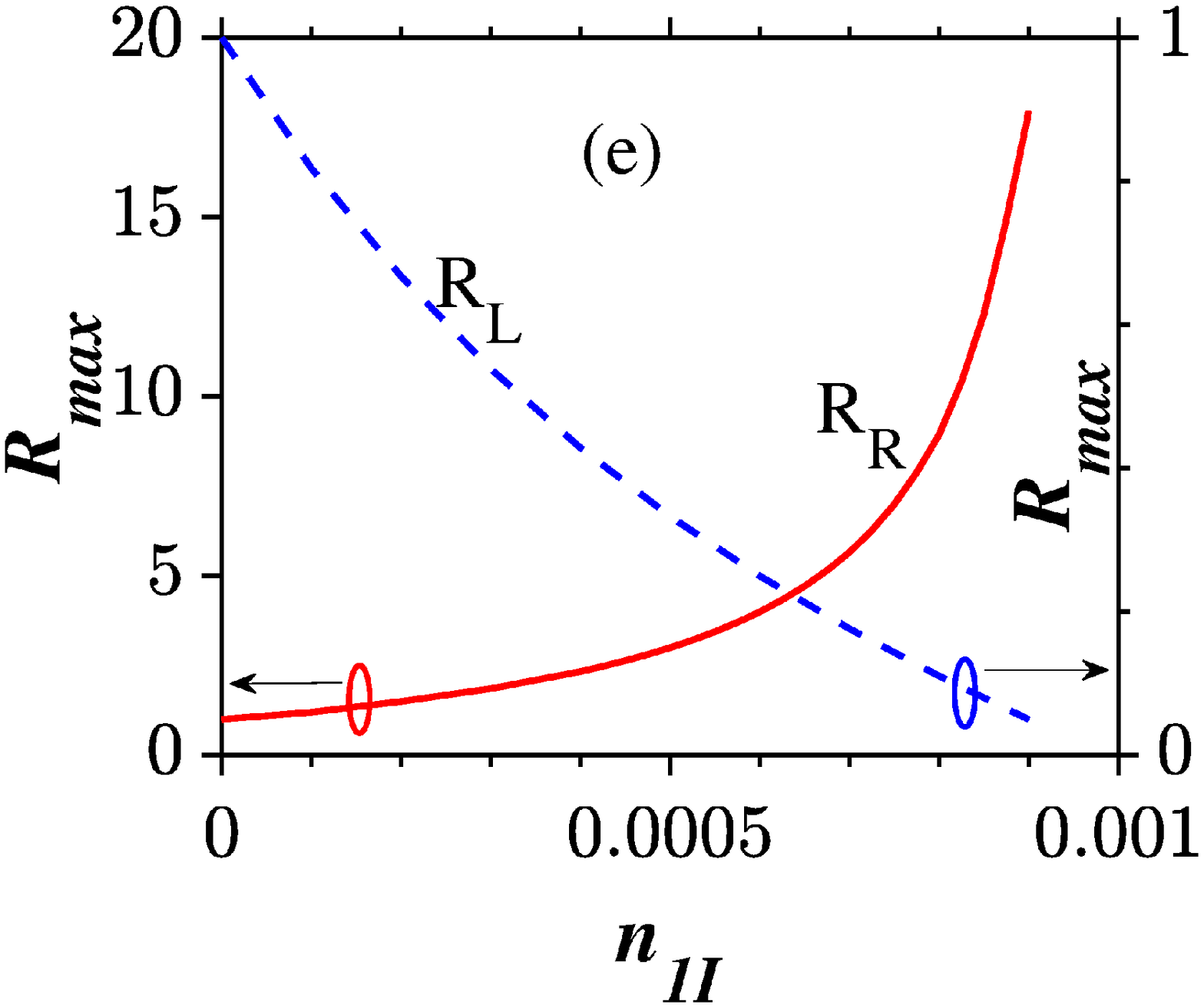}\includegraphics[width=0.5\linewidth]{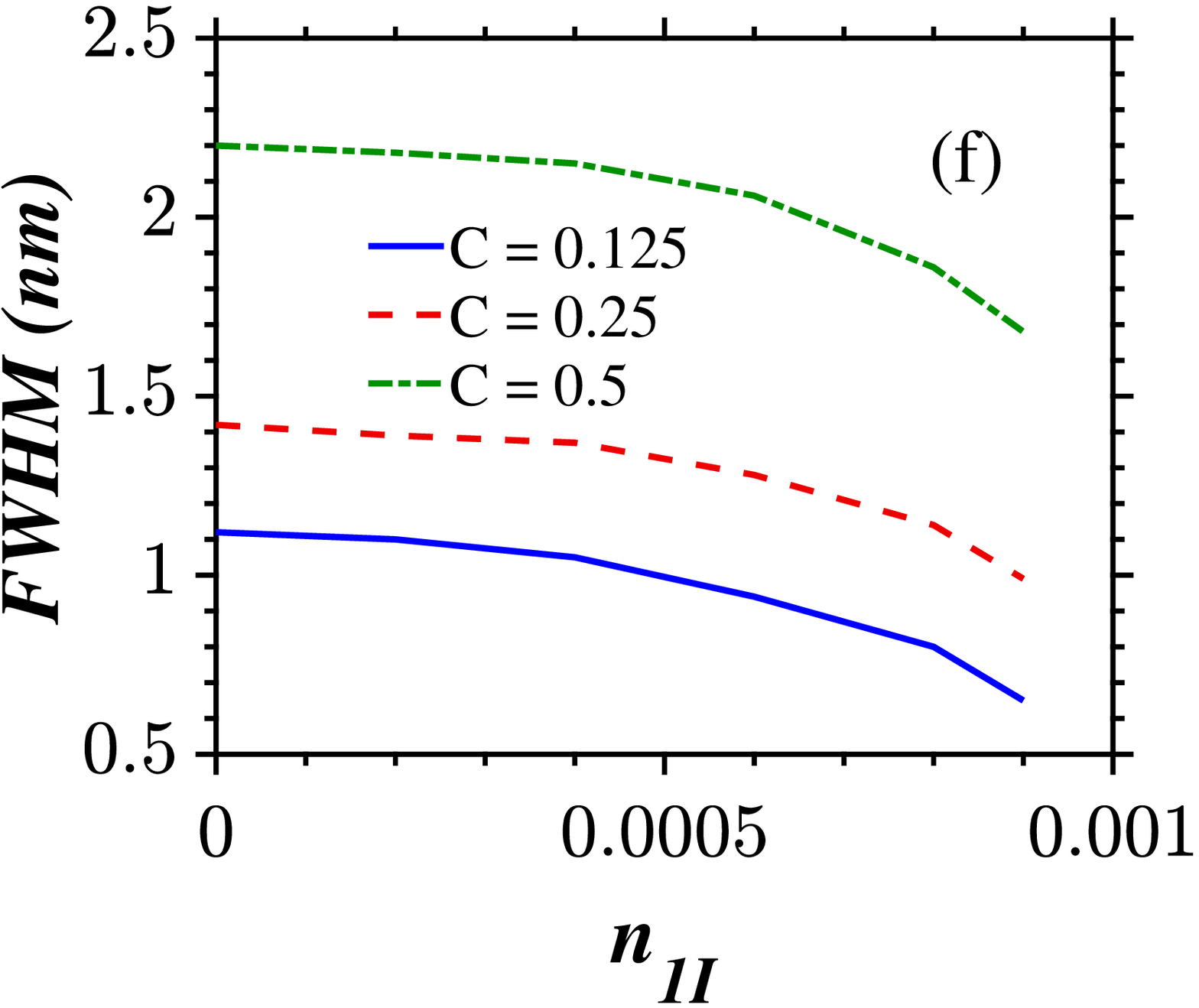}%\\\includegraphics[width=0.5\linewidth]{fig2g}\includegraphics[width=0.5\linewidth]{fig2h}
	\caption{Reflection and transmission characteristics of a chirped and apodized $\mathcal{PT}$-symmetric FBG (CAPT-FBG) in the unbroken regime with a  Gaussian apodization (G = 4). Top panels depict the role of $\mathcal{PT}$-symmetry at $C = 0.25$ nm/cm with (a) $n_{1I} = 0.0002$ and (b) $n_{1I} = 0.0006$. The middle panels depict the role of chirping at $n_{1I} = 0.0004$ with (c)  $C= 0.125$ nm/cm and (d) $C= 0.5$ nm/cm. Continuous variation of maximum reflection ($R_{max}$) against variation in gain-loss coefficient is plotted in (e) and it is independent of the value of $C$. The full width half maximum (FWHM) is plotted against continuously varying $n_{1I}$ for different values of $C$  in (f).} 
	\label{fig2}    
\end{figure}

\begin{figure}[hthb]
	\centering
	\includegraphics[width=0.5\linewidth]{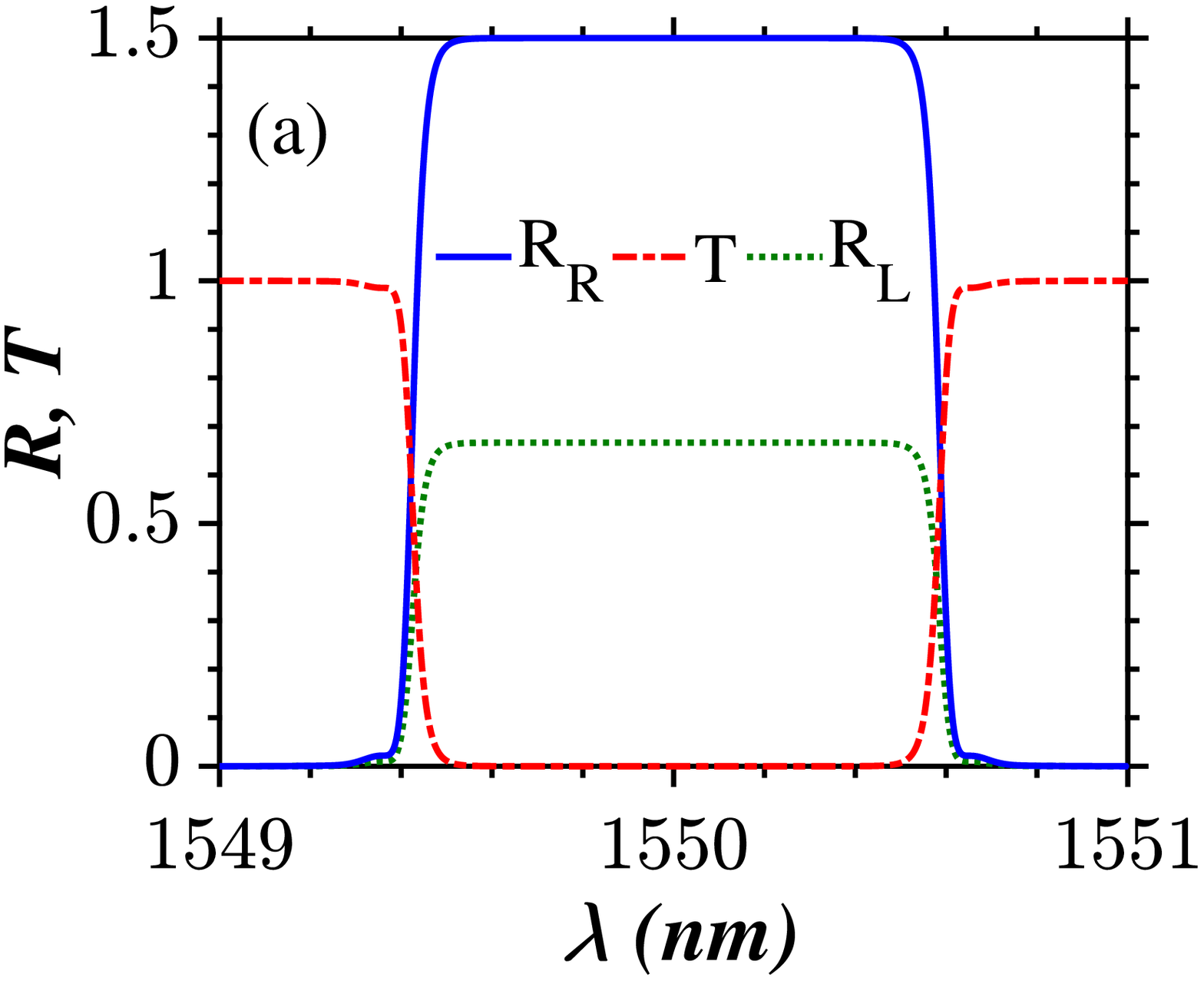}\includegraphics[width=0.5\linewidth]{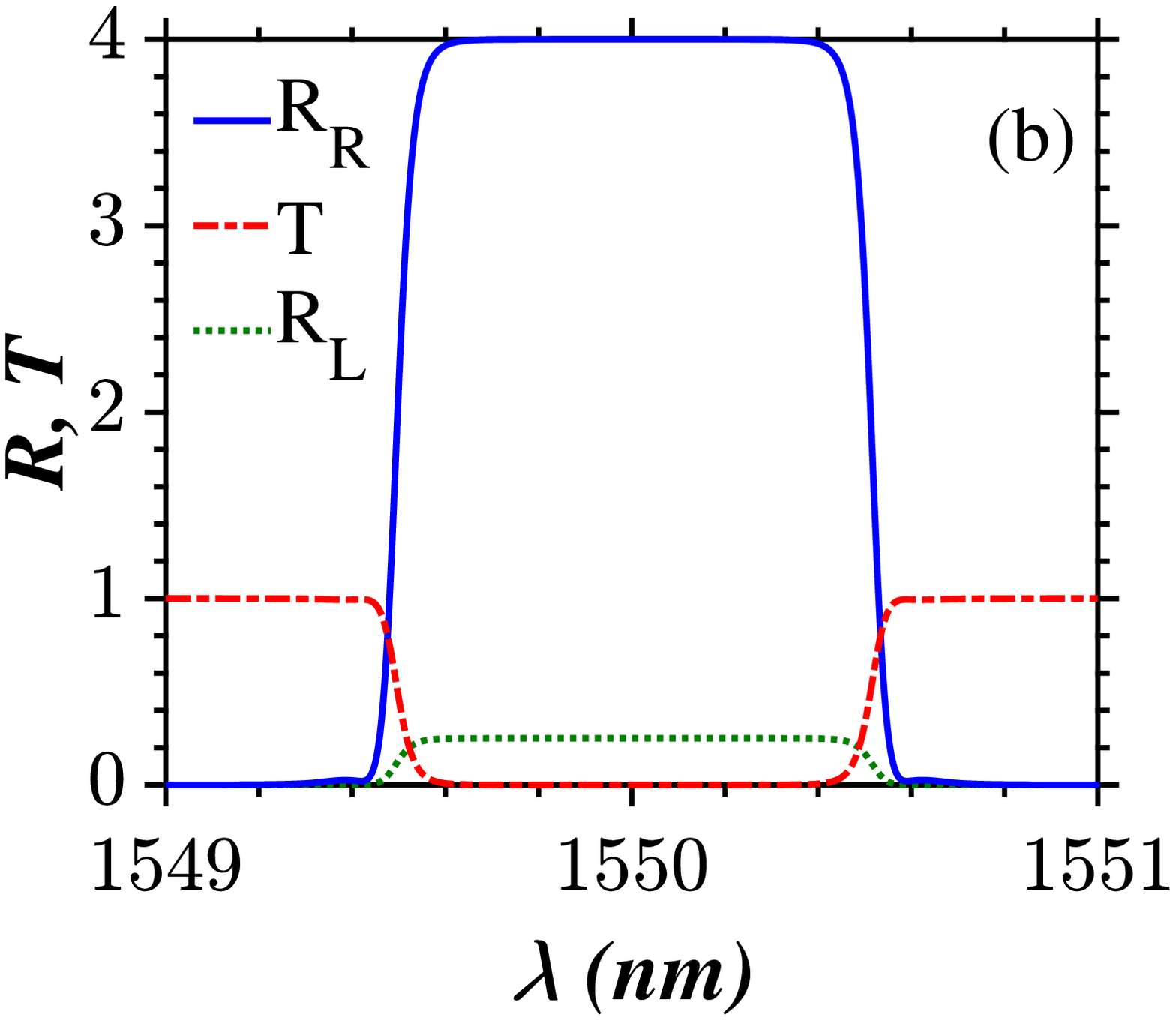}\\\includegraphics[width=0.5\linewidth]{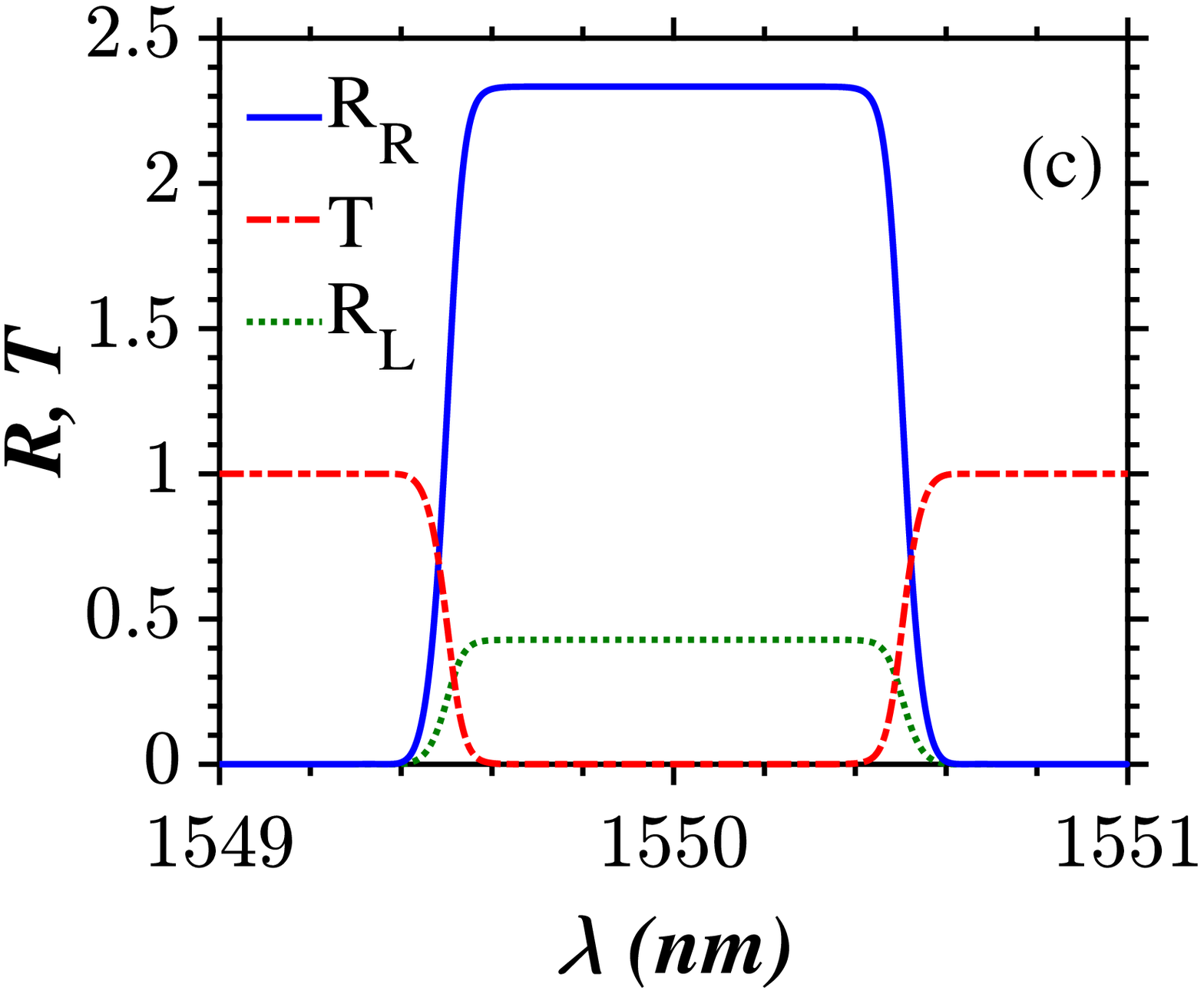}\includegraphics[width=0.5\linewidth]{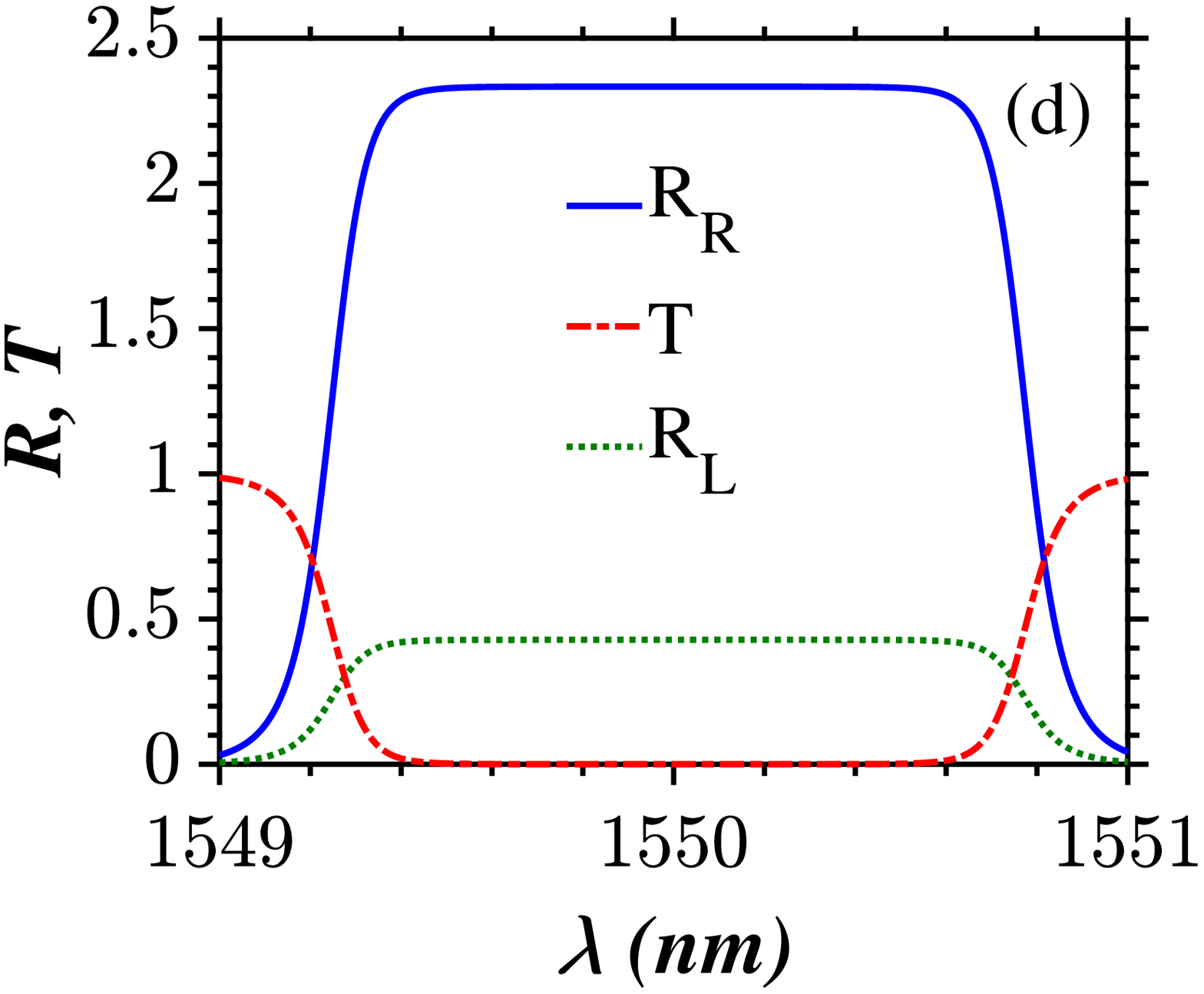}\\\includegraphics[width=0.5\linewidth]{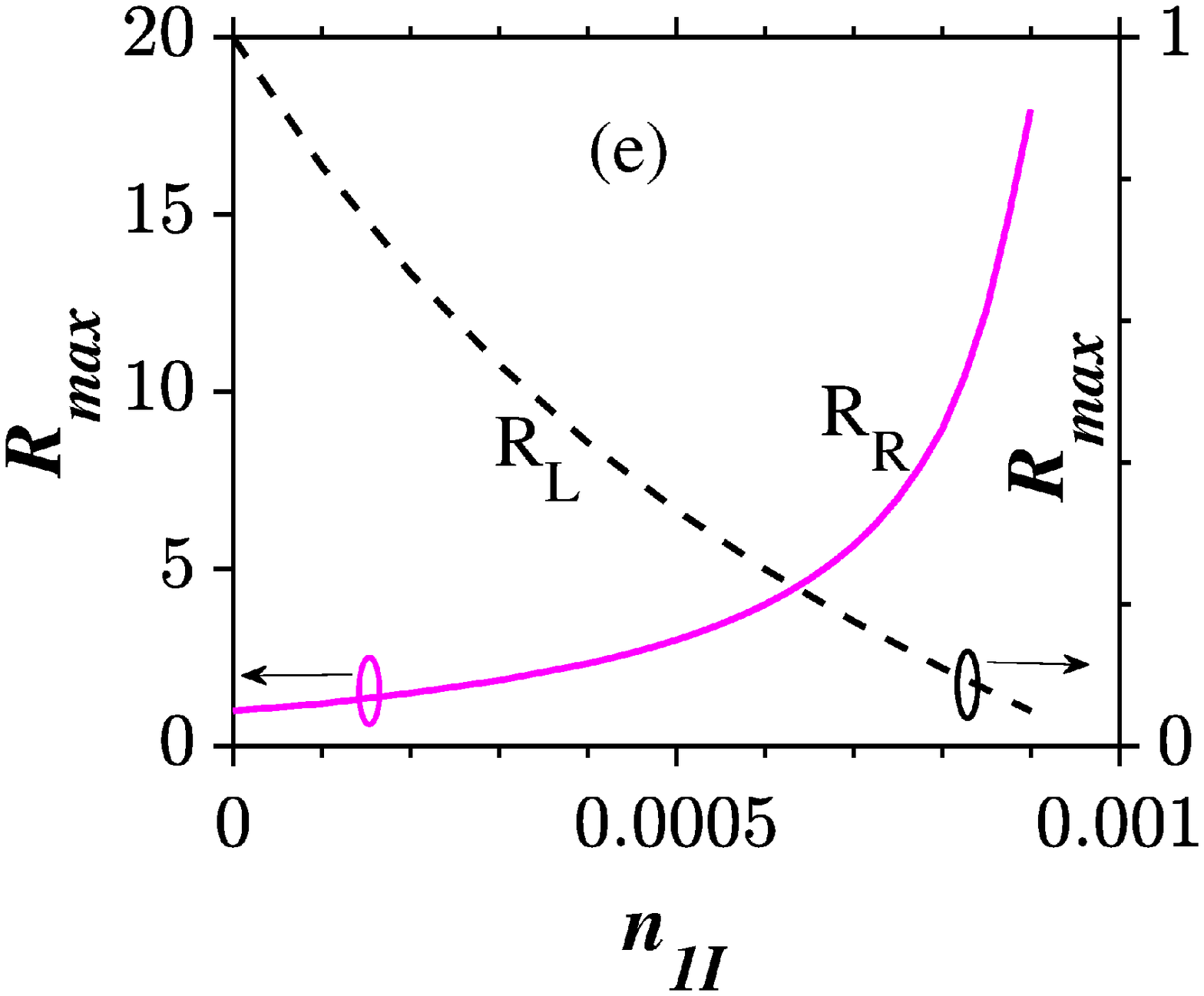}\includegraphics[width=0.5\linewidth]{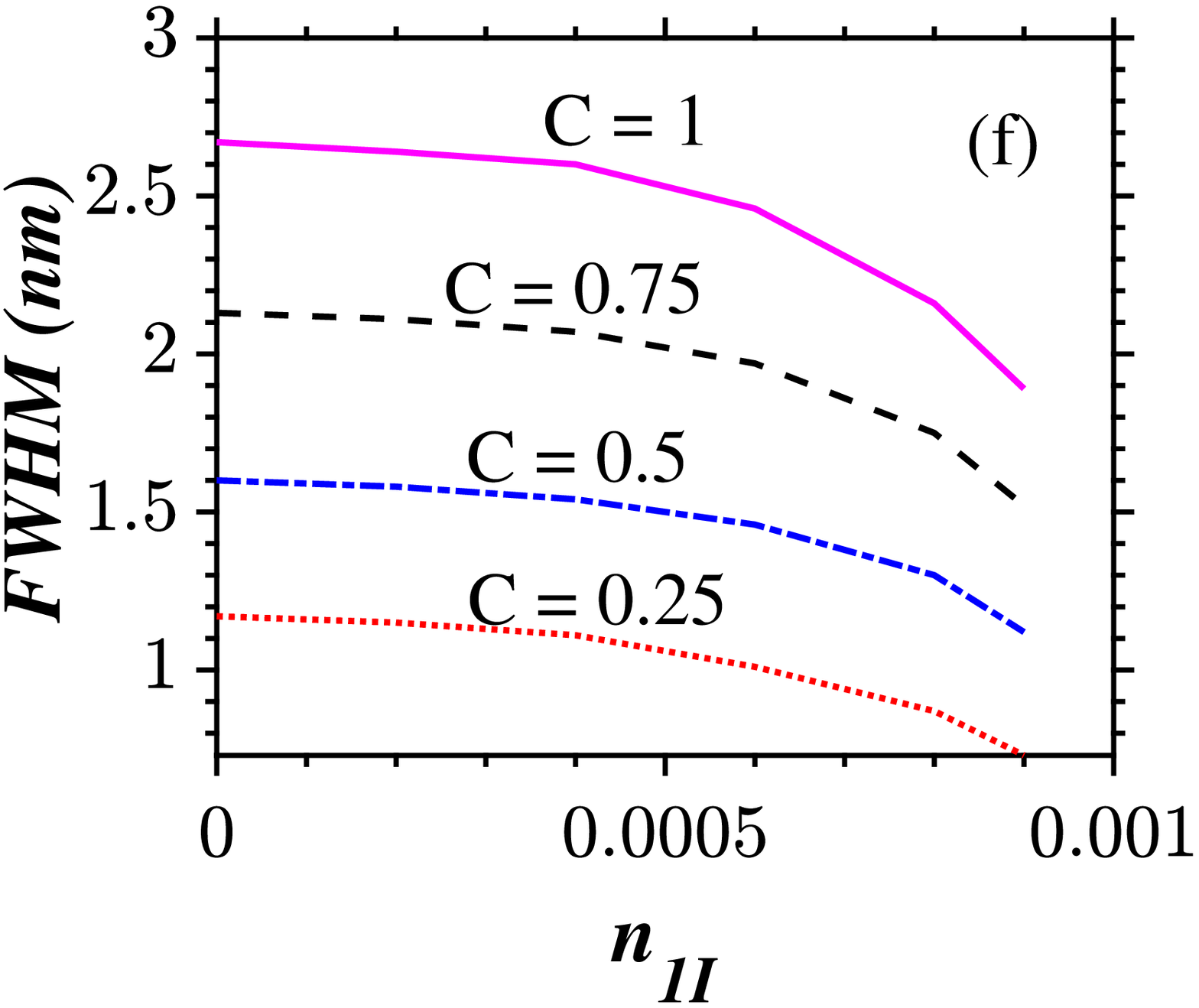}
	\caption{ Plots illustrating the same dynamics as in Fig. \ref{fig2} except it is plotted for the case of raised-cosine apodization. The wavelength range is selected between 1549 nm and 1551 nm in (a) -- (d).}
	\label{fig3}
\end{figure}

In Fig. \ref{fig3}, we observe that some weak reflections persist on the wings of the spectra. This issue can be addressed by employing a raised-cosine apodization profile which can eliminate the side lobes entirely on both sides of the central lobe as seen in Fig. \ref{fig3}(a). Having stated the above, we also note that the reflectivity (transmittivity) can be tuned by varying the parameter $n_{1I}$. Similar to the last subsection, the amplitude of reflection from the rear (front) end of the device gets amplified (reduced) when $n_{1I}$ is increased (decreased) as seen in Figs. \ref{fig3}(b) and \ref{fig3}(c). From Figs. \ref{fig2}(e) and \ref{fig3}(e), we conclude that the maximum reflectivity depends only on the value of $\mathcal{PT}$-symmetry and is independent of the nature of apodization. The bandwidth control is facilitated by varying the chirp (C) offered by the $\mathcal{PT}$-symmetric grating. In Fig. \ref{fig3}(c), the value of chirping is set to 0.125 nm/cm and thus we get a narrow spectrum and without ripples. Nevertheless, any further increase in the chirping contributes to the widening of the spectrum which gives an additional degree of freedom to enlarge the spectrum remarkably as identified in Figs. \ref{fig3}(d) and \ref{fig3}(f).  

\section{Group delay and dispersion characteristics of a CAPT-FBG}
\label{Sec:4}

A chirped FBG can induce dispersion as the light propagates inside it. It is desirable to have a linear delay without any ripples \cite{hill1994chirped} and hence apodization plays a central role in fine-tuning the delay characteristics of the device \cite{ennser1998optimization}. The delay may increase or decrease linearly according to the sign of chirp (C) and hence the sign of dispersion ($D$) also varies with the sign of $C$.  As mentioned earlier, studying these properties aids in the construction of applications such as delay lines \cite{lenz2001optical}, dispersion compensator \cite{litchinitser1997fiber} and so on. Kulishov \emph{et al.} investigated the delay characteristics of a  FBG in the absence of nonuniformities \cite{kulishov2005nonreciprocal}. The question that arises here is whether the addition of nonuniformities impacts the delay and dispersion properties of a $\mathcal{PT}$-symmetric FBG. With this aim, we now look into the group delay offered by the system under study.
\begin{figure}[t]
	\centering
	\includegraphics[width=0.5\linewidth]{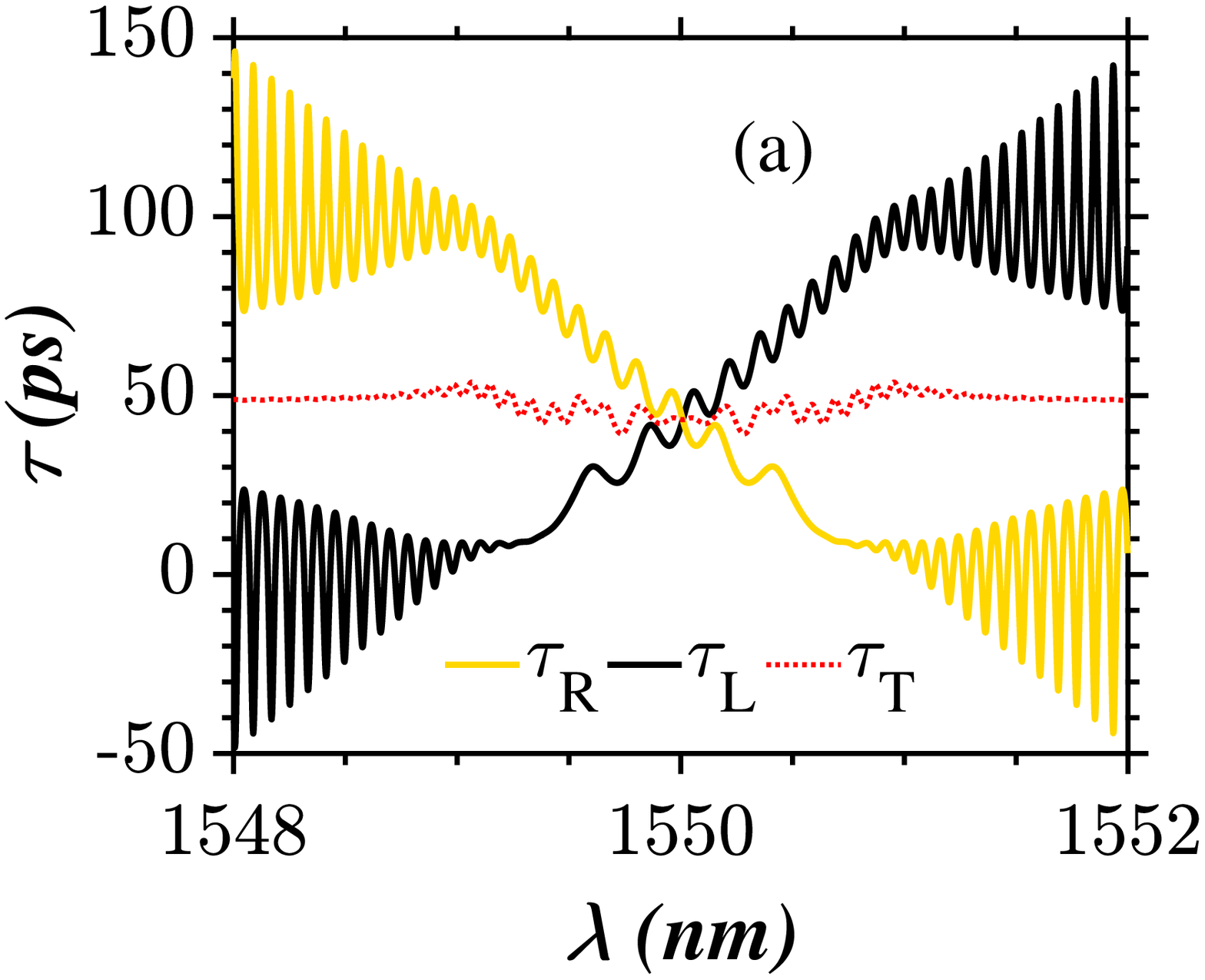}\includegraphics[width=0.5\linewidth]{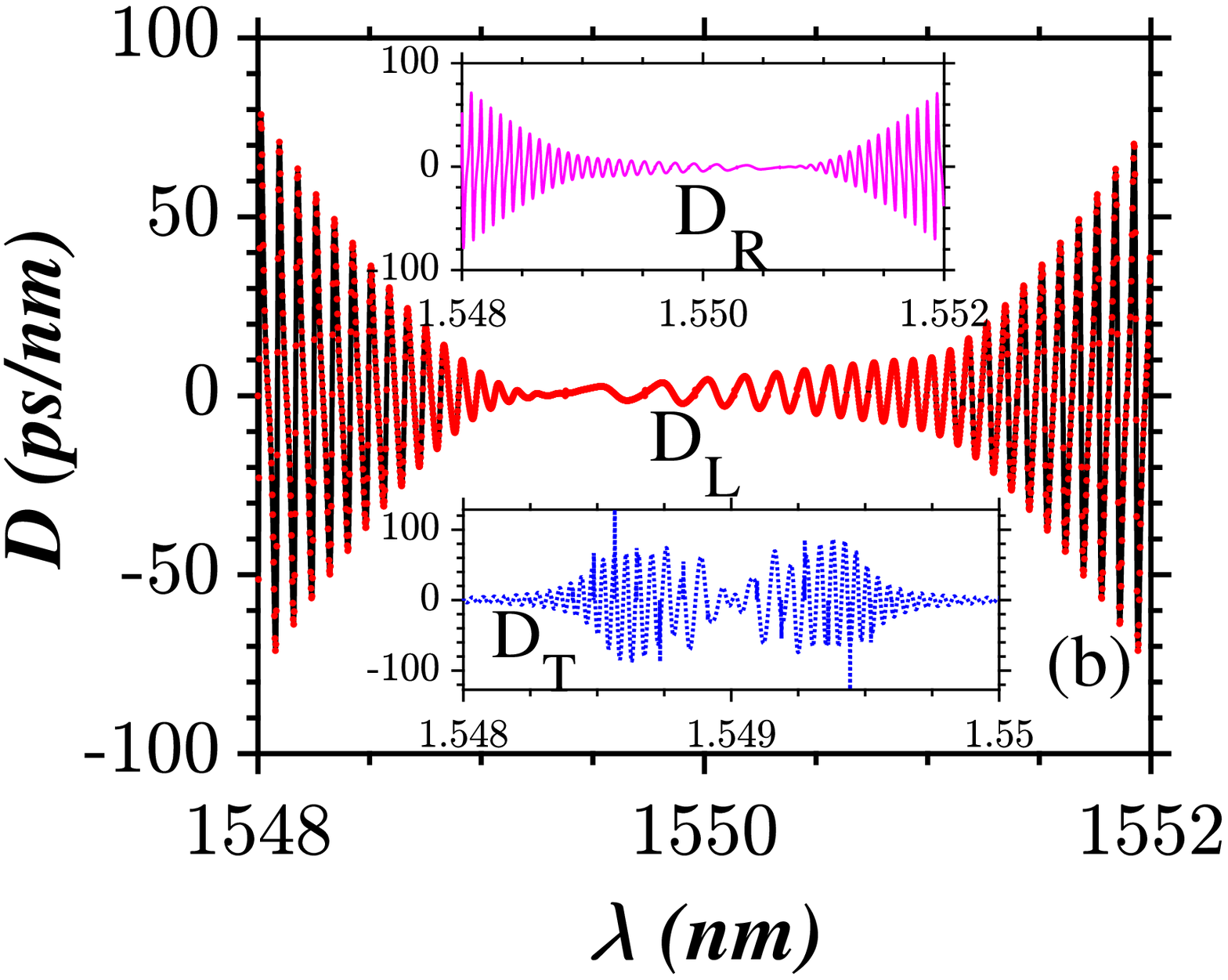}\\\includegraphics[width=0.5\linewidth]{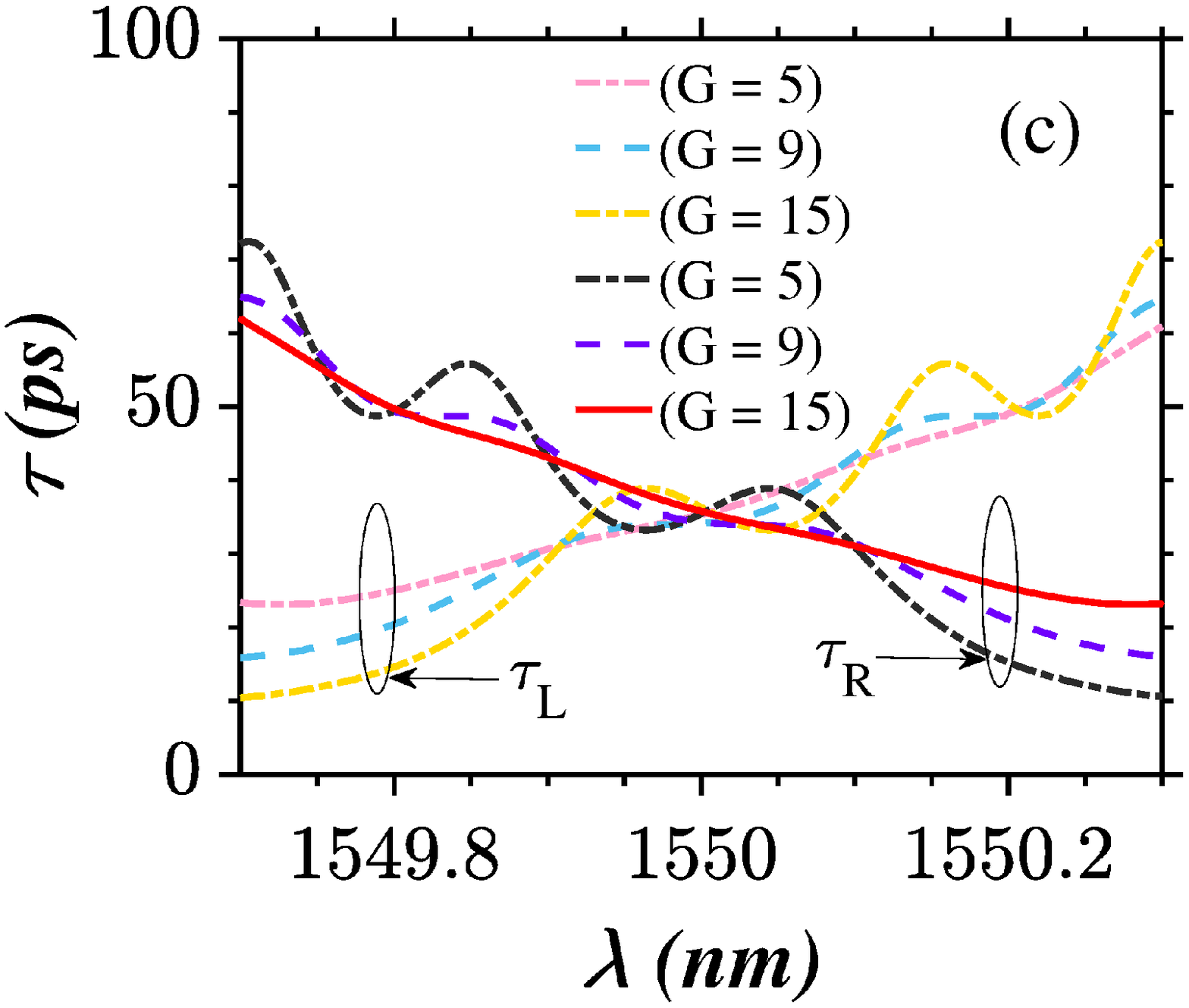}\includegraphics[width=0.5\linewidth]{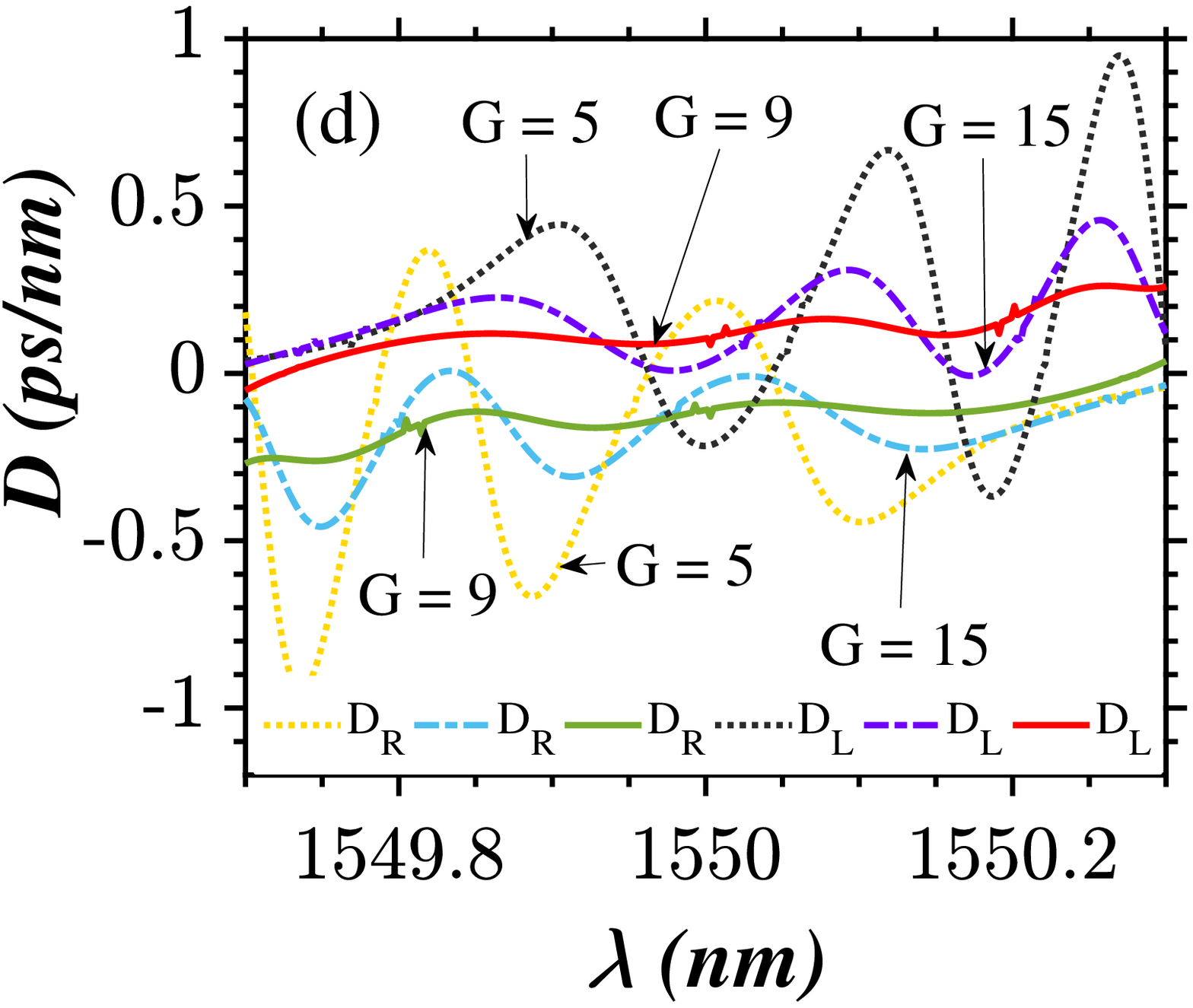}
	\caption{(a) Group delay and (b) dispersion characteristics of a chirped ($C = 0.25$ nm/cm) and Gaussian apodized (G = 4) $\mathcal{PT}$-symmetric FBG when the system is operated in the unbroken $\mathcal{PT}$-symmetric regime ($n_{1I} = 0.0008$). (c) and (d) Show the same dynamics with same parameters as in (a) and (b), respectively except that it is plotted for three different values of $G = 5$, 9, 15. Also, the x-axis is scaled between 1549.7 nm and 1550.3 nm in the bottom panel.}
	\label{fig4}
\end{figure}

\begin{figure}[t]	\includegraphics[width=0.5\linewidth]{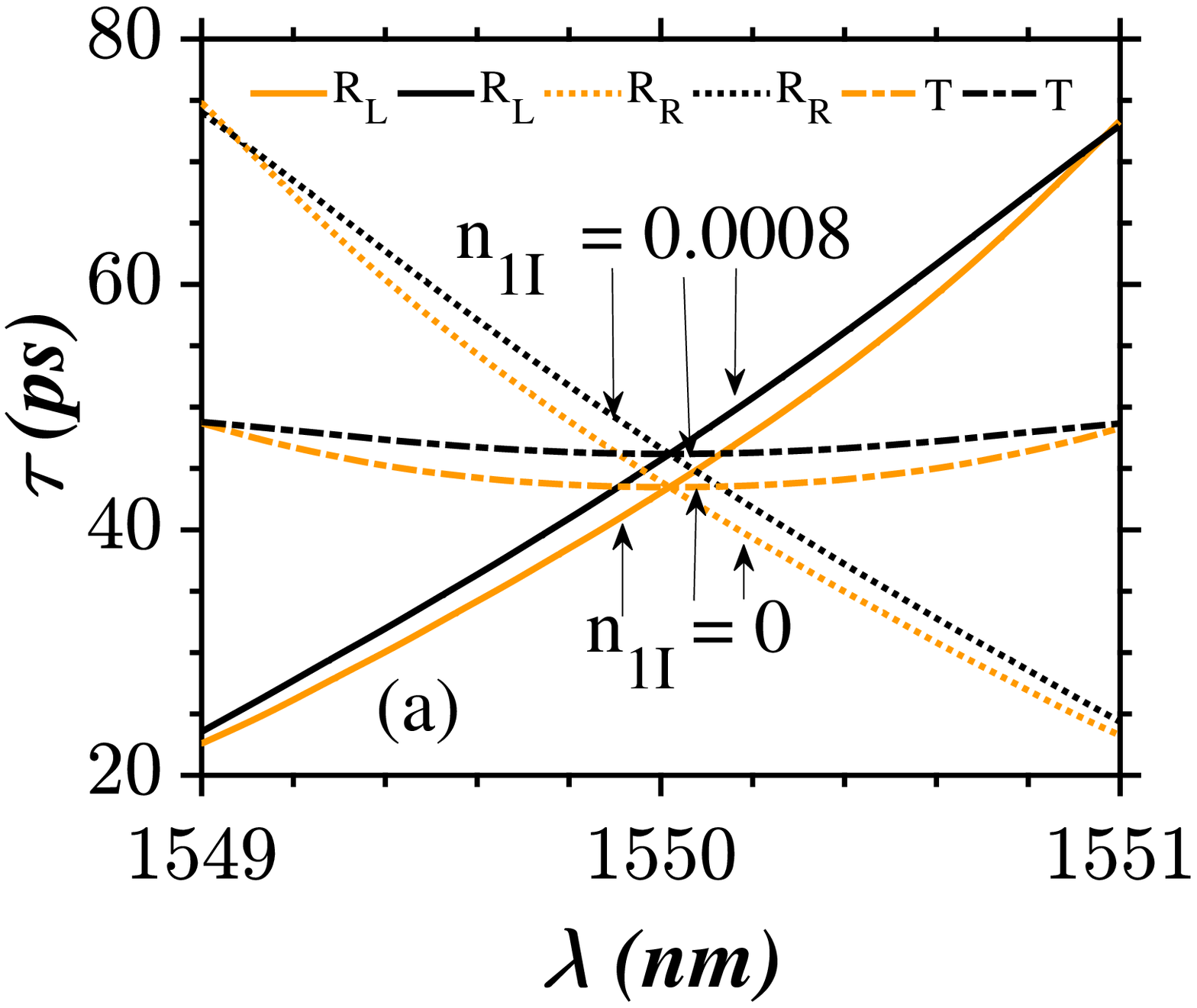}\includegraphics[width=0.5\linewidth]{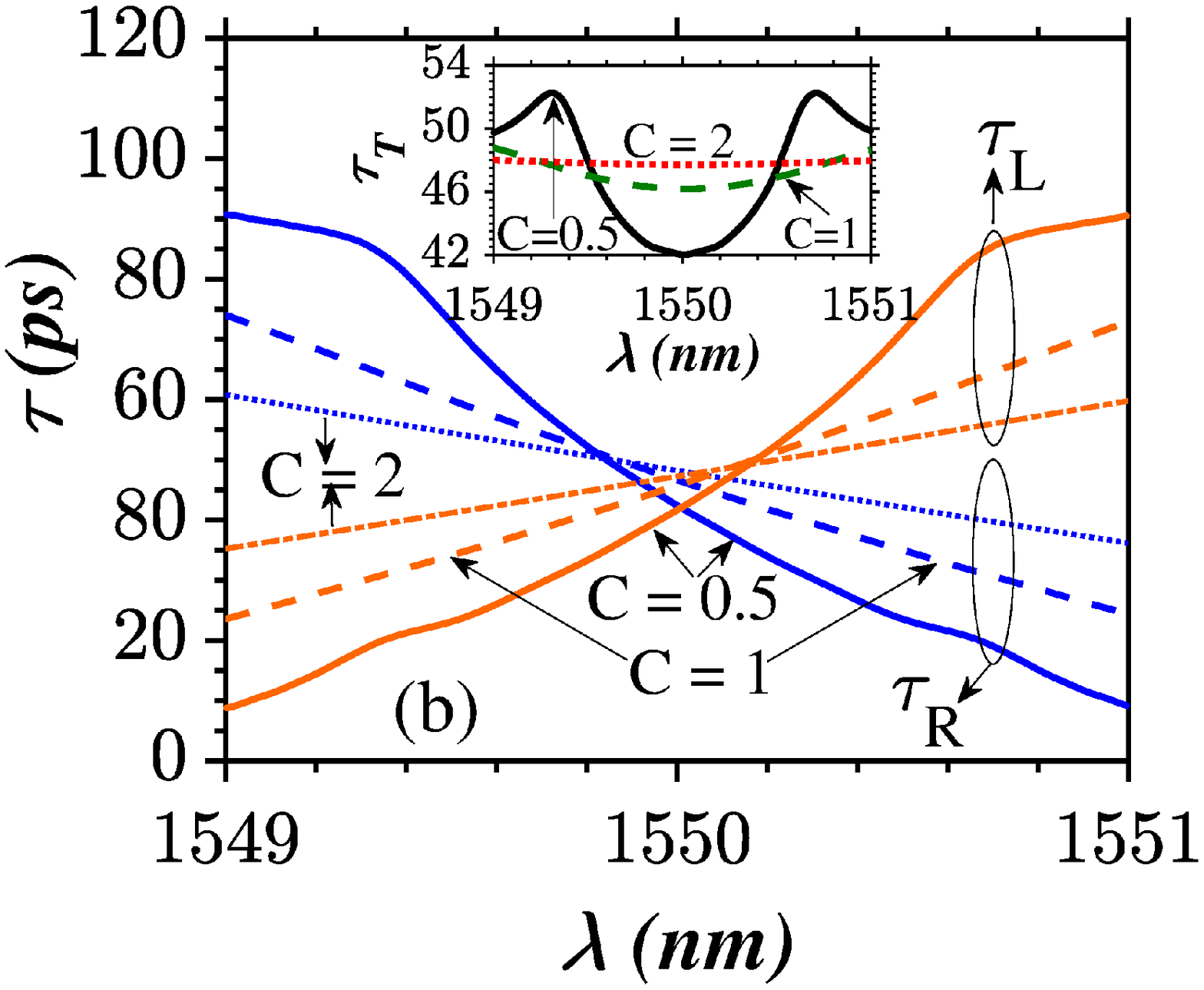}\\\includegraphics[width=0.5\linewidth]{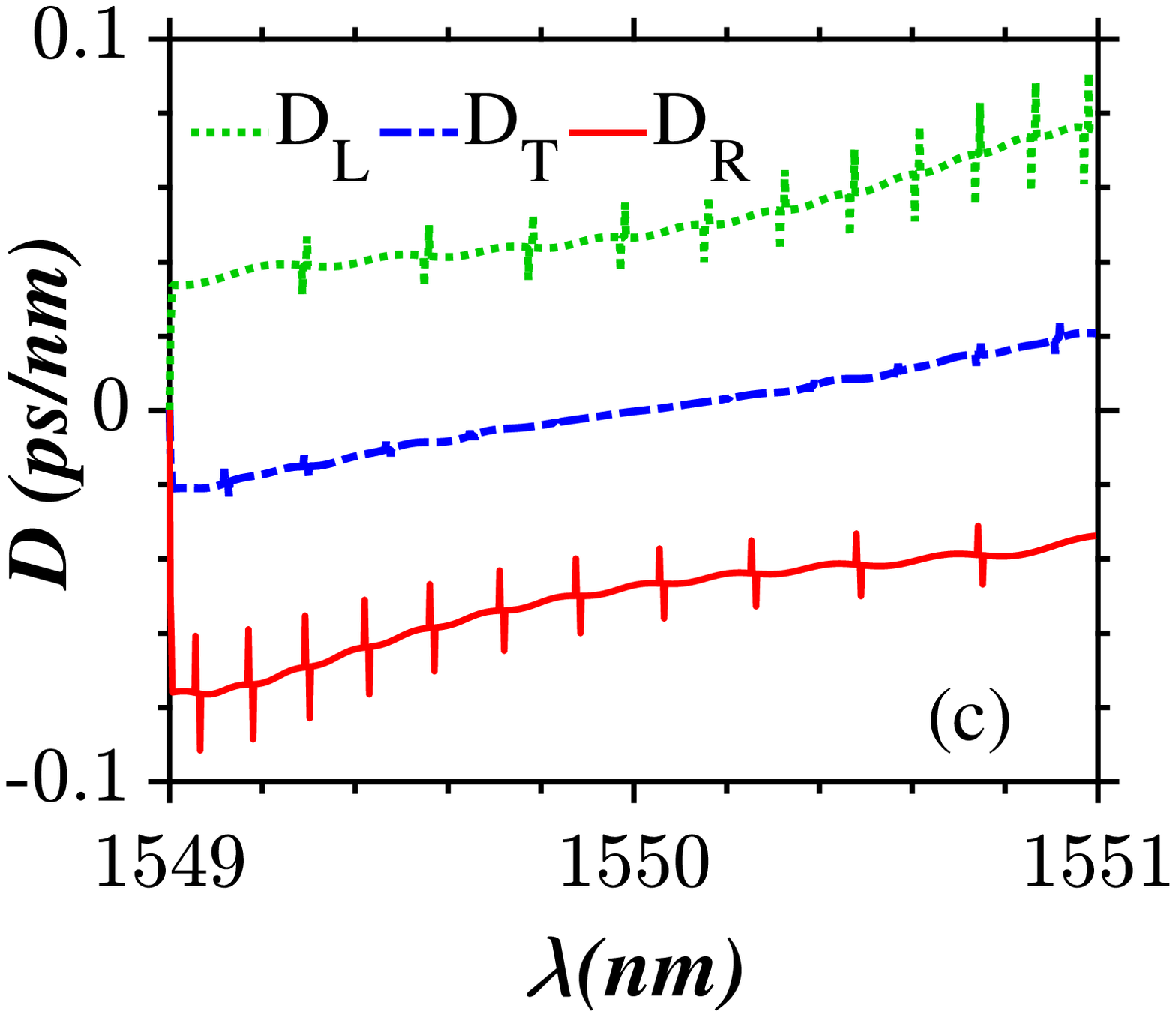}\includegraphics[width=0.5\linewidth]{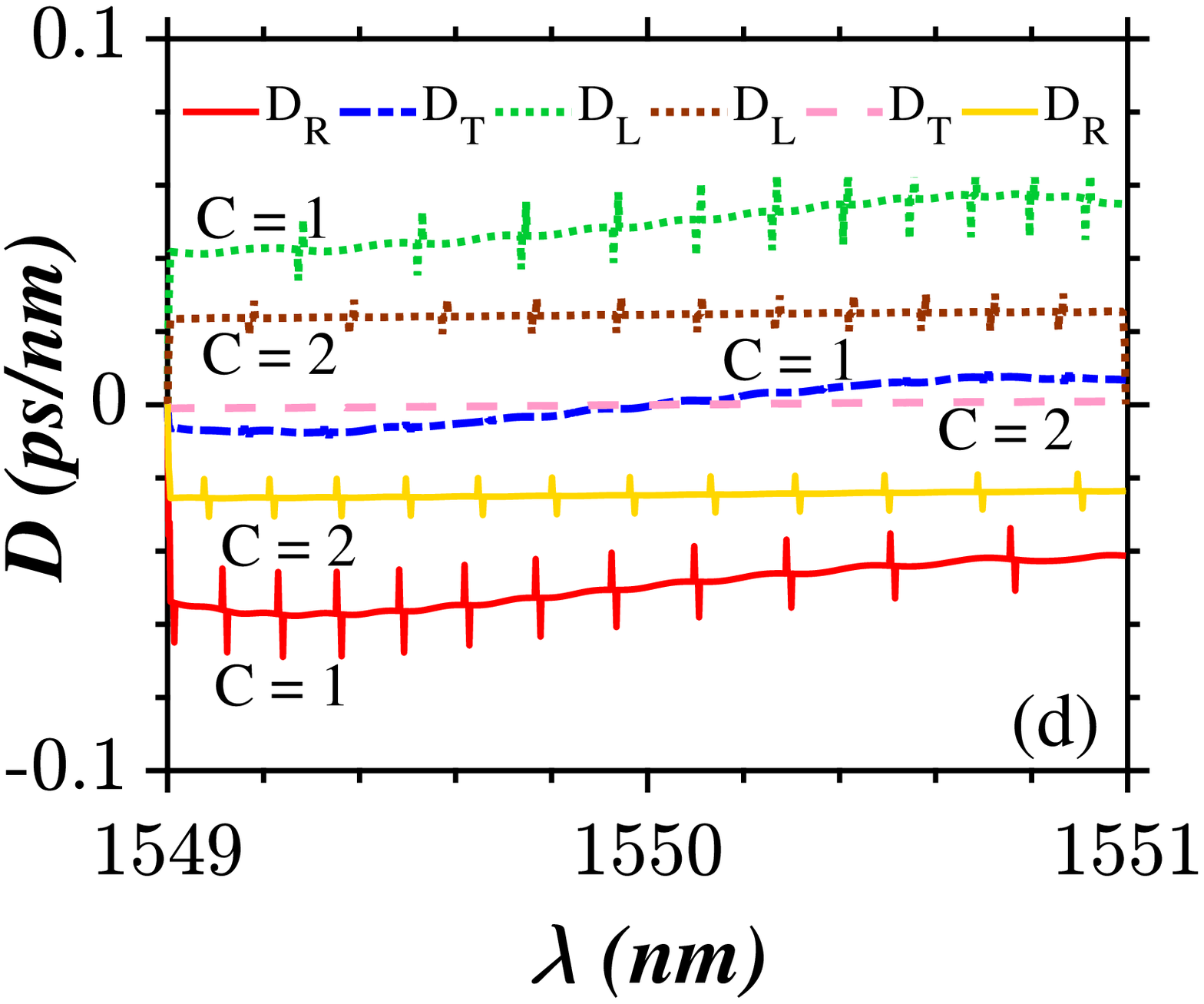}%\\\includegraphics[width=0.5\linewidth]{fig5e}\includegraphics[width=0.5\linewidth]{fig5f}
%	\\\includegraphics[width=0.5\linewidth]{fig7dc}
	\caption{ Delay and dispersion characteristics of a CAPT-FBG (raised-cosine). (a) Shows the role of $\mathcal{PT}$-symmetry on direction dependent delay ($\tau$)  plotted at $n_{1I}$ = 0 and 0.0008 with a chirp of $C = 1$ nm/cm. (b) Depicts the role of chirping ($C$) on delay ($\tau$)  at $n_{1I}$ = 0.0008. The dispersion characteristics in the absence of gain-loss  ($n_{1I}$ = 0)  is illustrated in (c). Influence of $\mathcal{PT}$-symmetry on dispersion (D) simulated at $C = 1$ and 2 nm/cm with a gain-loss parameter value of 0.0008 is shown in (d).}
	\label{fig5}
\end{figure}

\subsection{Gaussian apodization profile }
In Fig. \ref{fig4}(a), the group delay for the transmitted light ($\tau_T$)  is approximately constant around $50$ ps. The group delay for the light reflected from the front end ($\tau_{L}$) increases non-monotonically (with a lot of ripples even at the stop band) with an increase in the wavelengths, while the group delay from the rear end ($\tau_{R}$)  decreases in the same way with an increase in wavelength.Physically, these ripples in the delay characteristics arise as a consequence of the interference between the reflected signals from the grating edges and the nature of reflection within the stopband of the grating.
On either side of the Bragg wavelength (within the stopband), there are fluctuations in the dispersion observed in the same way as illustrated in Fig. \ref{fig4}(a). Outside the stop band, the dispersion is comparatively larger for the reflected lights ($D_L$ and $D_R$) unlike the dispersion of the transmitted light ($D_T$). In both delay and dispersion plots [see Figs. \ref{fig4}(a) and \ref{fig4}(b)], one can observe that there are a lot of ripples within the stop band which implies that one should opt for a better apodization profile like the raised-cosine one shown in Fig. \ref{fig5} while implementing applications like optical delay lines, and dispersion compensation.  Alternatively, such ripples can be reduced by using a Gaussian profile with a large Gauss width parameter \cite{pastor1996design,rebola2002performance}. From Fig. \ref{fig4}(c), we infer that the oscillations in the non-monotonically increasing (decreasing) delay characteristics get reduced when the Gauss width parameter is raised from $G = 5$ to $9$ and for $G = 15$, direction dependent delay characteristics with no ripples is obtained. Similarly, in Fig. \ref{fig4}(d) we conclude that it is possible to construct an efficient dispersion compensator, which is free from any oscillations in the stopband provided that the value of $G$ is sufficiently large.  It is reported that the Gaussian apodization profiles with large $G$ are found to offer an excellent sensitivity at higher values of chirping ($3-4$ nm/cm) in the infrared wavelengths \cite{8345713,rebola2002performance}.

\subsection{Raised-cosine apodization profile }
\begin{figure}[hthb]
 	\centering
 	\includegraphics[width=1\linewidth]{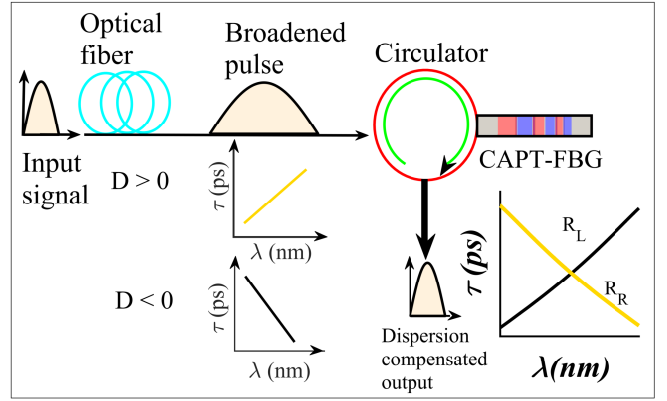}
 	\caption{Schematic of a dispersion compensator build using a CAPT-FBG}
 		\label{fig6}
 \end{figure}

 For the chirped and apodized FBG which is supposed to function as an efficient dispersion compensator the following criteria must be met: first, the delay characteristics of the device is expected to show monotonic increase or decrease with respect to increase in wavelength and must be free from any ripples. Second, these types of linear variations should be observed for a wide range of wavelengths so as to obtain a flat dispersion characteristics over a wide spectral range with high reflectivity \cite{hillet1994chirped,roman1993}. In this perspective, chirped FBGs with a raised-cosine apodization is the optimum choice to compensate dispersion accumulated in the input pulse as it travels through an optical fiber.The physical explanation for the absence of ripples in the delay is  due to the fact that there are no reflections from the ends of the grating for the interference phenomenon discussed in the previous section. As shown in Fig. \ref{fig5}(a), the delay is shifted to higher values with the inclusion of $\mathcal{PT}$-symmetry ($n_{1I} = 0.0008$) compared to the delay offered by a conventional FBG ($n_{1I} = 0$). It is very clear that the time delay for the transmitted light is $\tau_{T} \sim 50$ ps. The small number of fluctuations in the group delay plots which were visible in the presence of Gaussian apodization (see Fig. \ref{fig4}) is suppressed by employing a raised-cosine apodization profile as seen in Fig. \ref{fig5}(a). We also observe that the curves featuring $\tau_{L}$ and $\tau_{R}$ resemble ideal increasing and decreasing functions, respectively. Physically, this means that the optical signal launched from the front end of a CAPT-FBG (raised-cosine) is reflected such that only a small fraction of the short-wavelength light propagates and reaches the the other end of the grating, while most of the signals in the longer wavelength light arrive at the other end and vice-versa. As portrayed in Fig. \ref{fig5}(b), any increase (decrease) in chirping increases (decreases) the delay on the shorter wavelength side and decreases (increases) the delay in the longer wavelength sides (light solid, dashed, and dash-dotted lines). For the light launched from the right side the exact opposite phenomenon (dark solid, dashed, dash-dotted lines) occurs. Ideally, the dispersion curve should be flat over the entire stop band but this requirement is not fully satisfied by the FBG system shown in Fig. \ref{fig5}(c) which is plotted in the absence of $\mathcal{PT}$-symmetry ($n_{1I}$ = 0). But the curve resembling an ideal flat dispersion is obtained by the inclusion of $\mathcal{PT}$-symmetry ($n_{1I}$ = 0.0008). The mean dispersion value for left incidence [red (dark) solid lines] is measured to be $D = -0.045$ ps/nm, whereas for the right light incidence [green (light) dotted lines] the same magnitude of dispersion with opposite sign ($D = 0.045$ ps/nm) is obtained. Thus, it is possible to compensate both normal and anomalous dispersions emanating from the propagation of the signal inside the transport fiber by the use of appropriate CAPT-FBG (raised-cosine) with flat dispersion characteristics thanks to the notion of $\mathcal{PT}$-symmetry (See Fig. \ref{fig6}). Other important application of CAPT-FBG (raised-cosine) in the unbroken $\mathcal{PT}$-symmetry includes add and drop multiplexer-demultiplexer and delay lines.
 
\section{Effect of nonuniformities on the Spectra of a broken $\mathcal{PT}$-symmetric FBG }
\begin{figure}[t]
	\centering
	\includegraphics[width=0.5\linewidth]{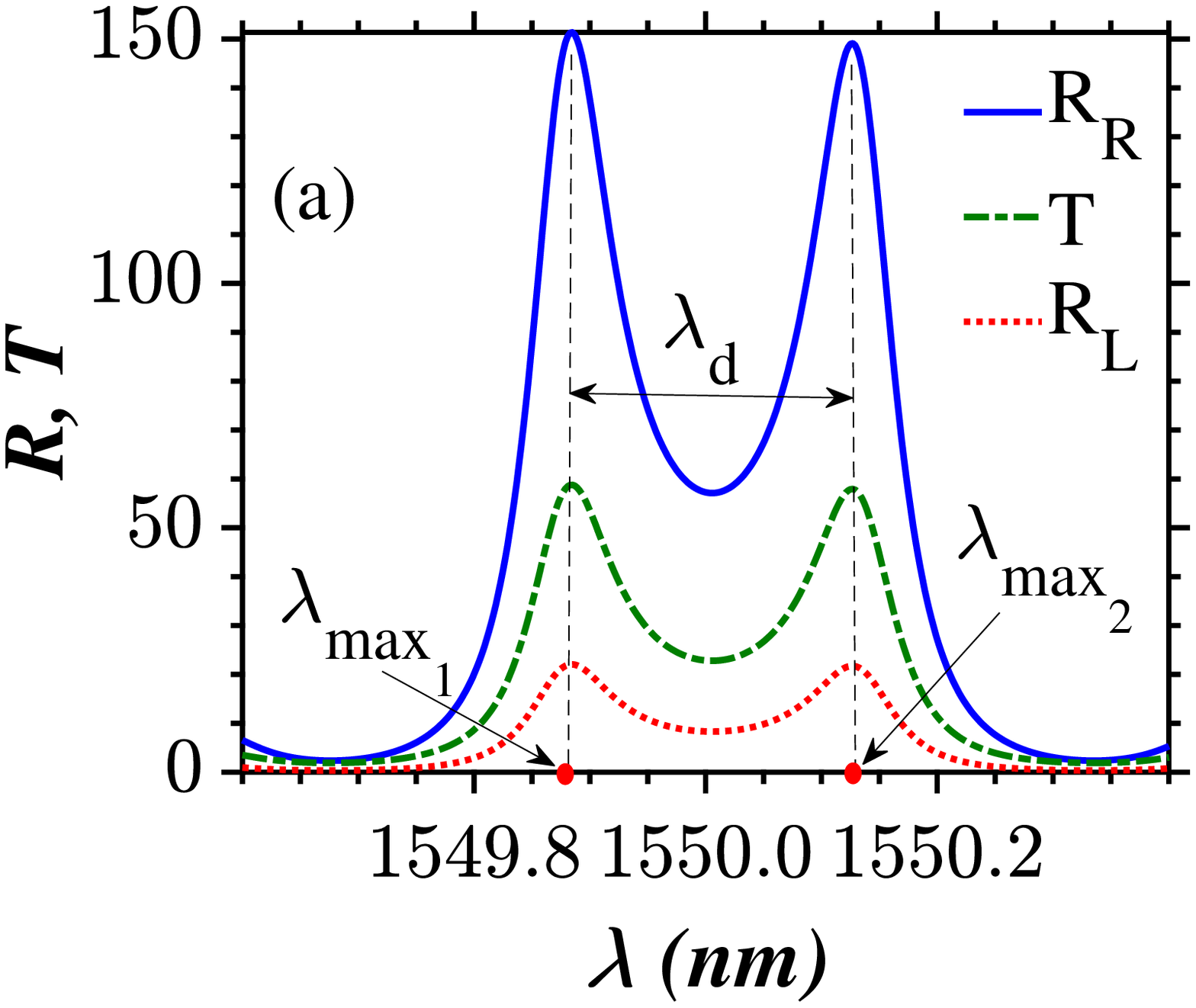}\includegraphics[width=0.5\linewidth]{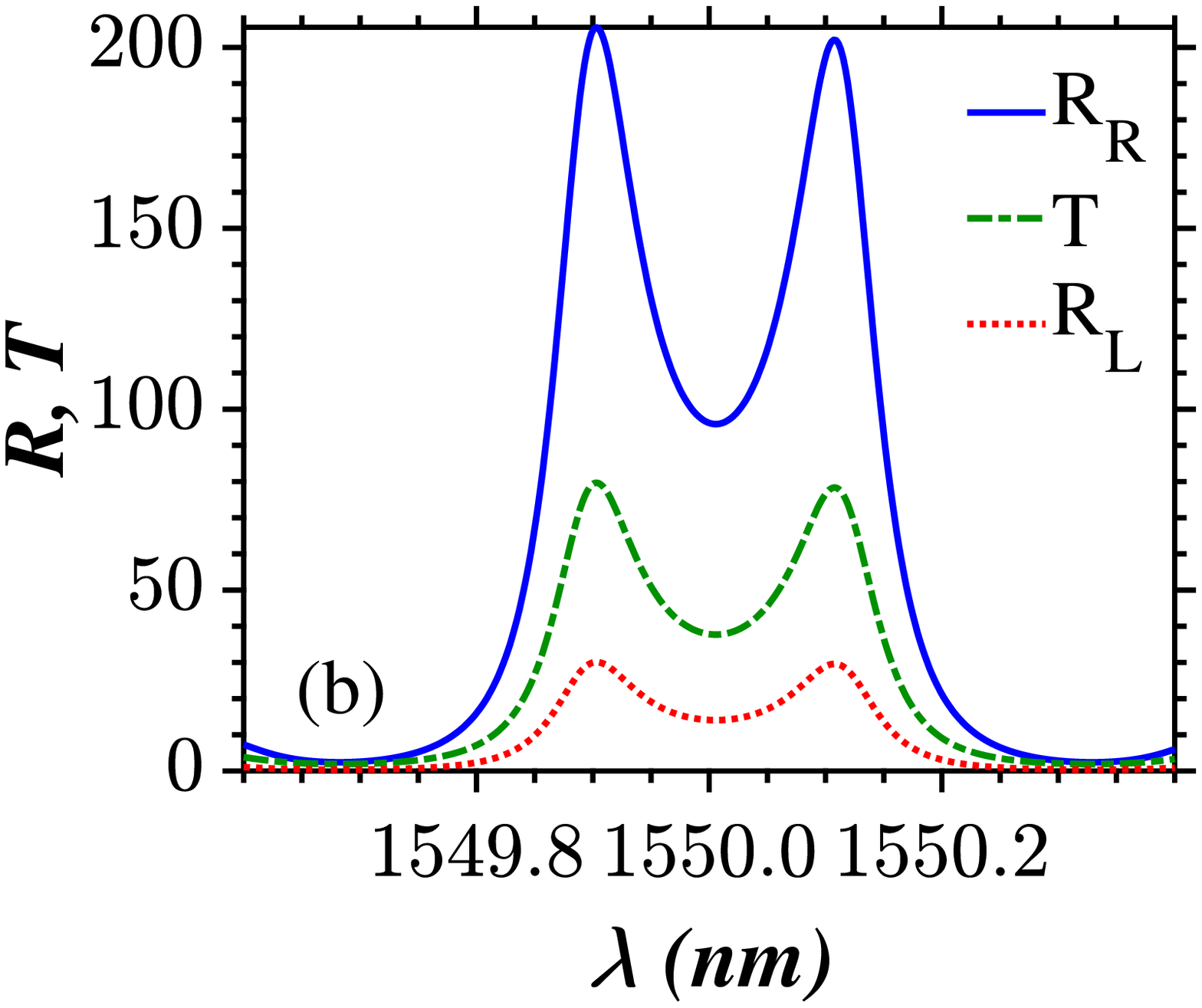}\\\includegraphics[width=0.5\linewidth]{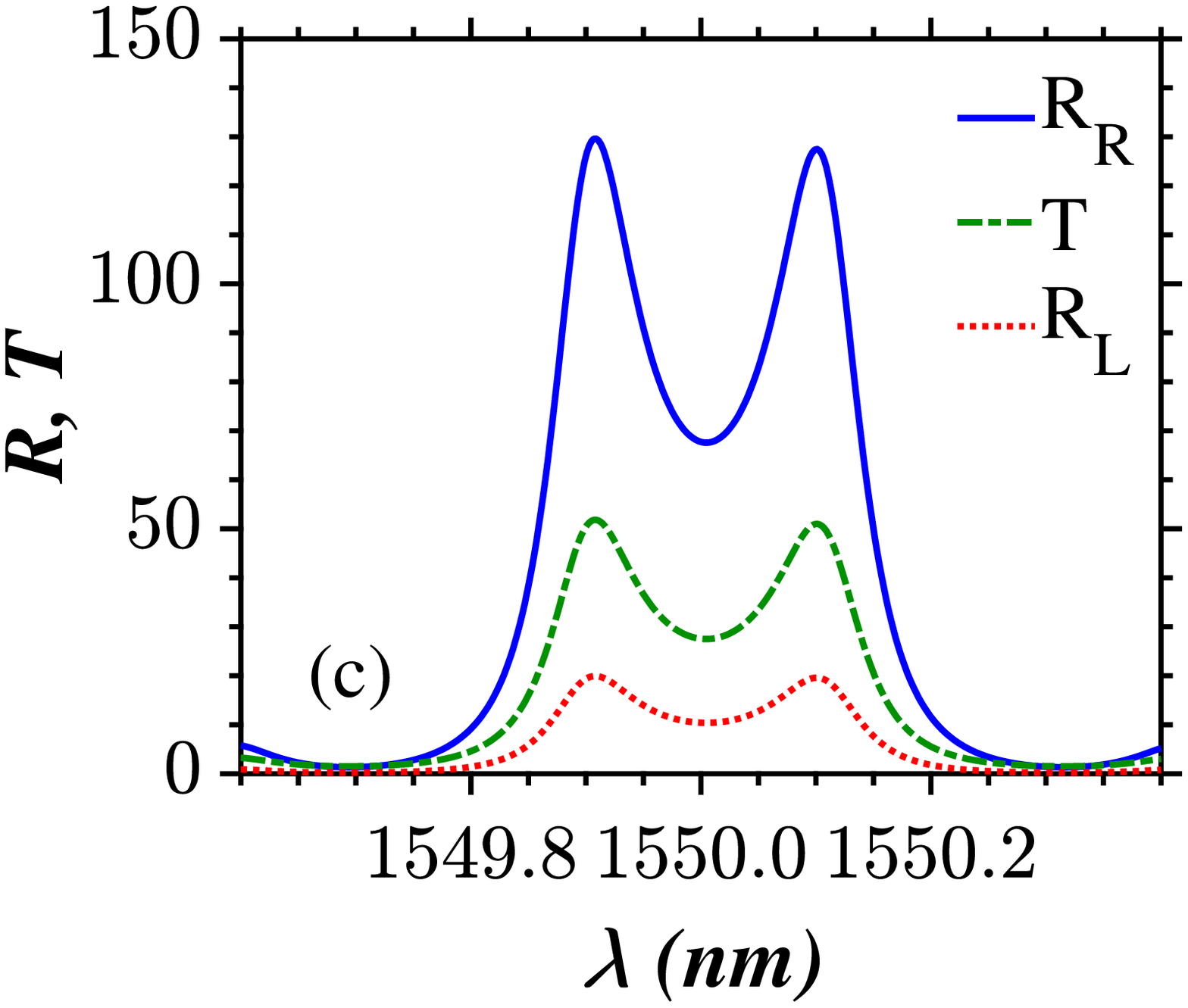}\includegraphics[width=0.5\linewidth]{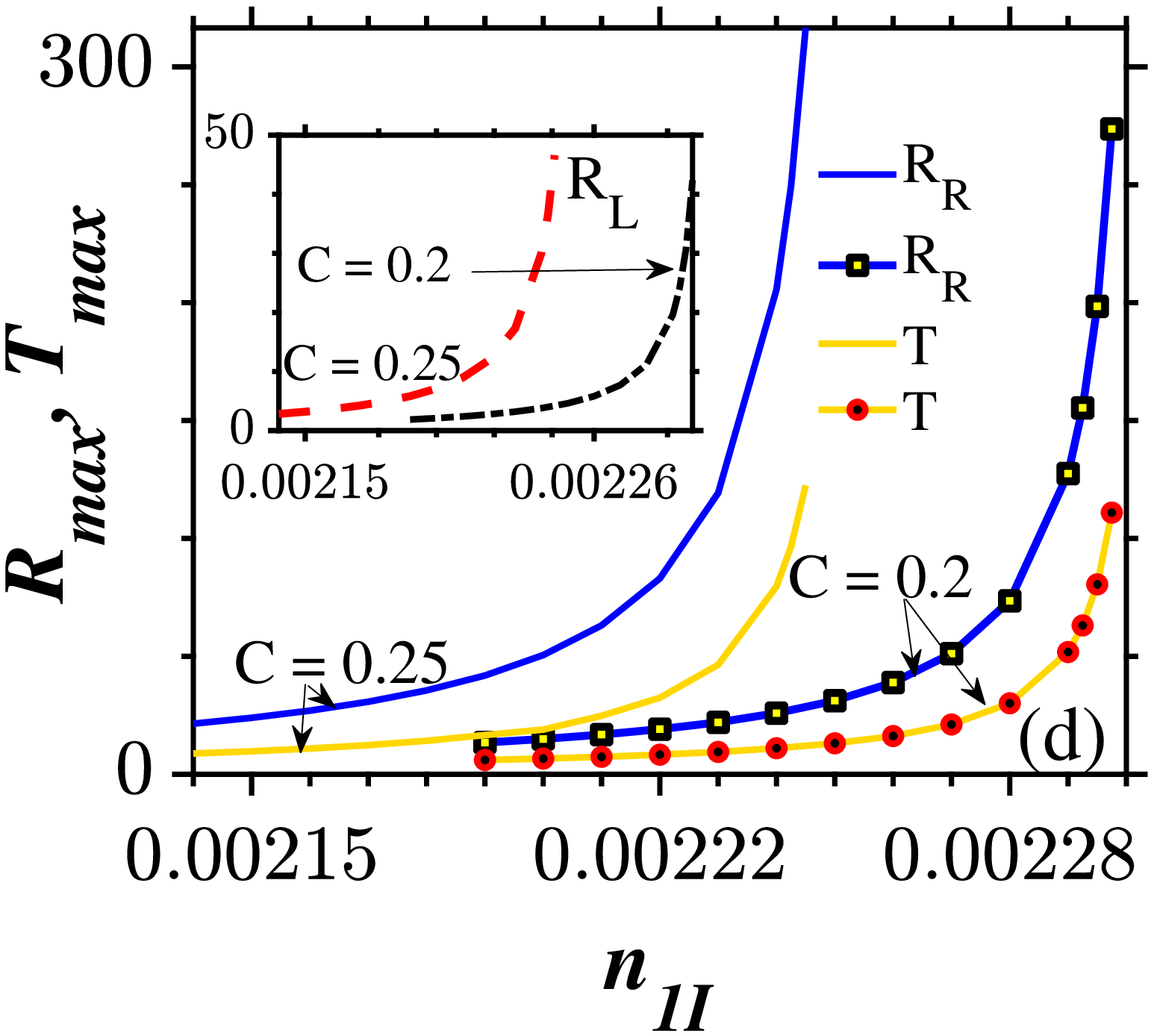}\\\includegraphics[width=0.5\linewidth]{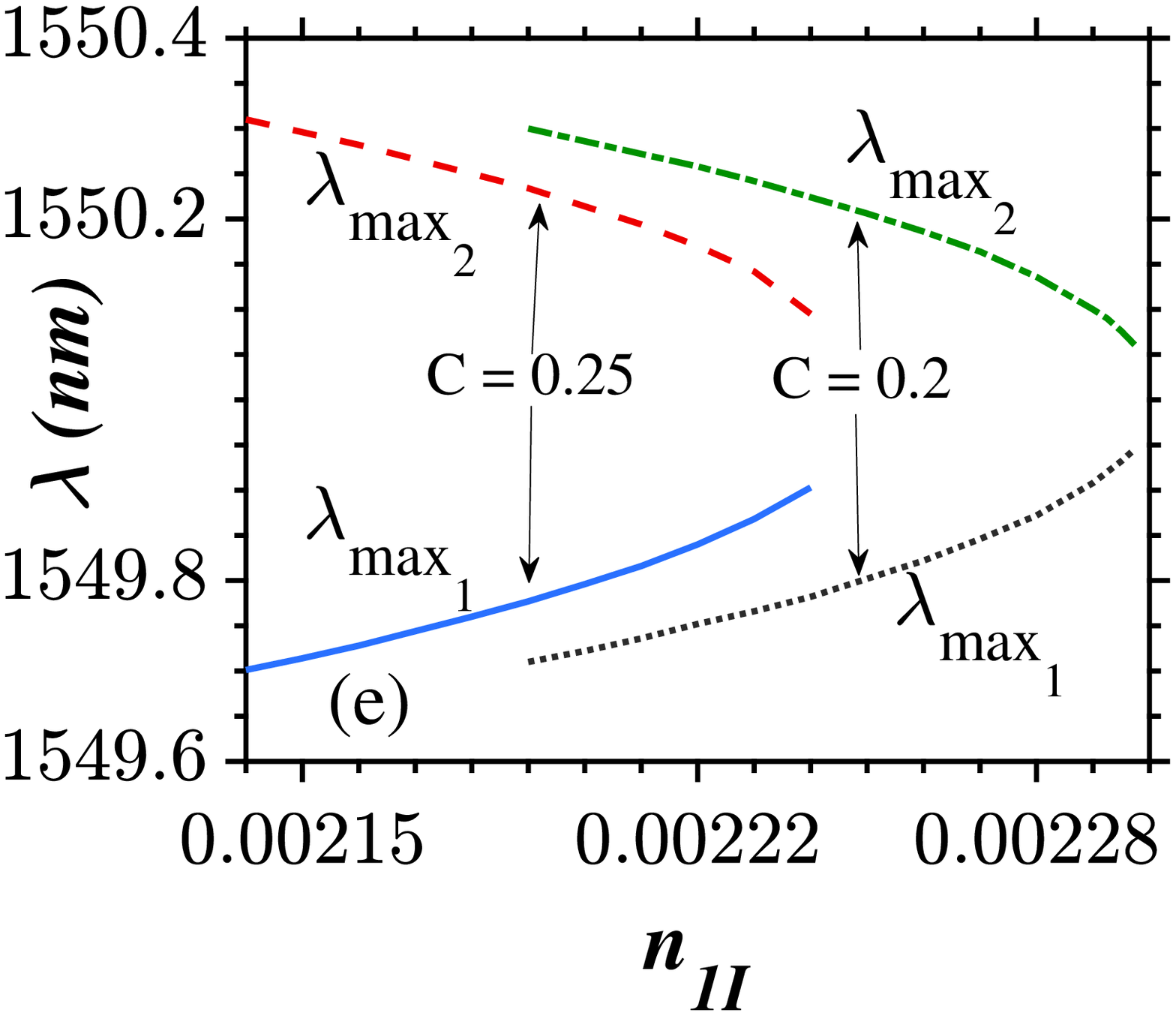}\includegraphics[width=0.5\linewidth]{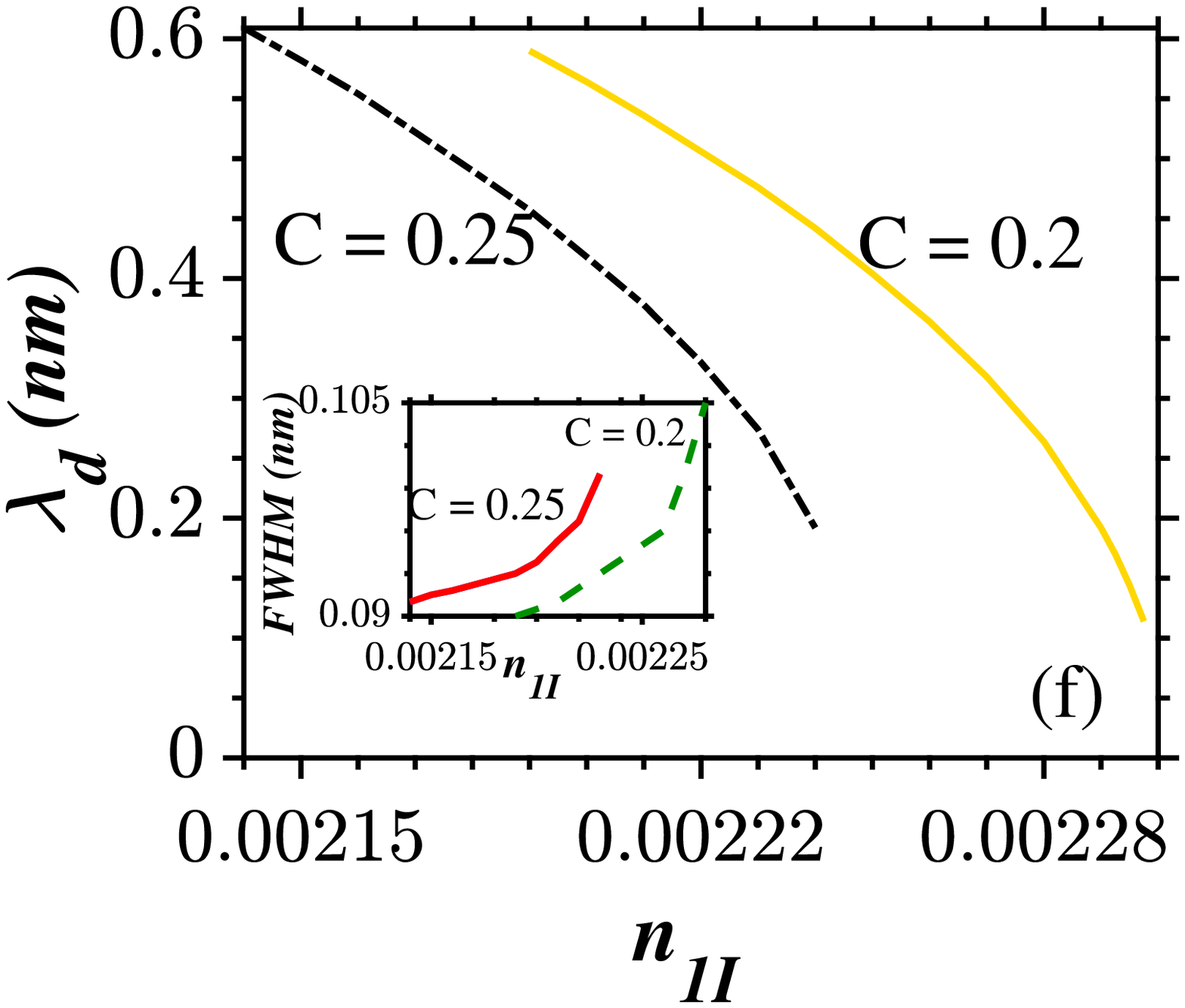}
	\caption{Transmission and reflection characteristics of CAPT-FBG (Gaussian) in the broken $\mathcal{PT}$-symmetric regime. (a) and (b) Show the role of gain-loss on the spectra simulated at $C = 0.25$ nm/cm with $n_{1I} = 2.22 \times 10^{-3}$ and $2.24 \times 10^{-3}$, respectively. (c) Illustrates the decrease in  reflectivity and transmittivity $R$ and $T$ of the spectra when the chirping is lowered to $C = 0.2$ nm/cm at $n_{1I} = 2.285 \times 10^{-3}$. Influence of  $n_{1I}$ at fixed values of chirping ($C = 0.25$ nm/cm and $0.2$ nm/cm) on  (d) the maximum reflectivity and transmittivity ($R_{max}$ and $T_{max}$), (e) wavelengths corresponding to maximum reflectivity and transmittivity ($\lambda_{{max}_{1}}$ and $\lambda_{{max}_{2}}$), and (f) spectral separation ($\lambda_d$) between the two wavelengths at which maximum reflectivity (transmittivity) occurs when $n_{1I}$ is varied. The inset in (f) shows full width maximum of the spectra with variation in the gain-loss at constant chirping.}
	\label{fig7}
\end{figure}
\begin{figure}[hthb]
	\centering
	\includegraphics[width=0.5\linewidth]{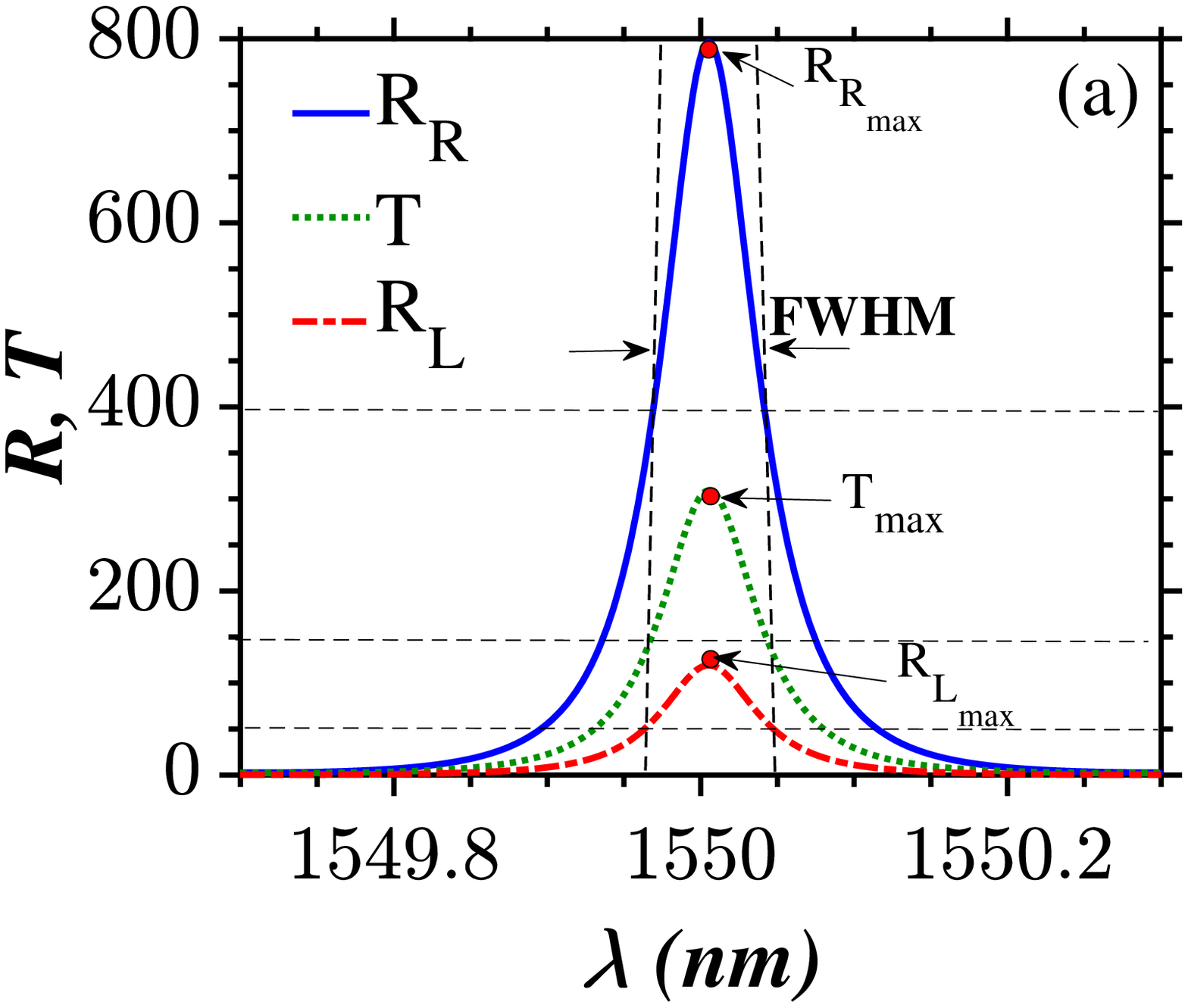}\includegraphics[width=0.5\linewidth]{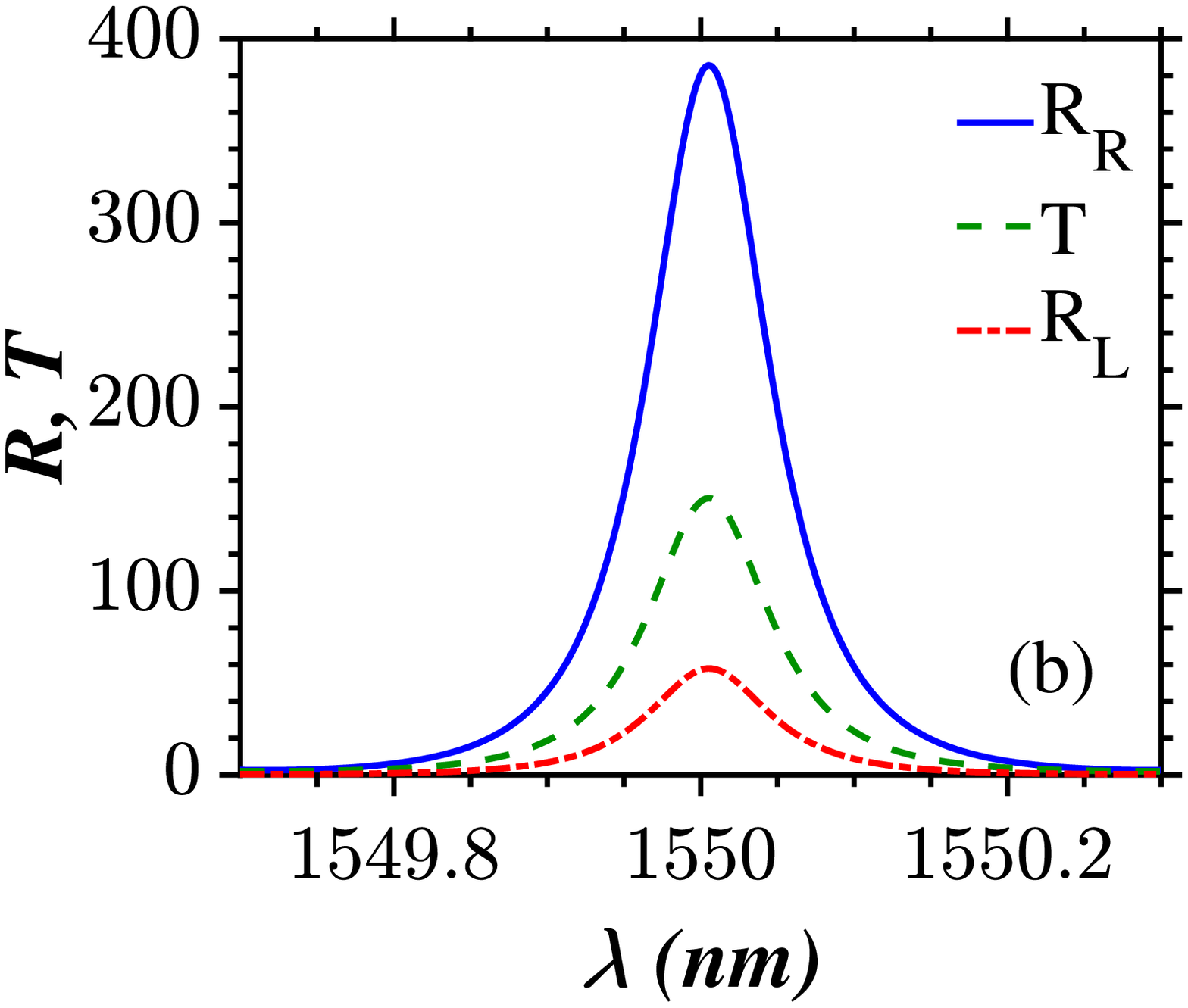}\\\includegraphics[width=0.5\linewidth]{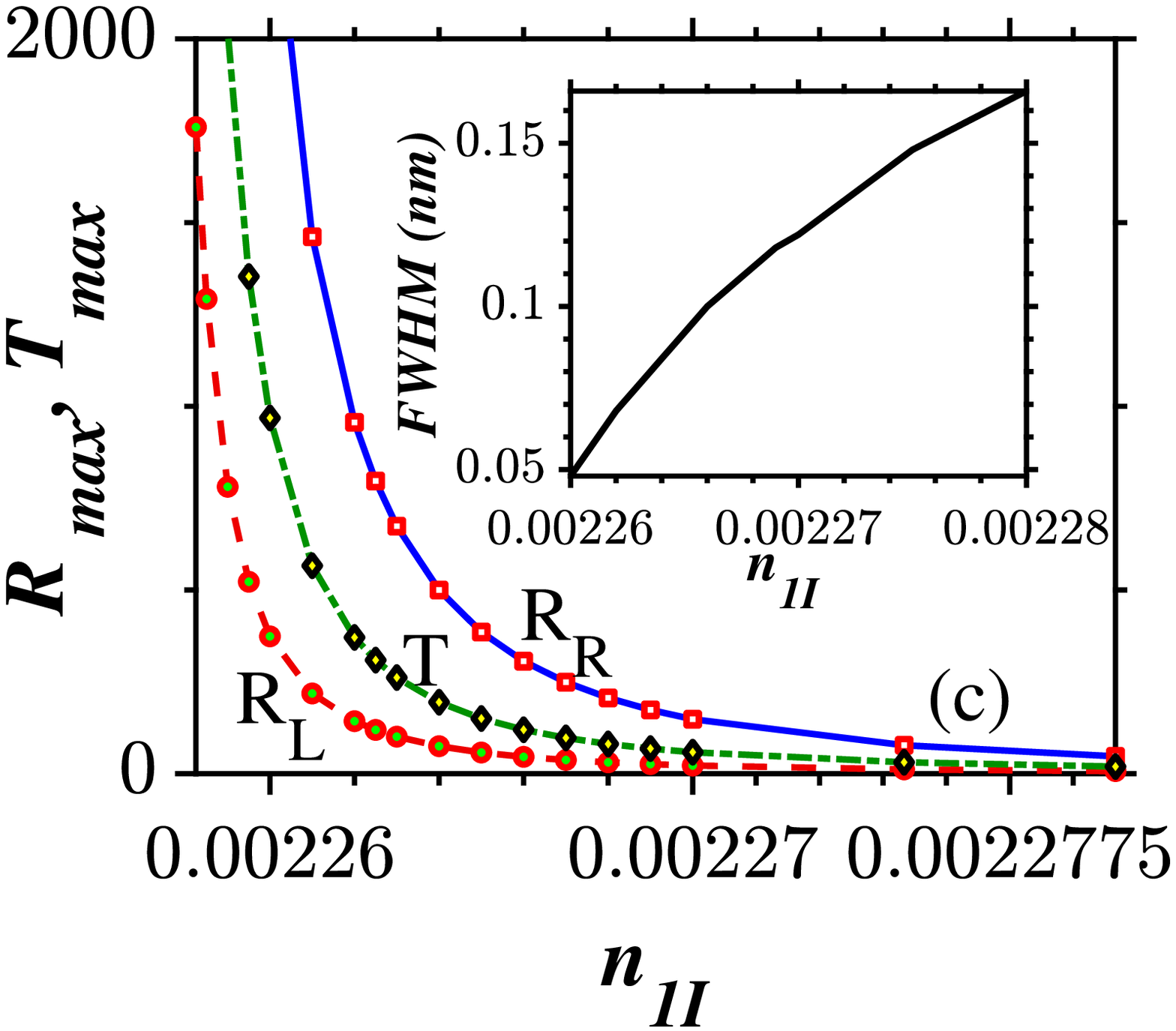}\includegraphics[width=0.5\linewidth]{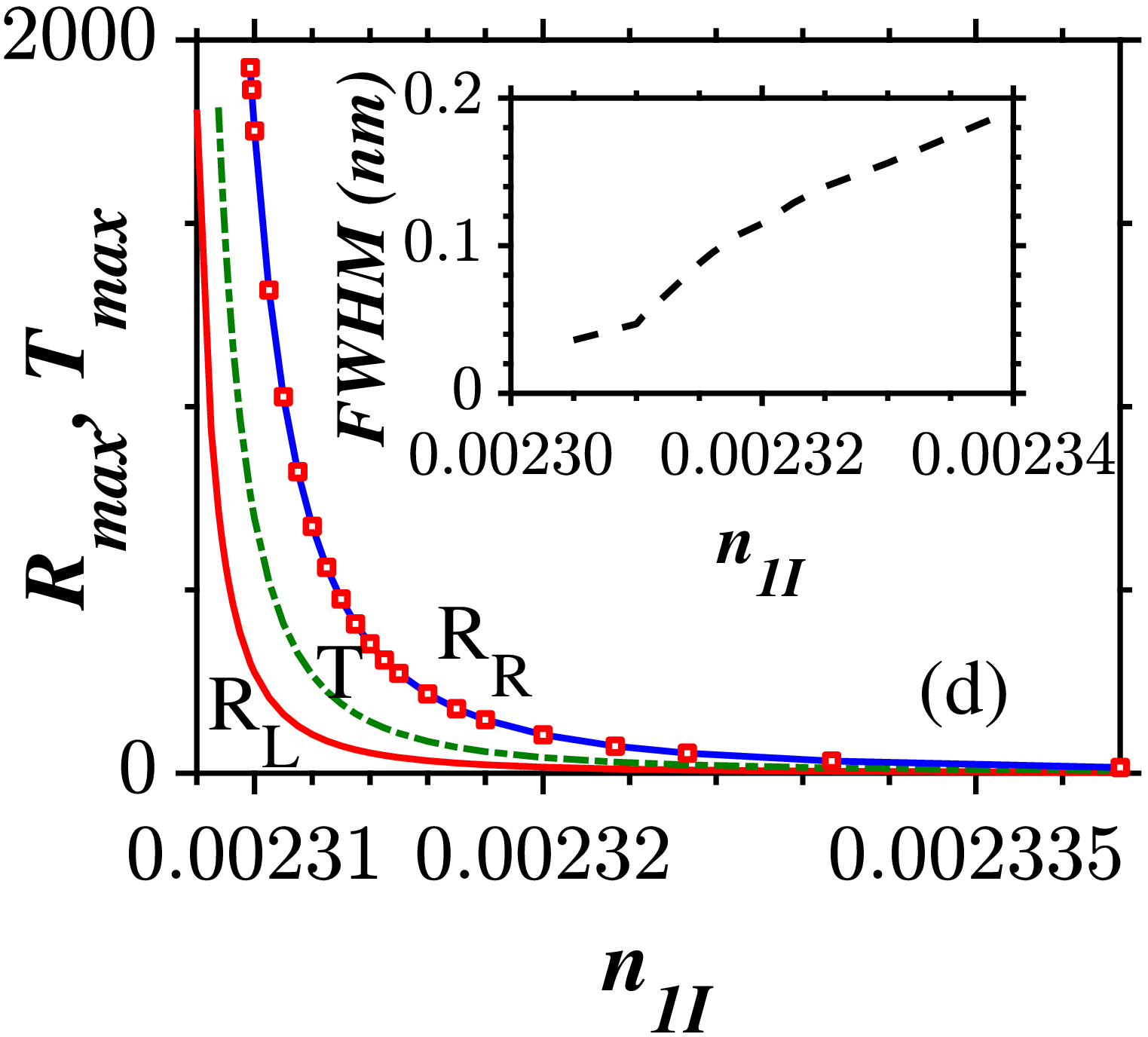}\\\includegraphics[width=0.5\linewidth]{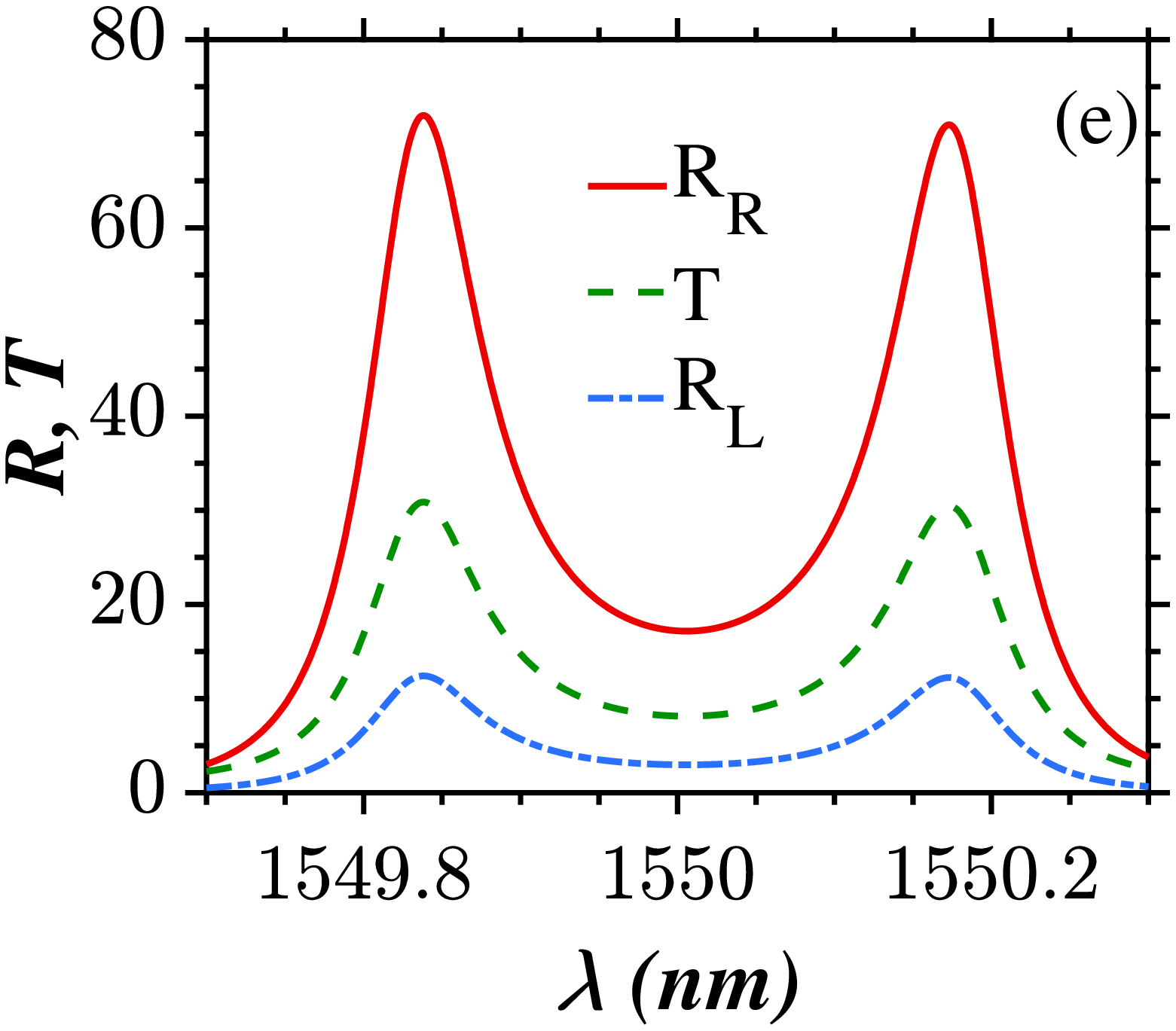}\includegraphics[width=0.5\linewidth]{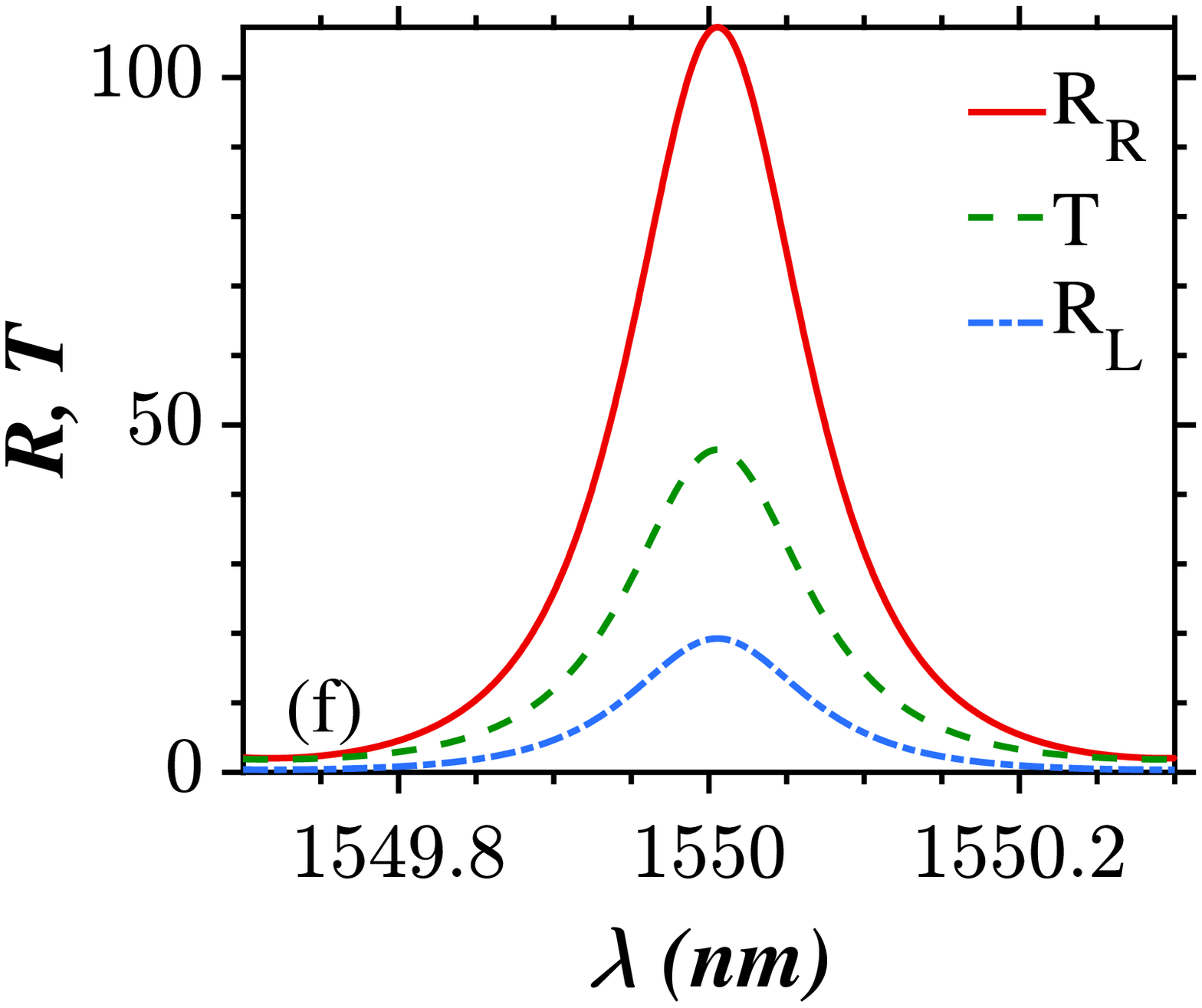}%\includegraphics[width=0.5\linewidth]{fig3f}
	\caption{Single mode lasing behavior in a broken CAPT-FBG (Gaussian). (a) Shows the spectra at $C = 0.25$ nm/cm and $n_{1I} = 2.2625 \times 10^{-3}$, (b)  Illustrates the decrease in reflectivity and transmittivity  when $n_{1I}$ is increased to $2.265 \times 10^{-3}$ and $C$ is same as (a). (c) and (d) Depict the continuous variation of maximum reflectivity and transmittivity  ($R_{max}$, $T_{max}$) against $n_{1I}$ at $C = 0.25$ nm/cm and 0.2 nm/cm, respectively. The insets in (c) and (d) show the corresponding variation in the full width half maximum of the broken CAPT-FBG spectra. (e) and (f) Show dual-mode amplification and single-mode lasing behavior in the broken CAPT-FBG spectra at $n_{1I} = 0.00242$ and $0.00247$, respectively, for a chirping of $C = 0.25$ nm/cm.}
	\label{fig8}
\end{figure}

\begin{figure}[t]
	\centering
	\includegraphics[width=0.5\linewidth]{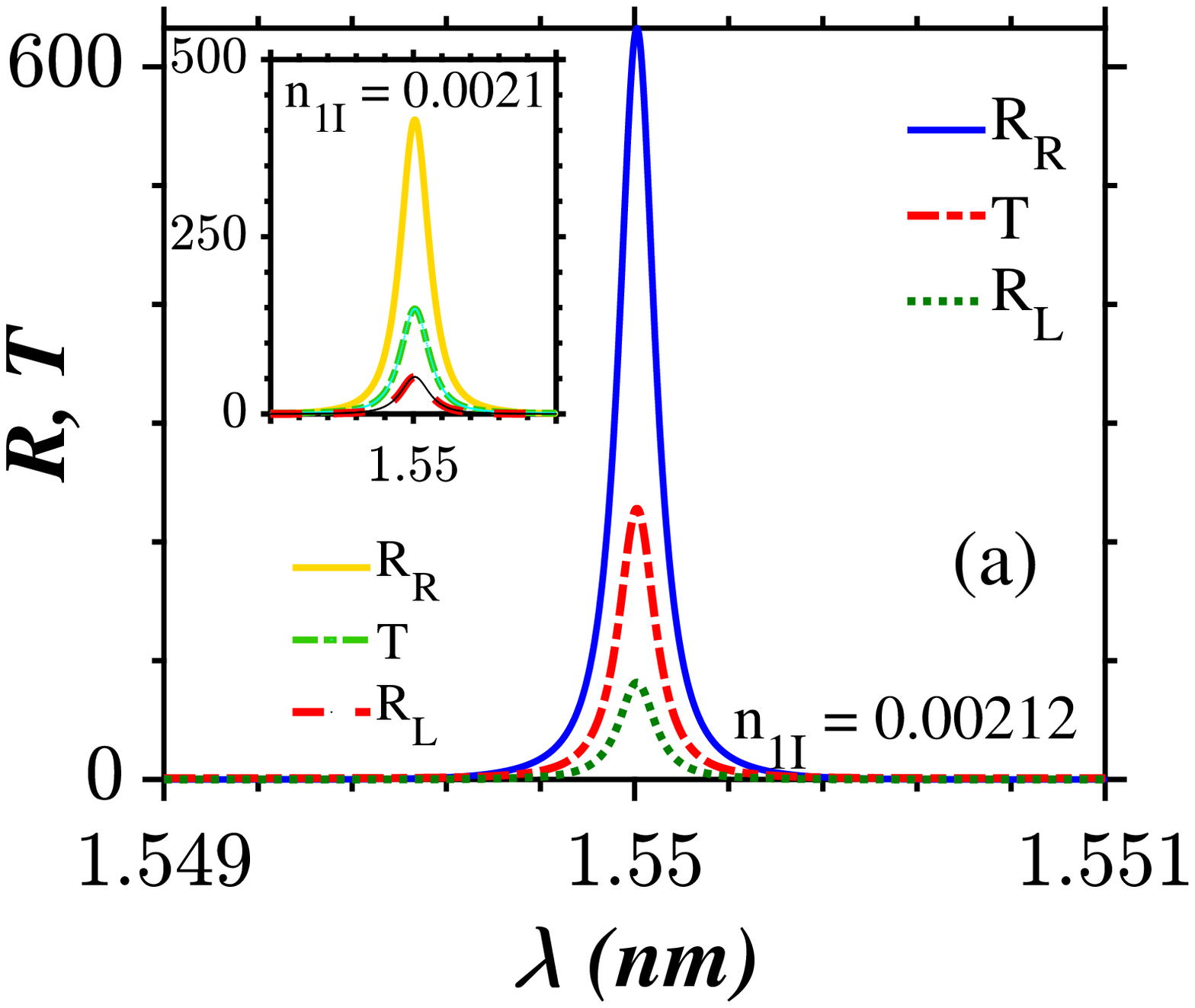}\includegraphics[width=0.5\linewidth]{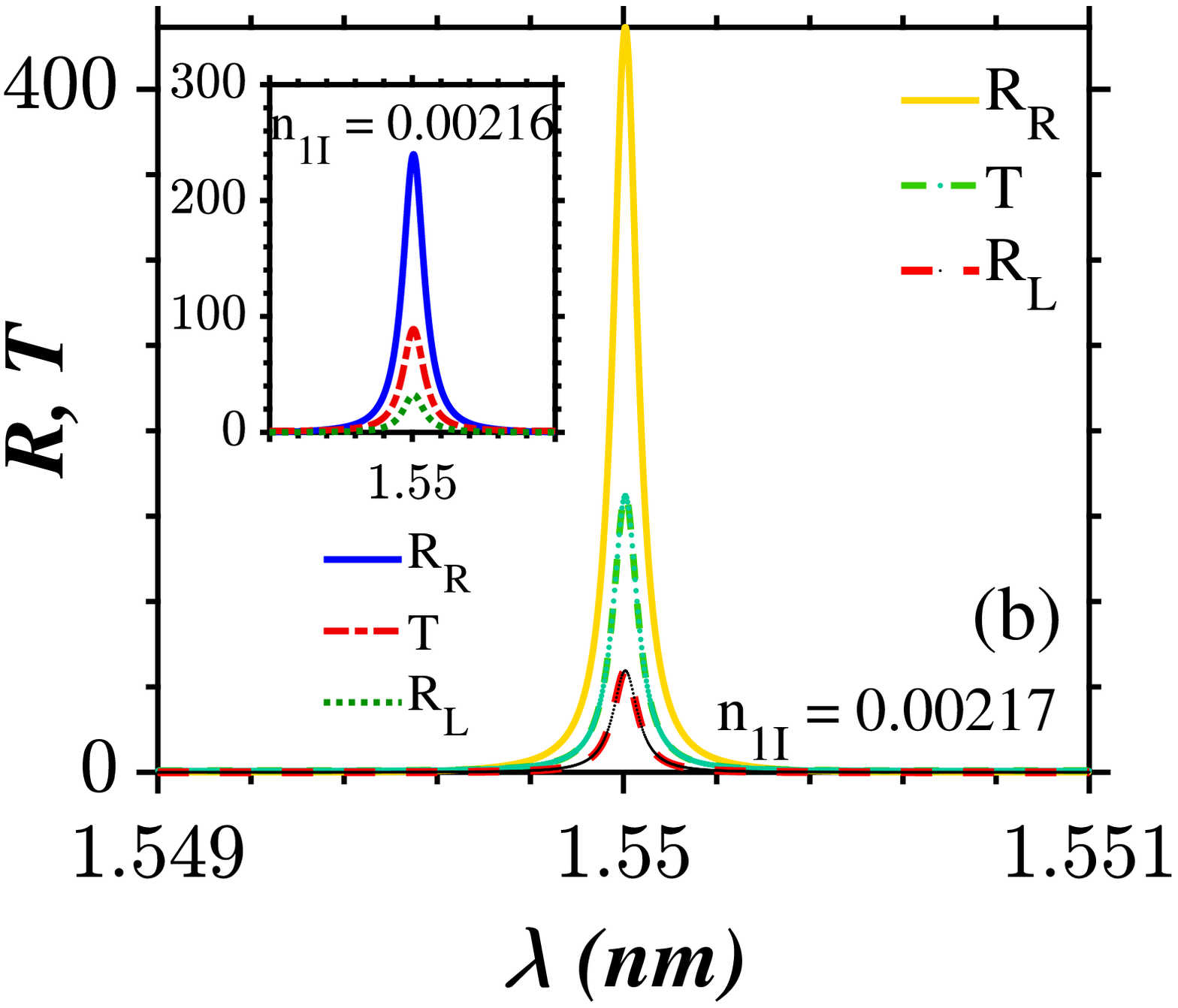}\\\includegraphics[width=0.5\linewidth]{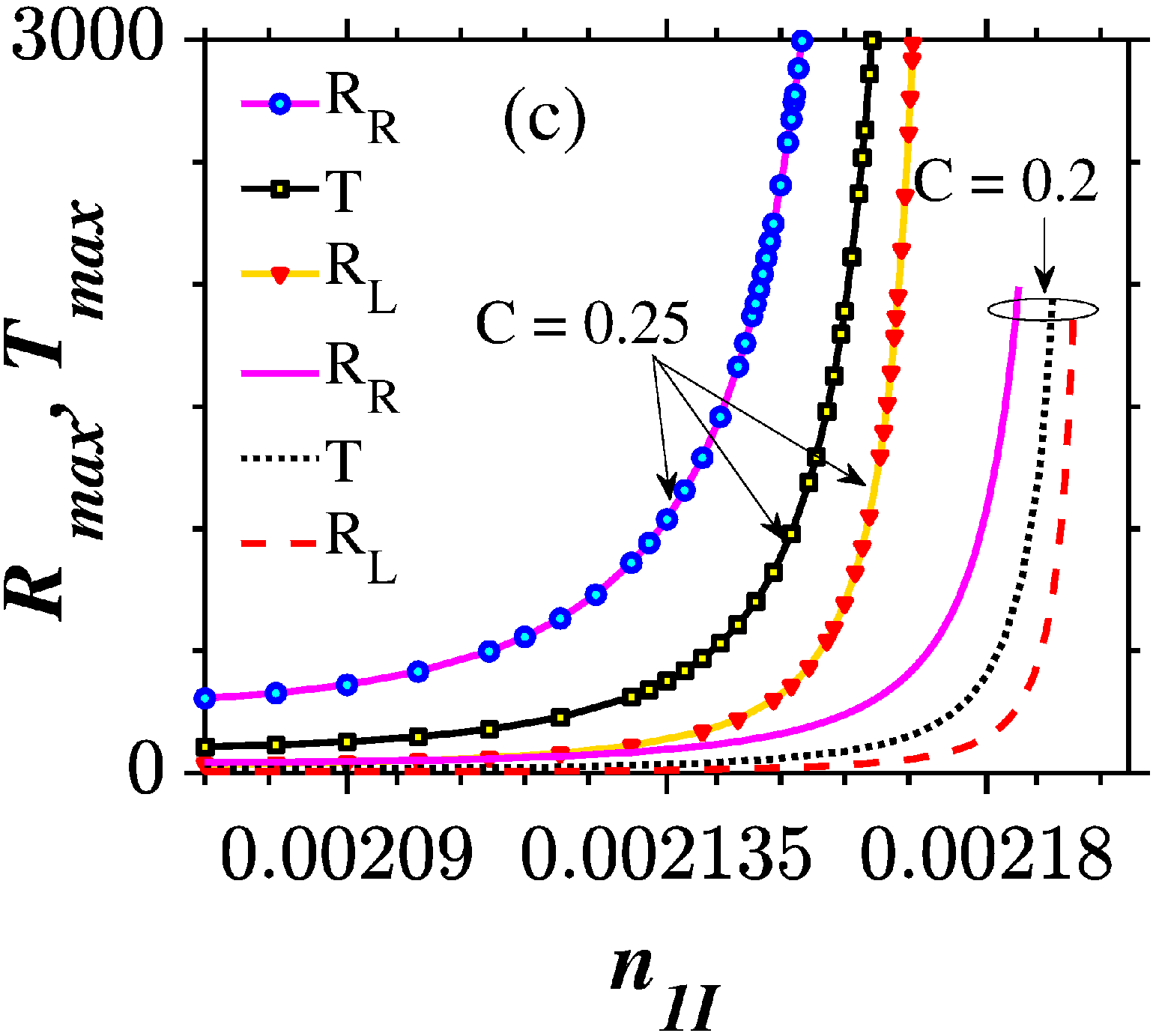}\includegraphics[width=0.5\linewidth]{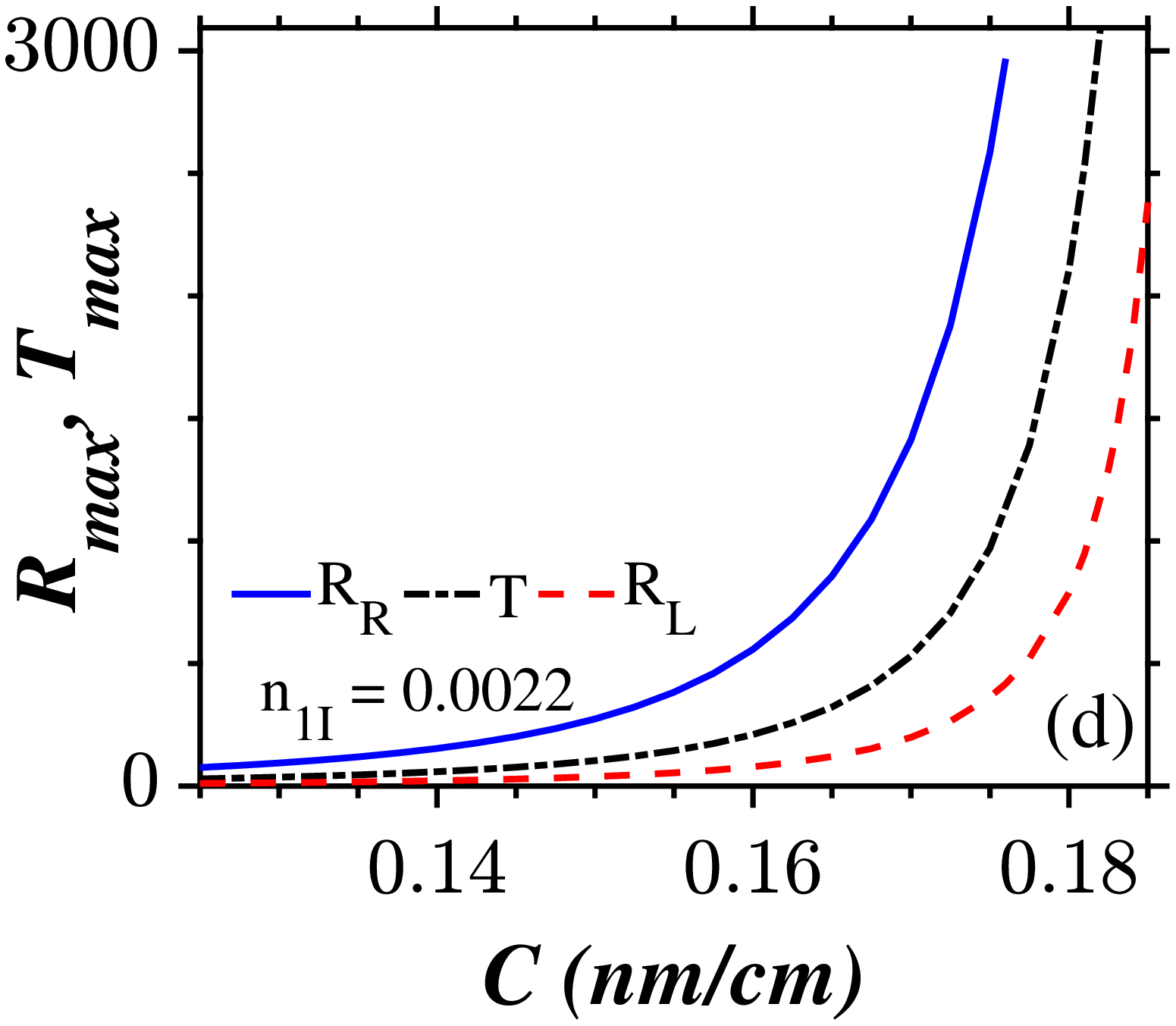}\\\includegraphics[width=0.5\linewidth]{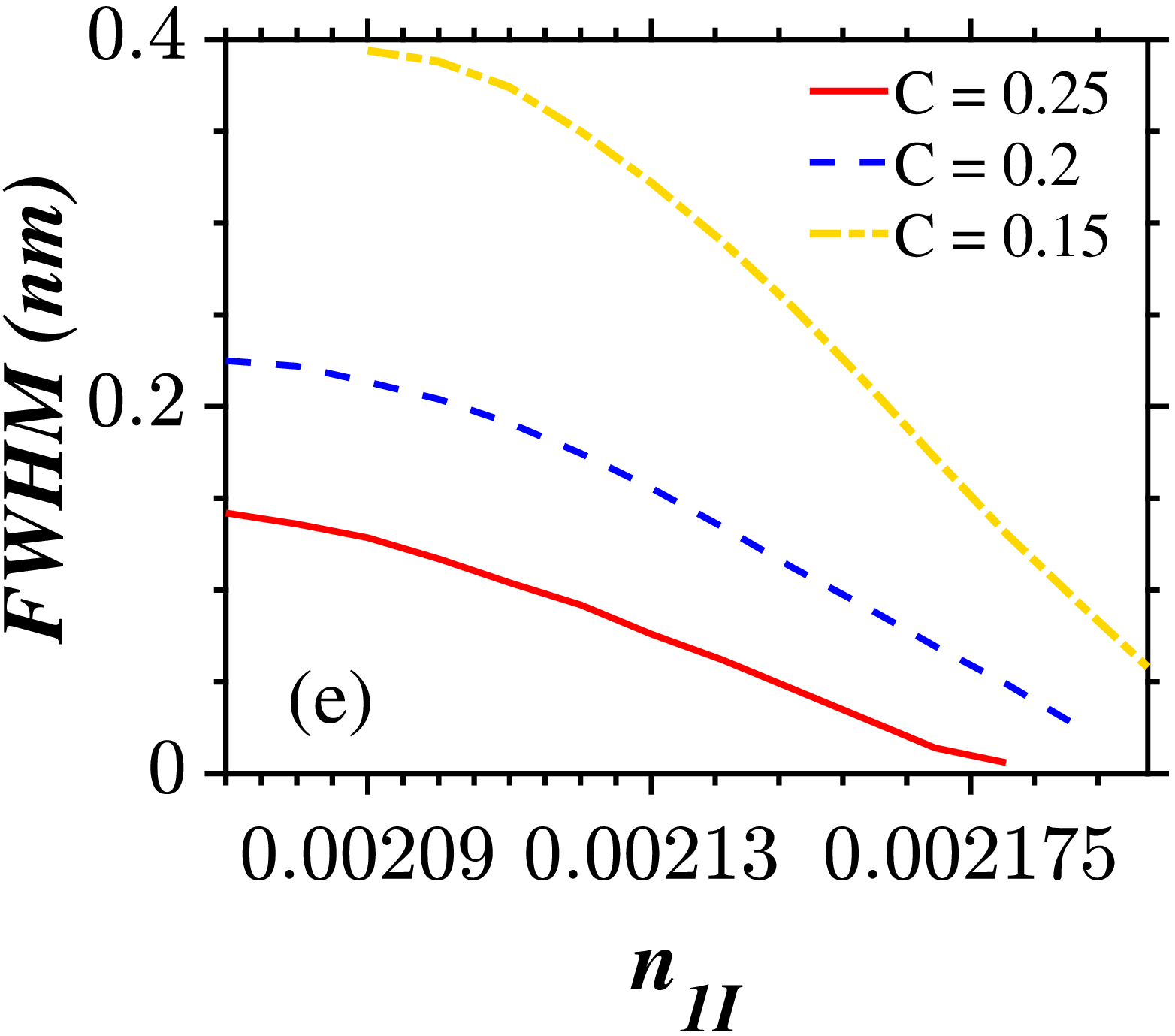}\includegraphics[width=0.5\linewidth]{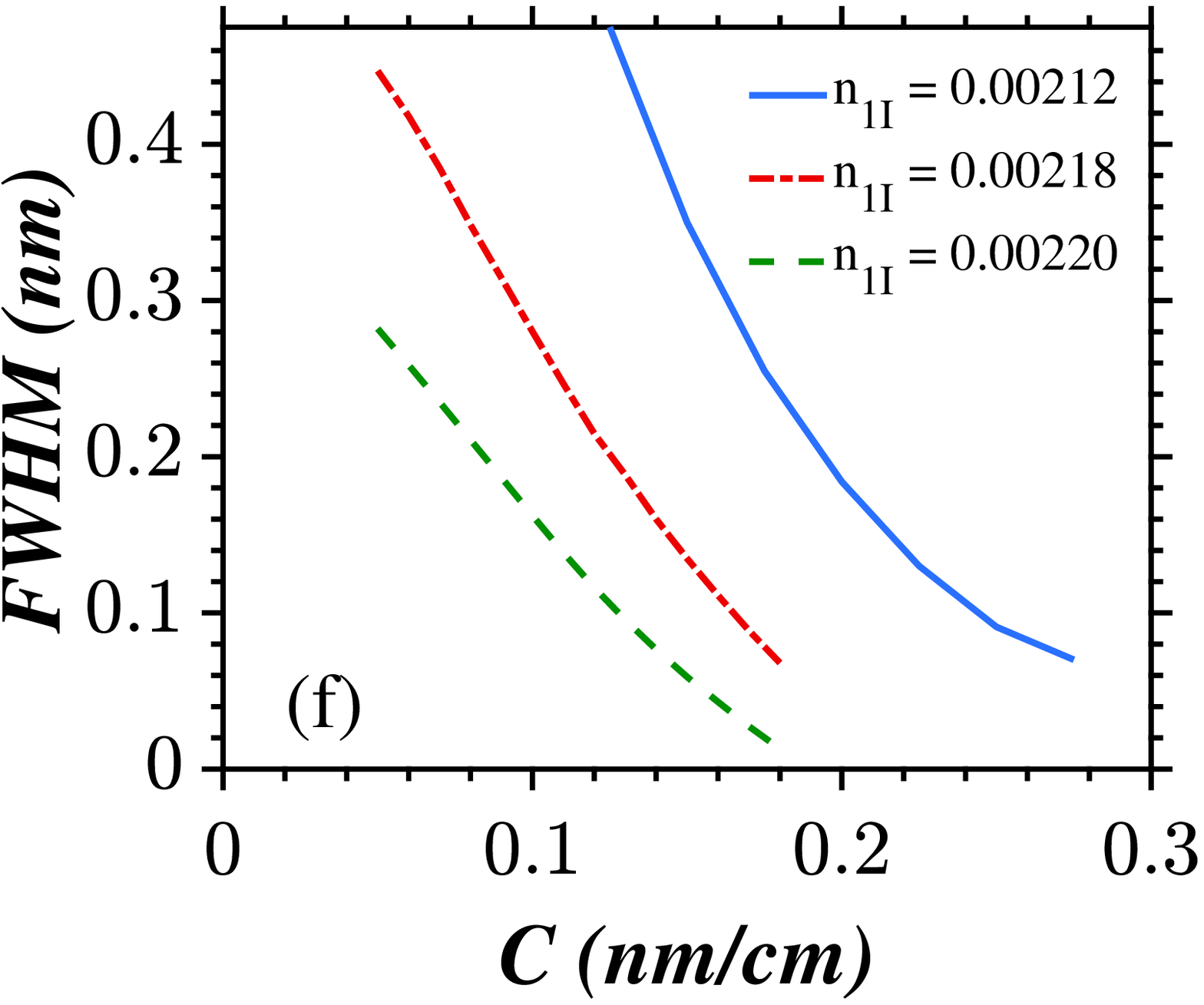}
	\caption{Single-mode lasing behavior of a broken $\mathcal{PT}$-symmetric chirped FBG with raised-cosine apodization. (a) and (b) Illustrate the spectral features at $C = 0.25$ nm/cm and for $C = 0.2$ nm/cm, respectively at different values of $n_{1I}$. The continuous variation of maximum reflectivity and transmittivity ($R_{max}$ and $T_{max}$) is plotted against (c) varying $n_{1I}$ and constant chirping ($C = 0.25$ nm/cm and $0.2$ nm/cm) and (d) varying chirping at a constant $n_{1I} = 2.2 \times 10^{-3}$. The corresponding variations in FWHM of the spectra against $n_{1I}$  and $C$ are shown in (e) and (f), respectively.} 
	\label{fig9}
\end{figure}

\begin{figure}[t]
	\centering
	\includegraphics[width=0.5\linewidth]{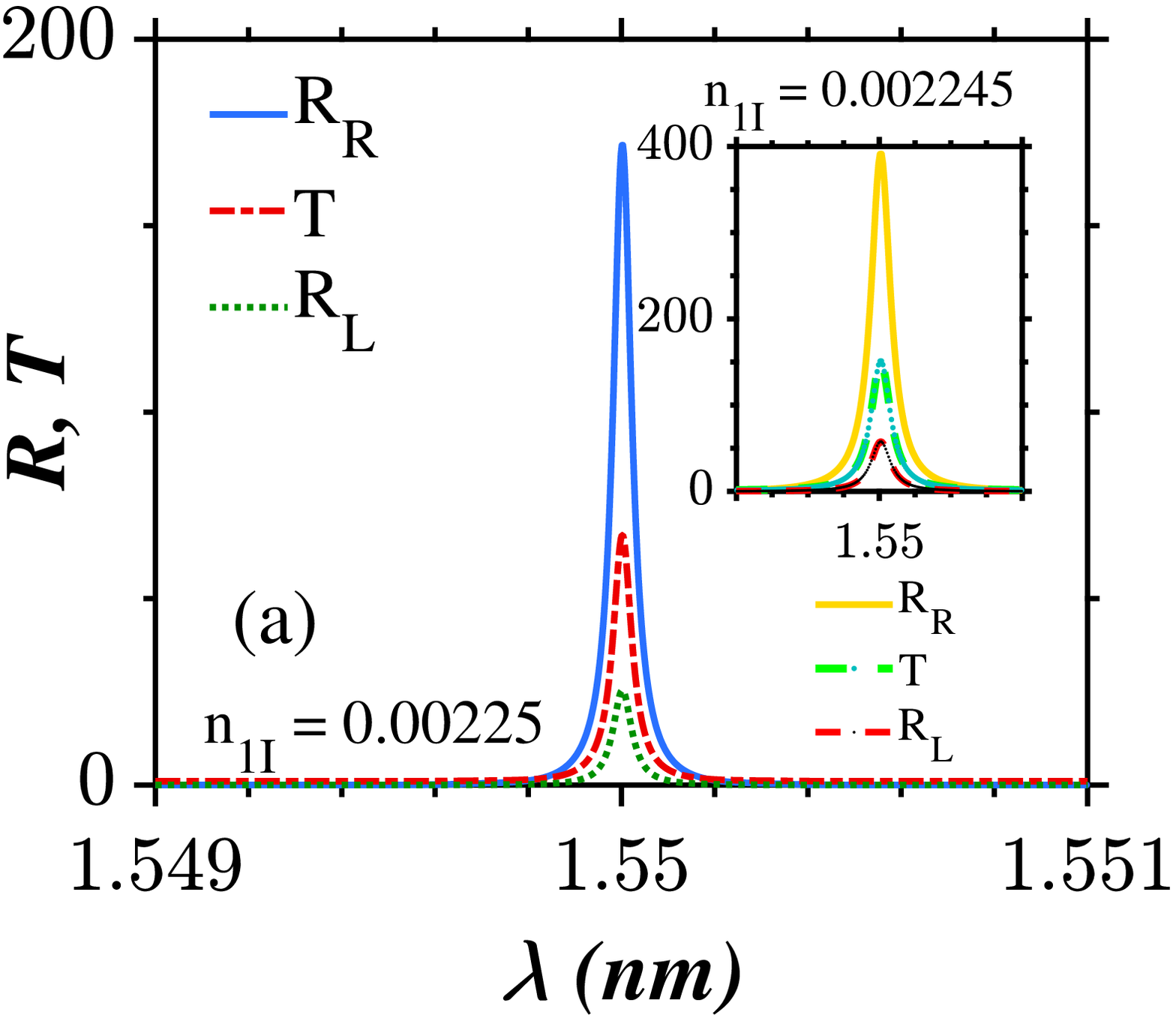}\includegraphics[width=0.5\linewidth]{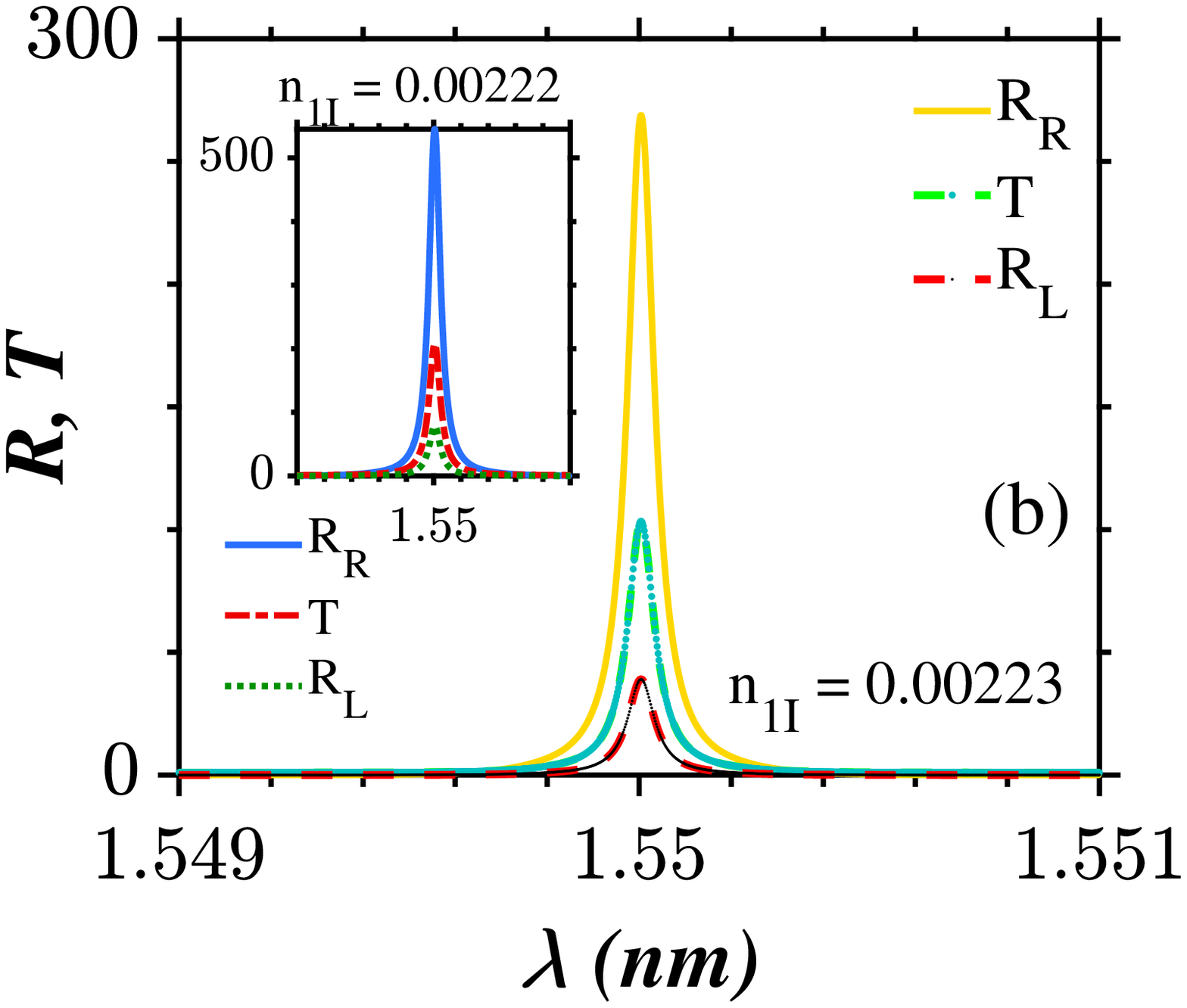}\\\includegraphics[width=0.5\linewidth]{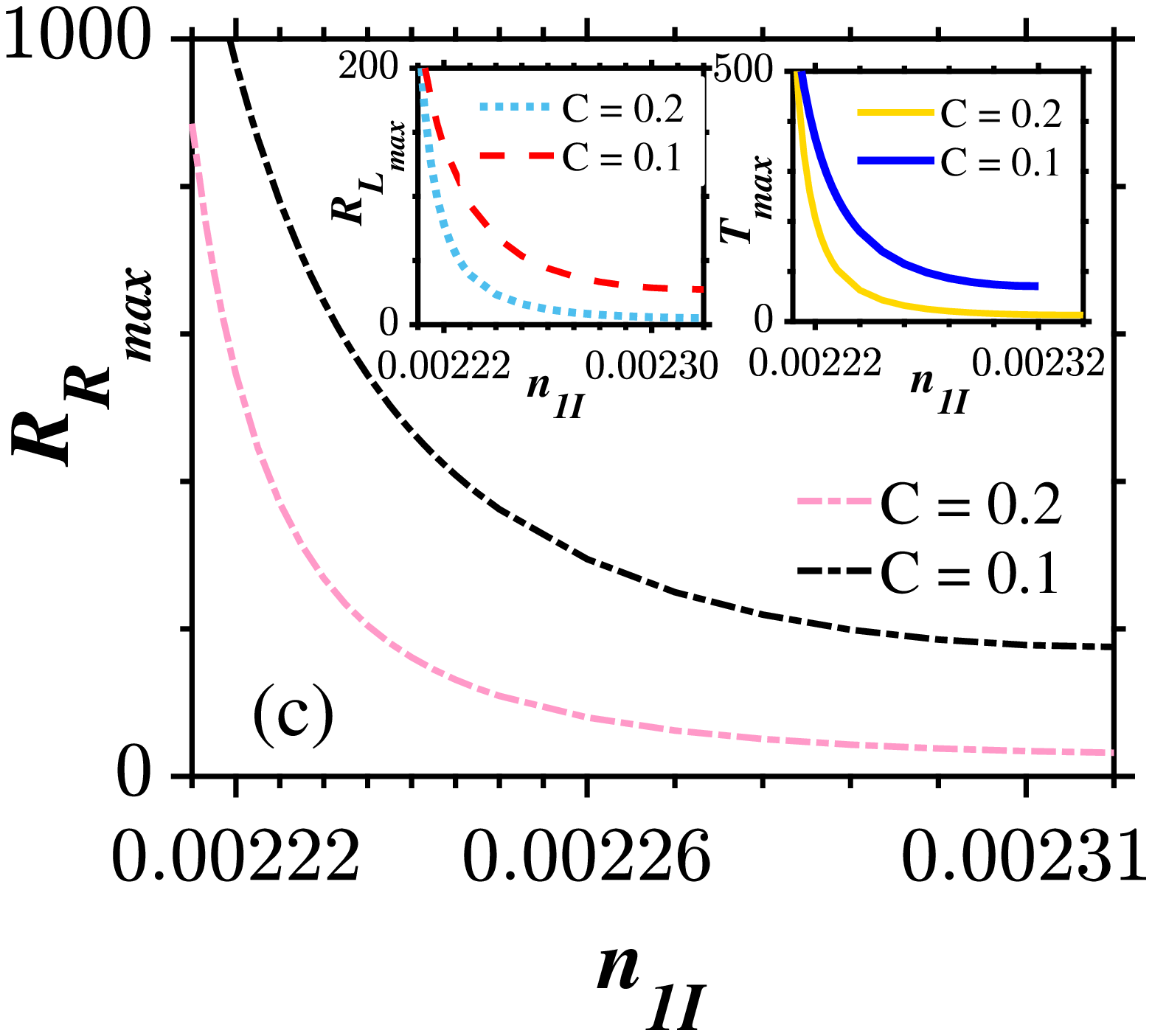}\includegraphics[width=0.5\linewidth]{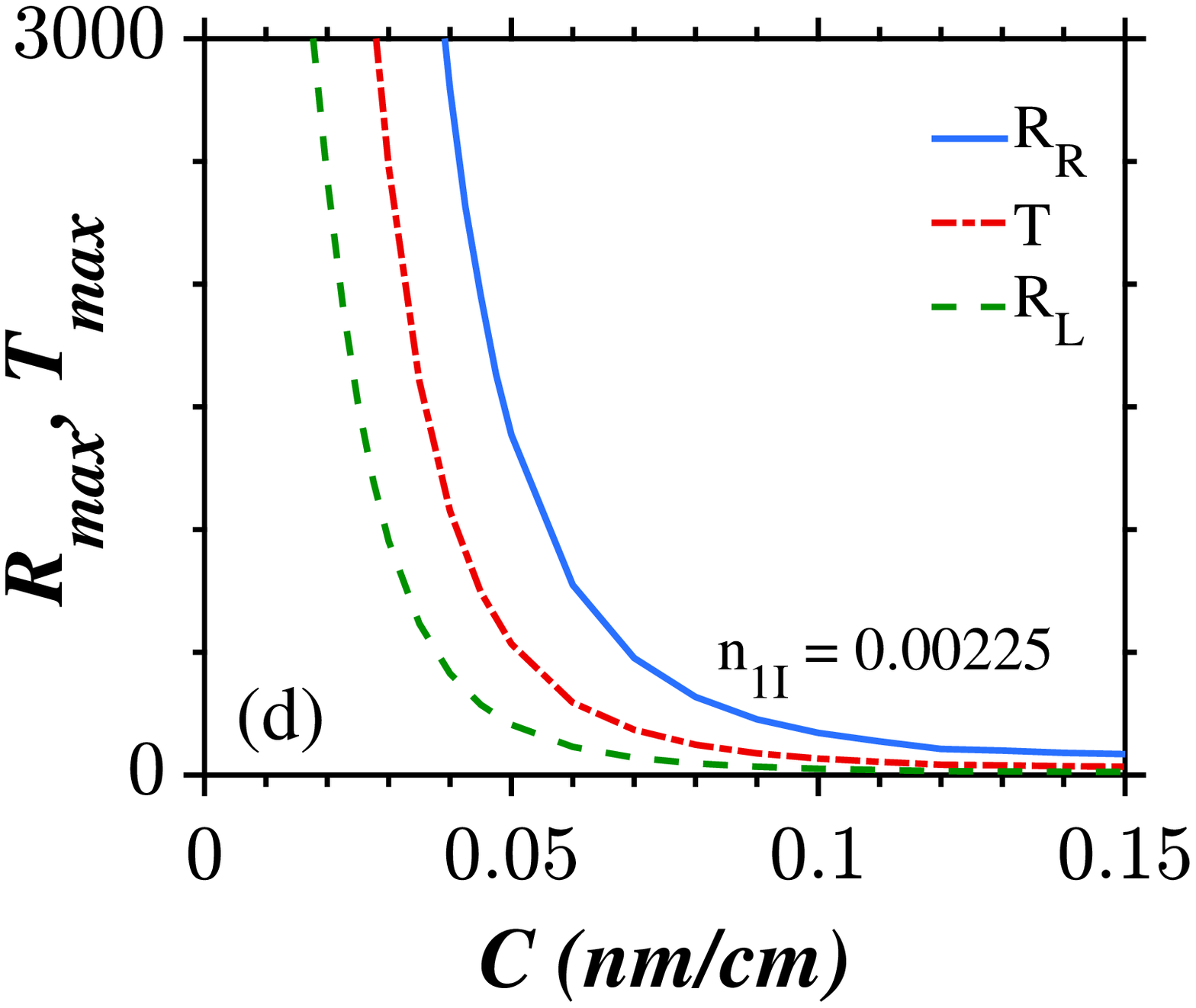}\\\includegraphics[width=0.5\linewidth]{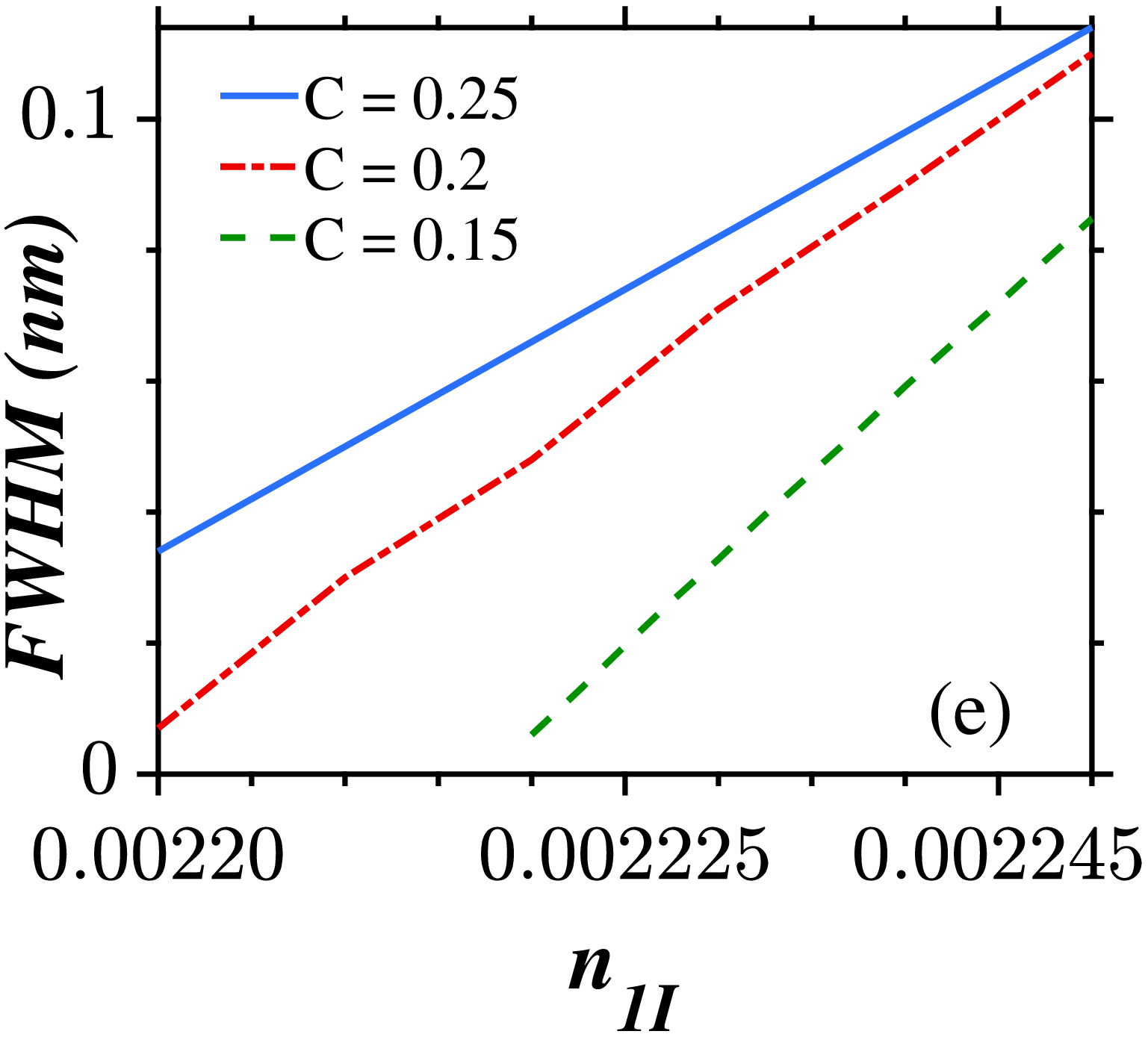}\includegraphics[width=0.5\linewidth]{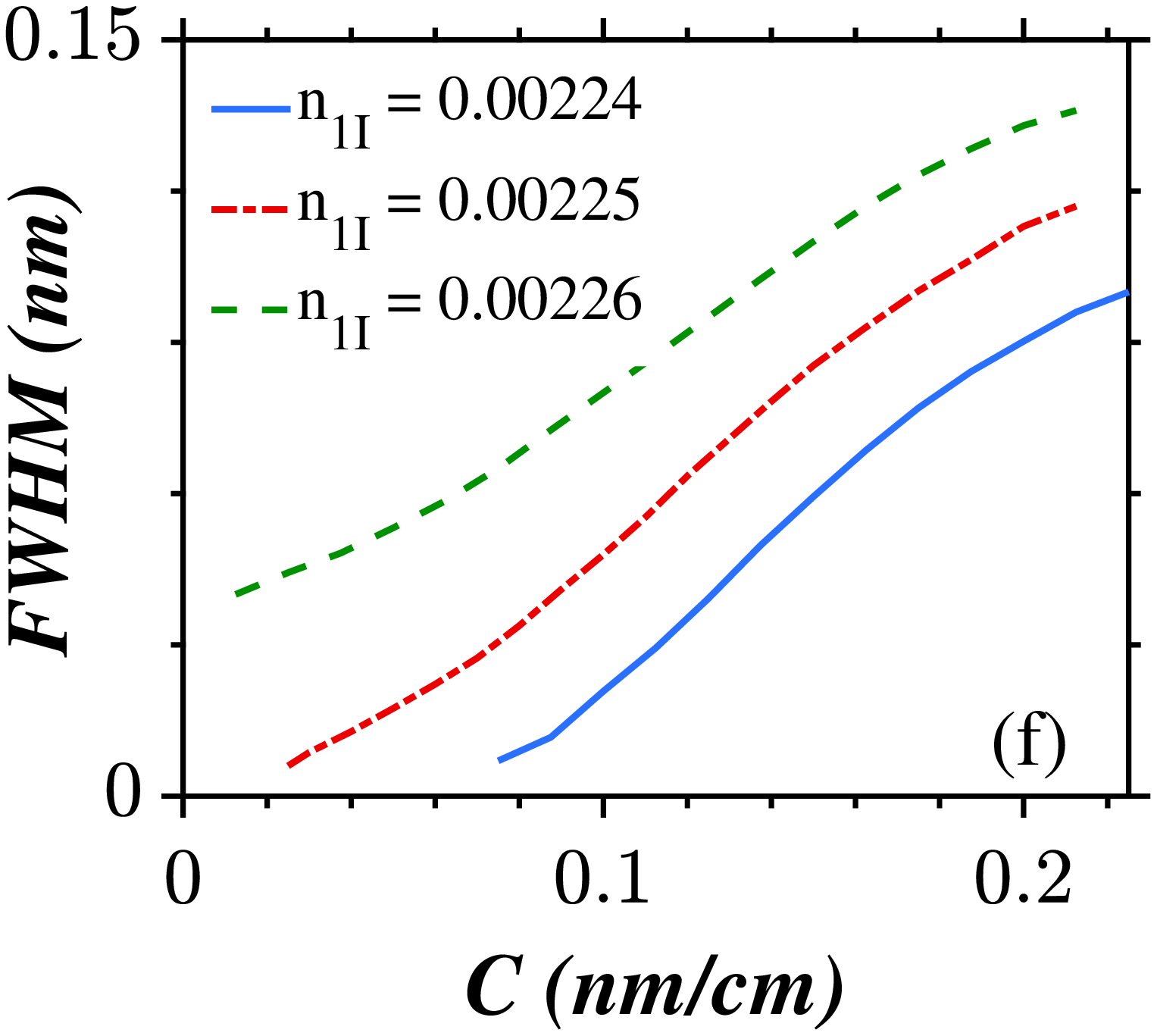}
	\caption{Single-mode lasing behavior of a broken $\mathcal{PT}$-symmetric chirped FBG with raised-cosine apodization. (a) and (b) Illustrate the spectral features at $C = 0.25$ nm/cm and 0.2 nm/cm, respectively, for different value of $n_{1I}$. The continuous variation of the maximum reflectivity and transmittivity ($R_{max}, T_{max}$) is plotted in (c) by varying $n_{1I}$ for constant chirping ($C = 0.2$ nm/cm and $0.1$ nm/cm) and (d) varying chirping at a constant $n_{1I} = 2.25 \times 10^{-3}$. The corresponding to variations in FWHM of the spectra against $n_{1I}$  and $C$ are shown in (e) and (f), respectively.} 
	\label{fig10}
\end{figure}
\label{Sec:5}
We recall that the FBG is said to be operating in the broken $\mathcal{PT}$-symmetric regime under the mathematical condition $n_{1I} > n_{1R} $. 
\subsection{Gaussian apodization profile}
It is worthwhile to mention that in a broken CAPT-FBG (Gaussian),  selective  amplification of certain wavelengths of the spectra can be achieved by varying the value of $n_{1I}$  and chirping ($C$). Under such circumstances, the system may behave like a Fabry-P\'{e}rot cavity \cite{erdogan1997fiber}. Thus it allows the amplification of certain wavelengths and the suppression of remaining wavelengths of the reflected (transmitted) spectra as shown in Figs. \ref{fig7} and \ref{fig8}. The undesirable amplification from the sidelobes of the spectra can be eliminated with ease by passing the output through a bandpass FBG filter and thus we scale down the plots in Fig. \ref{fig7} between $1549.6$ nm and $1550.4$ nm. In contrast to the previous sections (unbroken $\mathcal{PT}$-symmetry), in the present case of broken $\mathcal{PT}$-symmetry the reflectivity (R) [transmittivity (T)] can be controlled by changing the value of the chirp or gain and loss. Due to the nature of apodization, the amplification at the Bragg wavelength of the spectra is suppressed (but not to zero), but  the wavelengths lying on the shorter and longer sides of the Bragg wavelengths of the spectra experience a symmetric (resonant) amplification behavior pertaining to  blue and red shifts of the resonant frequency (wavelength) of the spectrum within the stopband as shown in Figs. \ref{fig7}(a) -- \ref{fig7}(c).  Recently, such amplification behavior was reported in a metamaterial waveguide with gain and loss \cite{tennant2019distributed,govindarajan2019}. This once again proves that the $\mathcal{PT}$-symmetric FBG is closely associated with anti-directional coupler structures with gain and loss. From Figs. \ref{fig7}(a), \ref{fig7}(b), and \ref{fig7}(d), we conclude that with increase in $n_{1I}$ there is a significant growth in reflectivity ($R$) and transmittivity ($T$) at a given value of chirping parameter ($C$). As shown in Figs. \ref{fig7}(c) and \ref{fig7}(d), when the chirping is decreased (increased)  the maximum reflectivity and transmittivity of the spectra ($R_{max}$ and $T_{max}$) gets reduced (raised). Both $n_{1I}$ and $C$ influence the wavelength ($\lambda_{{max}_{1}}$ and $\lambda_{{max}_{2}}$) at which  maximum reflectivity and transmittivity occur in the spectra as shown in Fig. \ref{fig7}(e). Also, the FWHM of the spectra grows with an increase in $n_{1I}$ as shown in the inset of Fig. \ref{fig7}(f). The wavelengths ($\lambda_{{max}_{1}}$ and $\lambda_{{max}_{2}}$) at which maximum reflectivity ($R_{max}$) or transmittivity ($T_{max}$) occurs in the spectra are shifted towards the Bragg wavelength. To illustrate the spectral response further, we introduce another useful parameter $\lambda_d$ which indicates the difference in wavelengths between $\lambda_{{max}_{1}}$ and $\lambda_{{max}_{2}}$. It is inferred from Fig. \ref{fig7}(f) that $\lambda_d$ gets shrunk with an increase in the value of $n_{1I}$. Physically, this means that with an increase in the value of $n_{1I}$ the wavelengths which were experiencing a larger gain at lower values of $n_{1I}$ now experience comparably lesser gain and wavelengths closer to the Bragg wavelength will acquire larger gain. Any further increase in the value of $n_{1I}$ leads to a narrow band single-mode lasing behavior of the spectra centered at the Bragg wavelength as depicted in Fig. \ref{fig8}. Interestingly, any increase in the value of $n_{1I}$ decreases the maximum reflectivity ($R_{max}$) and transmittivity ($T_{max}$) of the spectra in the single mode lasing regime as shown in Figs. \ref{fig8}(a), \ref{fig8}(b) and \ref{fig8}(c) which is in contrast to the dual-mode amplification regime. At the same time, the FWHM of the spectra increases with an increase in $n_{1I}$ as shown in the inset of Fig. \ref{fig8}(c). The decrease (increase) in the chirping tends to push the magnitude of $n_{1I}$ to higher (lower) values where $R_{max}$ and $T_{max}$ occurs as shown in Figs. \ref{fig8}(c) and \ref{fig8}(d). For instance, at a chirp of $C = 0.2$ nm/cm and $n_{1I} = 0.00226$ the system exhibits amplification at two different wavelengths, whereas the same system at $C = 0.25$  will exhibit single-mode lasing behavior [see Figs. \ref{fig7}(d) and \ref{fig8}(d)]. The FWHM of the spectra keeps on increasing until a particular value of $n_{1I}$ and beyond that any increase will $n_{1I}$ will break the single-mode lasing behavior in the spectra and it degenerates to produce amplification at the resonances outside the Bragg wavelength. The occurrence of alternate regimes of single-mode lasing behavior and mode selective amplification in the broken CAPT-FBG spectra is observed for a wide range of $n_{1I}$ for a given value of chirping as depicted in Figs. \ref{fig8}(e) and \ref{fig8}(f).

\subsection{ Raised-cosine apodization profile}

In Fig. \ref{fig9}(a), we observe that the amplification at other wavelengths of the spectra is totally inhibited except that the amplification  is now experienced only by a single wavelength ($1550$ nm) thanks to the presence of a raised-cosine apodization profile. Physically, this means that the resonance can occur only at the Bragg wavelength and this wavelength experiences a larger gain compared to other wavelengths and shifting of resonances with variations in gain is prohibited by the nature of apodization.  This phenomenon is observed for a wide range of $n_{1I}$ and chirping. In the previous section, we discussed about two distinct regimes (dual and single-mode) of the broken CAPT-FBG spectra depending on the range of $n_{1I}$ in which the system is operated. In the presence of raised-cosine apodization, there is no such dual mode amplification behavior. Instead, we have two distinct regimes of single-mode lasing behavior. One of the regimes (see Fig. \ref{fig9}) is characterized by increase in reflectivity and transmittivity and narrowing of the full width half maximum of the spectra with any increment in the device parameters ($C$ and $n_{1I}$) and the other regime (see Fig. \ref{fig9}) features decrement in the  reflectivity and transmittivity  and broadening of FWHM of the  spectra with an increase in the value of the system parameters. Similar to the conclusions drawn in the last section, the appearance of two distinct single-mode lasing regimes is cyclic in nature with variations in $n_{1I}$.  When $n_{1I}$ increases, we observe that the reflectivity and transmittivity grows at the given value of chirping as shown in Figs. \ref{fig9}(a) -- \ref{fig9}(c).  Nevertheless, in the other regime shown in Figs. \ref{fig10}(a) -- \ref{fig10}(c), any increment in $n_{1I}$ decreases the magnitude of the maximum reflectivity and transmittivity ($R_{max}$ and $T_{max}$). Decrement in the value of chirping reduces (increases) the  reflectivity and transmittivity in Fig. \ref{fig9}(d) [Fig. \ref{fig10}(d)]. Thus, controlling $R_{max}$ and $T_{max}$ in the single-mode lasing behavior of a CAPT-FBG (raised-cosine) is feasible in two different ways: first, by varying the chirp parameter and second, by varying the gain and loss. Also, it  is possible to control the full width half maximum of the spectra in two distinctive directions just like the  amplitude control as shown in Figs. \ref{fig9}(e), \ref{fig9}(f), \ref{fig10}(e) and \ref{fig10}(f). 
Thus the system offers unique ways to control the spectral characteristics in the broken $\mathcal{PT}$-symmetric regime compared to the unbroken $\mathcal{PT}$-symmetric regime.

\section{Unidirectional reflectionless wave transport with grating nonuniformities }
\label{Sec:6}
\begin{figure}[hthb]
	\centering
	\includegraphics[width=0.5\linewidth]{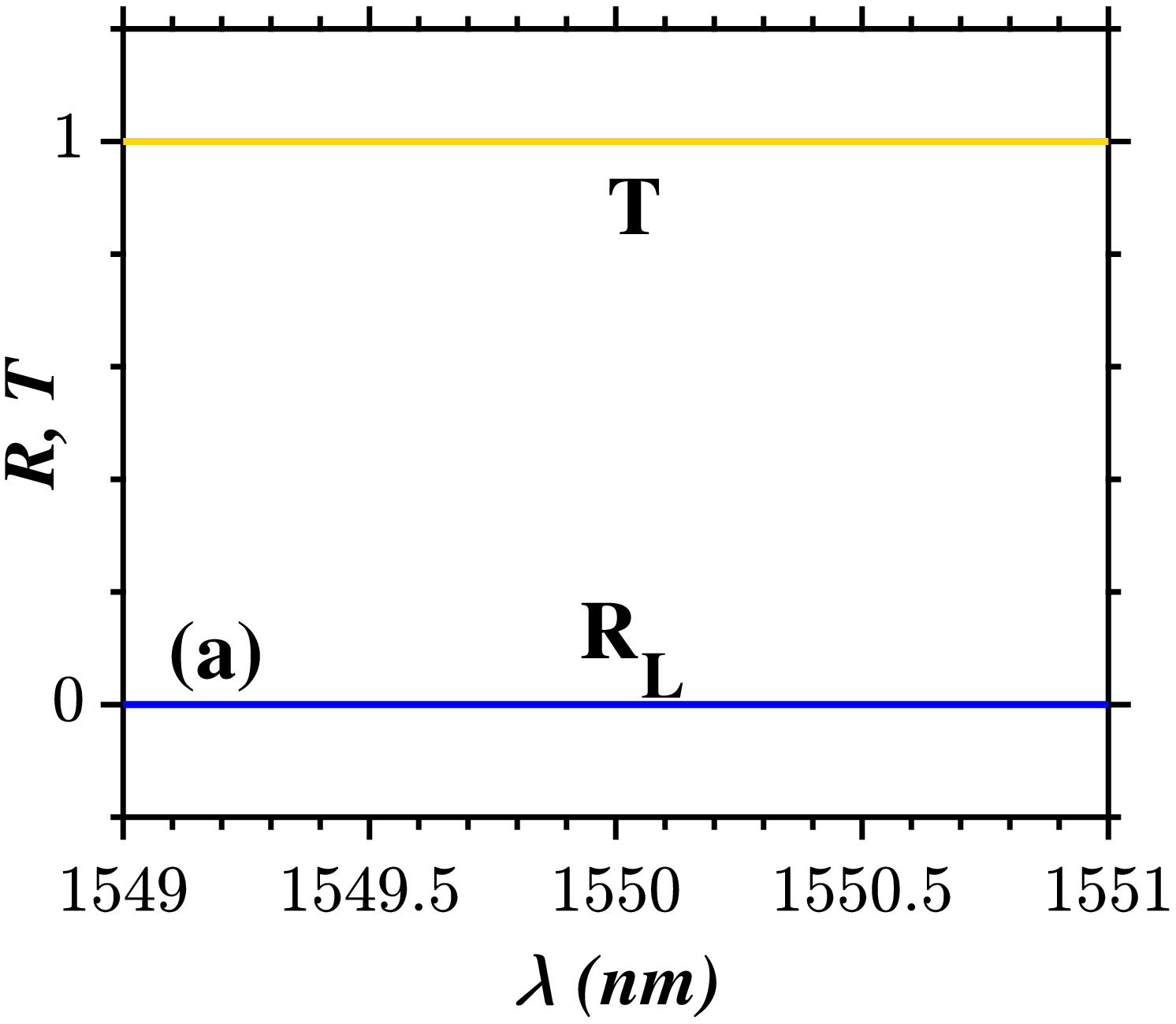}\includegraphics[width=0.5\linewidth]{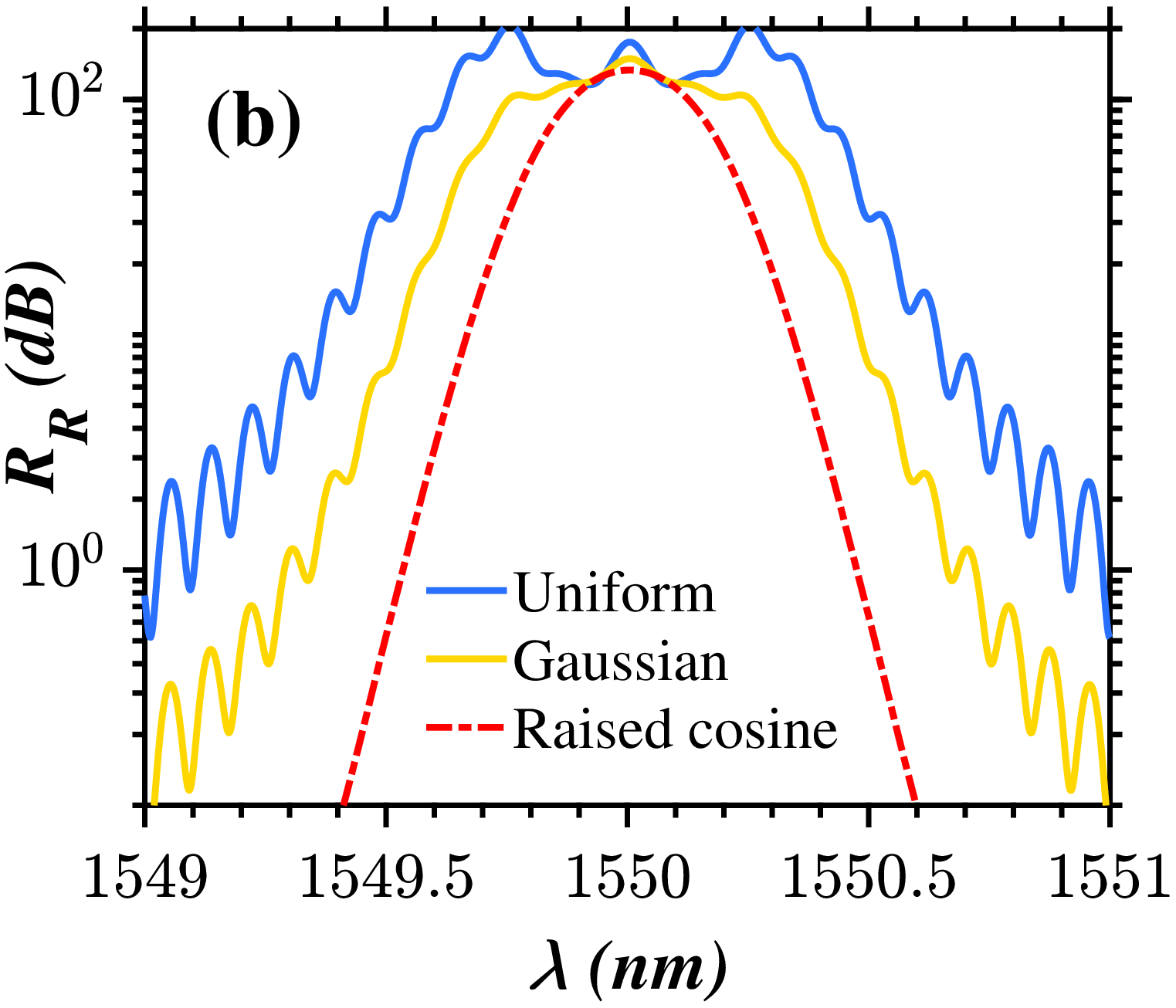}\\\includegraphics[width=0.5\linewidth]{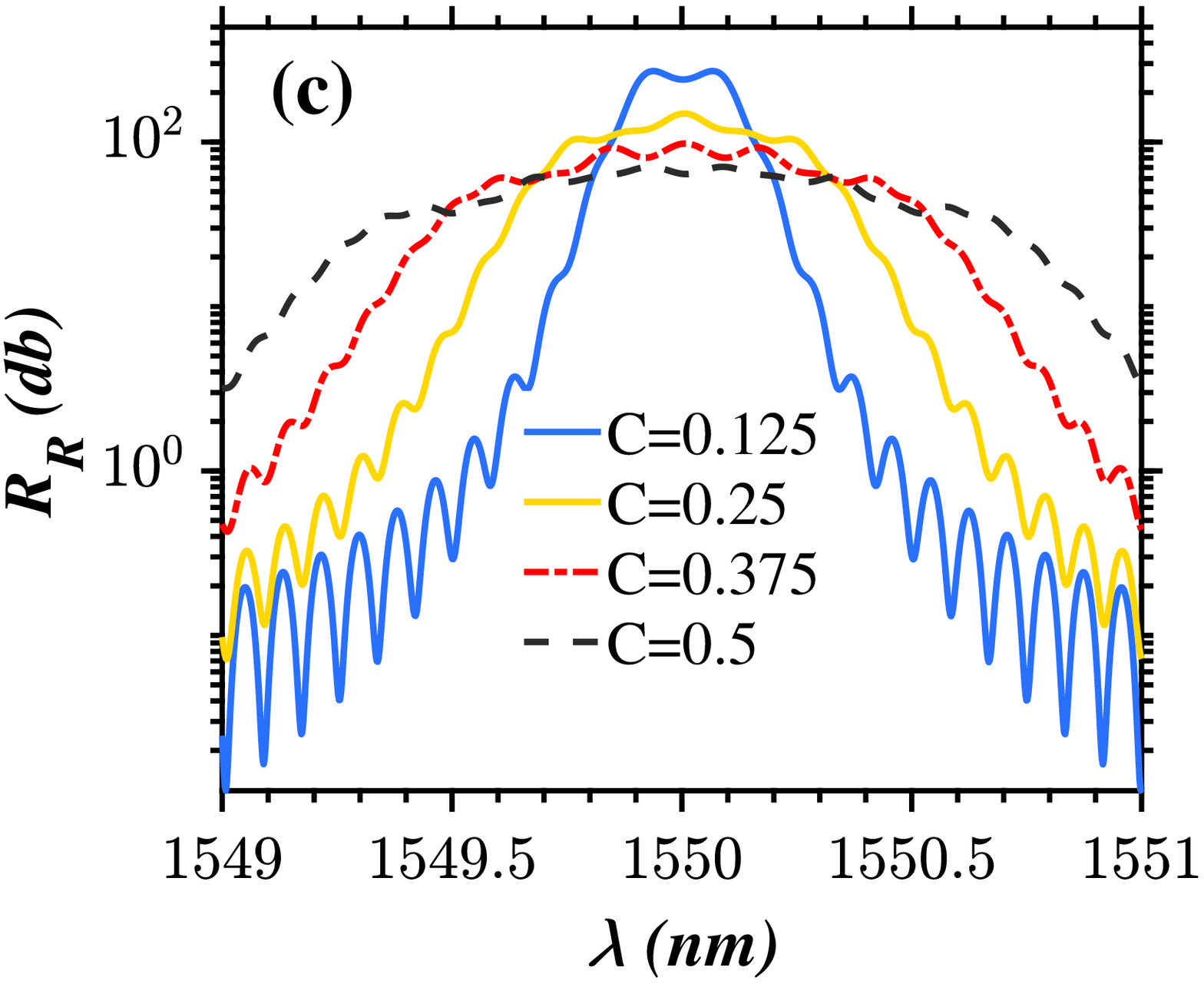}\includegraphics[width=0.5\linewidth]{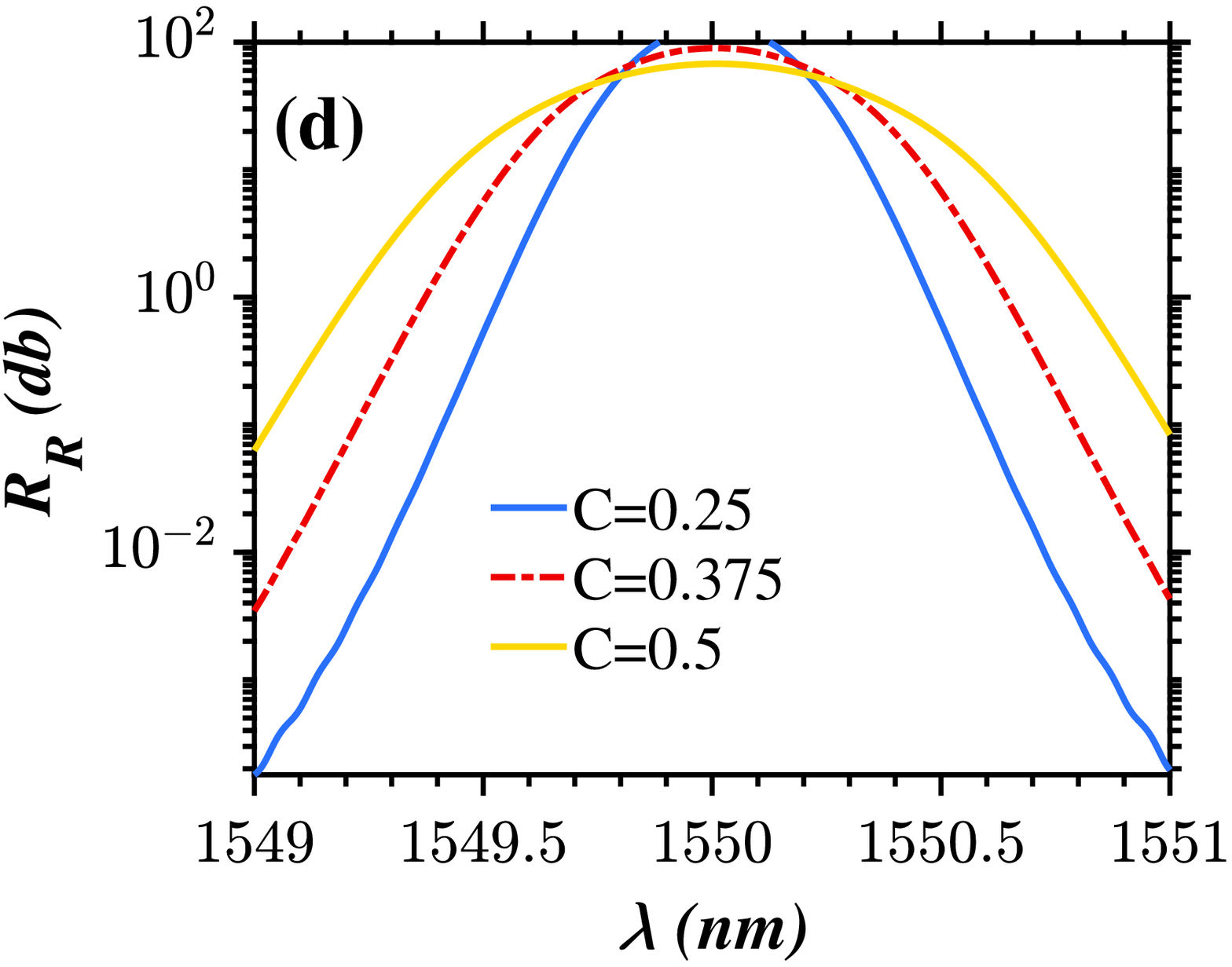}\\\includegraphics[width=0.5\linewidth]{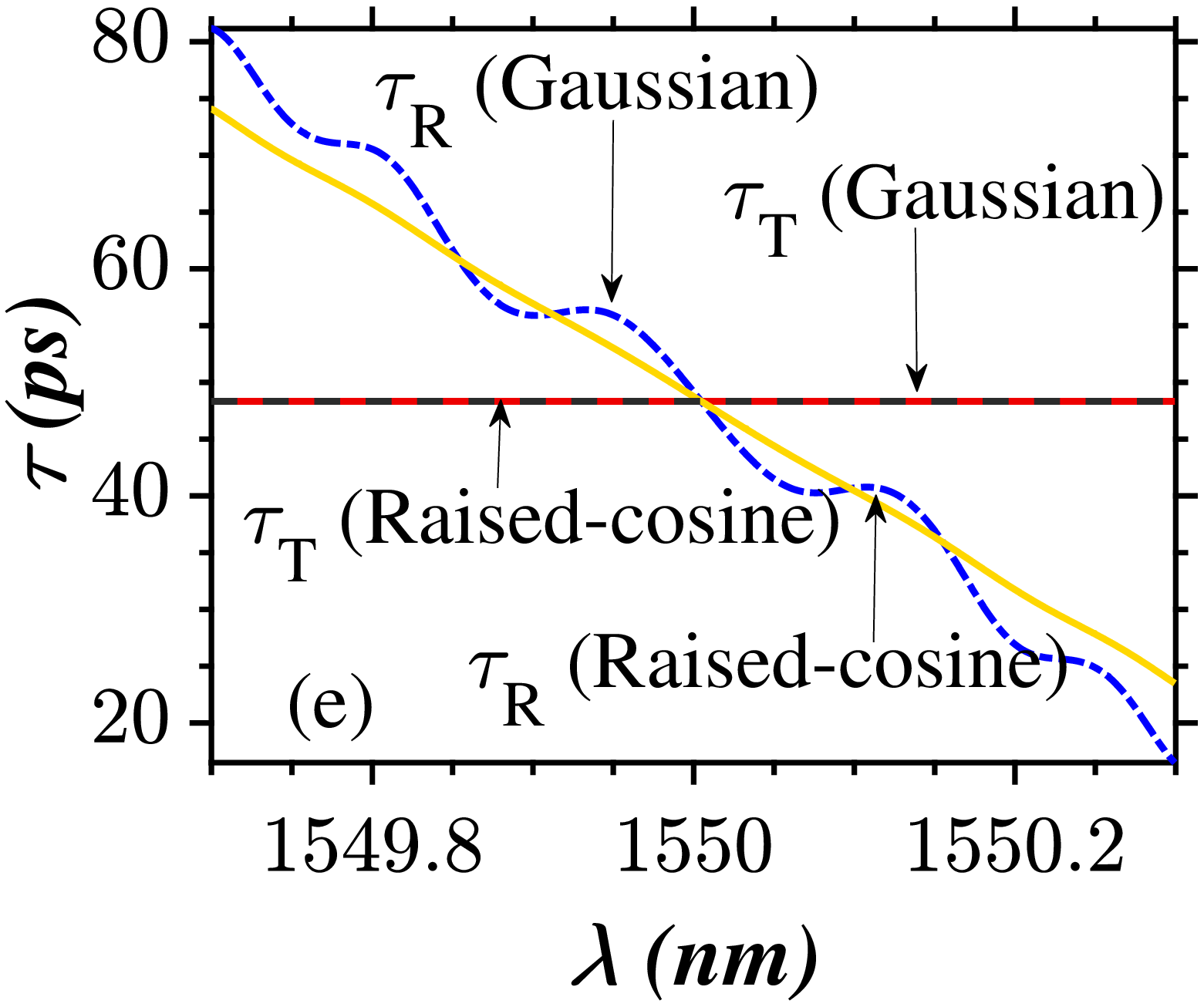}%\includegraphics[width=0.5\linewidth]{fig11f}
	\caption{Exceptional point dynamics exhibited by a chirped and apodized $\mathcal{PT}$-symmetric FBG ($n_{1R}$ = $n_{1I}$ = 0.001). The phenomenon of unidirectional invisibility is illustrated in (a). The reflected spectra (in dB) for the right incidence for different apodization profiles with $C = 0.25$ nm/cm is depicted in (b). (c) and (d) Depict the same under variations in $C$ in the presence of Gaussian apodization and raised-cosine apodization profiles respectively. (e) Depicts the variation in delay for the two apodization profiles when $C = 0.25$ nm/cm.}
	\label{fig11}	
\end{figure}

When $n_{1R} = n_{1I}$, the system is said to be operating at the exact $\mathcal{PT}$-symmetric phase. As a consequence of operating at the exceptional point, the FBG reveals a peculiar phenomenon of unidirectional reflectionless transmission which was termed as \textit{unidirectional invisibility} \cite{lin2011unidirectional, yoon2018time}  that results in an ideal light transmission on one side  ($T = 1$) and ($R = 0$) which is unrealistic in the context of conventional systems. This concept is well established in the framework of $\mathcal{PT}$-symmetric linear as well as nonlinear gratings \cite{lin2011unidirectional,lupu2016tailoring}. From our investigations, we confirm that unidirectional invisibility can persevere even in the presence of chirping and apodization as depicted in Fig. \ref{fig11}(a). On the contrary, the reflection spectrum for the right incidence is perturbed by the presence of chirping and apodization. This is evident from Figs. \ref{fig11}(b) -- \ref{fig11}(d). The effect of the nonuniformities of the grating is similar to the unbroken $\mathcal{PT}$-symmetric case discussed earlier except for an additional growth in the amplitude of the reflected light as a result of an increase in the value of $n_{1R}$. When Gaussian apodization is used, ripples are seen all over the spectra due to the presence of chirping as shown in Fig. \ref{fig11}(c). The convenient way is to make use of a raised cosine apodization window which can completely reduce the side lobes and ripples in the spectra when chirping is sufficient. We would like to stress that a chirp of 0.375 nm/cm or greater can produce a large band spectrum without any side lobes which is one of the notable outcomes of our study as illustrated in Fig. \ref{fig11}(d). In Fig. \ref{fig11}(e), we can see that the delay in the transmitted signal is constant around $48$ ps and this is due to presence of chirping and by increasing or decreasing the value of chirping the delay can be tuned in the transmission and thus can be used as delay lines. Similar to unbroken $\mathcal{PT}$-symmetry, $\tau_R$ features a lot of oscillations in the presence of Gaussian apodization  whereas a monotonically decreasing delay can be obtained by employing a raised-cosine apodization profile.

\section{Conclusions}
\label{Sec:8}
In this article, we have presented a detailed study on the spectra exhibited by a $\mathcal{PT}$-symmetric FBG in the presence of nonuniformities of the grating such as chirping and apodization. The system offers some unique features in  the presence of $\mathcal{PT}$-symmetry. We found that the spectral bandwidth varies as the chirping term is varied, and reflectivity and transmittivity get altered as the gain/loss term is tuned, while the apodization suppresses the sidelobes partially or completely 
in the unbroken $\mathcal{PT}$-symmetric regime. We also found that the role of chirping on delay characteristics depends on the direction of incidence and it is possible to obtain a flatter dispersion curve with the addition of $\mathcal{PT}$-symmetry compared to the conventional system. The spectral response of the broken $\mathcal{PT}$-symmetric FBG features dual-mode amplification and single-mode lasing behavior as a consequence of the interplay between the other system parameters in the presence of Gaussian apodization. Also, the system exhibits single-mode lasing behavior in its spectral characteristics for a wide range of system parameters in the presence of raised-cosine apodization. From the simulations, we have confirmed that the phenomenon of unidirectional invisibility is independent of the nonuniformities of the grating and the reflected light corresponding to the right light incidence is influenced by the variations in the nonuniformities of the grating. To sum up, the proposed system can offer multi-functionalities against the variations in nonuniformities of the grating together with $\mathcal{PT}$-symmetry and can yield  applications like direction dependent dispersion compensation, optical delay lines and demultiplexers in light-wave communication systems.

\section*{Acknowledgments}
SVR is indebted for the financial assistantship provided by an Anna University through Anna Centenary Research Fellowship (CFR/ACRF-2018/AR1/24). AG was supported by Science and Engineering Research Board (SERB), Government of India through a National Postdoctoral Fellowship (Grant No. PDF/2016/002933). ML is supported by SERB through a Distinguished Fellowship (Grant No. SB/DF/04/2017) in which A. G. is a Visiting Scientist. A.G. and M.L. also
thank Council of Scientific and Industrial Research (Grant No.03/1331/15/EMR-II).

\end{document}